\documentclass[11pt,twoside,a4paper,cmspaper,final,collab]{cms-tdr}
\def\svnVersion{004c946}\def\svnDate{2026/04/21}\def\cmsCernNoTag{CERN-EP-2025-240}\def\cmsCernDate{\today}\def\cmsMessage{Published in the Journal of High Energy Physics as \href{http://dx.doi.org/10.1007/JHEP04(2026)086}{\doi{10.1007/JHEP04(2026)086}.}}
\begin{document}\cmsNoteHeader{TOP-23-002}

\newlength{\cmsTabSkip}
\setlength{\cmsTabSkip}{1ex}
\newcommand{\il}{\ensuremath{138\fbinv}\xspace}
\newcommand{\fb}{\unit{fb}}
\newcommand{\ttg}{\ensuremath{\ttbar\PGg}\xspace}
\newcommand{\twg}{\ensuremath{\PQt\PW\PGg}\xspace}
\newcommand{\tw}{\ensuremath{\PQt\PW}\xspace}
\newcommand{\pp}{\ensuremath{\Pp\Pp}\xspace}
\newcommand{\mll}{\ensuremath{m(\Pell\Pell)}\xspace}
\newcommand{\mllg}{\ensuremath{m(\Pell\Pell\PGg)}\xspace}
\newcommand{\Zg}{\ensuremath{\PZ\PGg{+}\text{jets}}\xspace}
\newcommand{\Wg}{\ensuremath{\PW\PGg{+}\text{jets}}\xspace}
\newcommand{\Zj}{\ensuremath{\PZ{+}\text{jets}}\xspace}
\newcommand{\Wj}{\ensuremath{\PW{+}\text{jets}}\xspace}
\newcommand{\chiso}{\ensuremath{I_{\text{ch}}}\xspace}
\newcommand{\sieie}{\ensuremath{\sigma_{i\eta i\eta}}\xspace}
\newcommand{\Ra}{{\text{A}}\xspace}
\newcommand{\Rb}{{\text{B}}\xspace}
\newcommand{\Rc}{{\text{C}}\xspace}
\newcommand{\Rd}{{\text{D}}\xspace}
\newcommand{\Rx}{{\text{X}}\xspace}
\newcommand{\Na}{\ensuremath{N_{\Ra}}\xspace}
\newcommand{\Nb}{\ensuremath{N_{\Rb}}\xspace}
\newcommand{\Nc}{\ensuremath{N_{\Rc}}\xspace}
\newcommand{\Nd}{\ensuremath{N_{\Rd}}\xspace}
\newcommand{\Nx}{\ensuremath{N_{\Rx}}\xspace}
\newcommand{\RmisID}{\ensuremath{R_{\text{misID}}}\xspace}
\newcommand{\ratio}{\ensuremath{R_{\PGg}}\xspace}
\newcommand{\ratioexp}{\ensuremath{R^{\text{exp}}_{\PGg}}\xspace}
\newcommand{\Dphill}{\ensuremath{\Delta\phi(\Pell,\Pell)}\xspace}
\newcommand{\DRgl}{\ensuremath{\DR(\PGg,\Pell)}\xspace}
\newcommand{\DRglmin}{\ensuremath{\text{minimum }\DR(\PGg,\Pell)}\xspace}
\newcommand{\DRgt}{\ensuremath{\text{min. }\DR(\PGg,\PQt)}\xspace}
\newcommand{\DRgtt}{\ensuremath{\DR(\PGg,\ttbar)}\xspace}
\newcommand{\muR}{\ensuremath{\mu_{\mathrm{R}}}\xspace}
\newcommand{\muF}{\ensuremath{\mu_{\mathrm{F}}}\xspace}
\newcommand{\hdamp}{\ensuremath{h_{\mathrm{damp}}}\xspace}
\newcommand{\fullcorr}{\ensuremath{\checkmark}\xspace}
\newcommand{\partcorr}{\ensuremath{\sim}\xspace}
\newcommand{\nocorr}{\ensuremath{\times}\xspace}
\newcommand{\mtop}{\ensuremath{m_{\PQt}}\xspace}
\newcommand{\mZ}{\ensuremath{m_{\PZ}}\xspace}
\newcommand{\mttbar}{\ensuremath{m(\ttbar)}\xspace}
\newcommand{\DY}{\ensuremath{\text{DY}{+}\text{jets}}\xspace}
\newcommand{\boldtheta}{\ensuremath{\boldsymbol{\theta}}\xspace}
\newcommand{\Irel}{\ensuremath{I_{\mathrm{rel}}}\xspace}
\newcommand{\dytops}{\ensuremath{\Delta\abs{y}(\PQt,\PAQt)}\xspace}
\newcommand{\toppt}{\ensuremath{\pt(\PQt_{1})}\xspace}
\newcommand{\llpt}{\ensuremath{\pt(\Pell_{1})}\xspace}
\newcommand{\gammapt}{\ensuremath{\pt(\PGg)}\xspace}

\cmsNoteHeader{TOP-23-002}

\title{Inclusive and differential measurements of the \texorpdfstring{\ttg}{ttg} cross section and the \texorpdfstring{$\ttg/\ttbar$}{ttg/tt} cross section ratio in proton-proton collisions at \texorpdfstring{$\sqrt{s}=13\TeV$}{sqrt(s)=13 TeV}}

\date{\today}

\abstract{Inclusive and differential cross section measurements of top quark pair (\ttbar) production in association with a photon (\PGg) are performed as a function of lepton, photon, top quark, and \ttbar kinematic observables, using data from proton-proton collisions at $\sqrt{s}=13\TeV$, corresponding to an integrated luminosity of \il, collected at the CERN LHC with the CMS detector. Events containing two leptons (electrons or muons) and a photon in the final state are considered. The fiducial cross section of \ttg is measured to be $137\pm8\fb$, in a phase space including events with a high momentum, isolated photon. The fiducial cross section of \ttg is also measured to be $56\pm5\fb$ when considering only events where the photon is emitted in the production part of the process. Both measurements are in agreement with the theoretical predictions, of $126\pm19\fb$ and $57\pm5\fb$, respectively. Differential measurements are performed at the particle and parton levels. Additionally, inclusive and differential ratios between the cross sections of \ttg and \ttbar production are measured. The inclusive ratio is found to be $0.0133\pm0.0005$, in agreement with the standard model prediction of $0.0127\pm0.0008$. The top quark charge asymmetry in \ttg production is also measured to be $-0.012\pm0.042$, compatible with both the standard model prediction and with no asymmetry.}

\hypersetup{
pdfauthor={CMS Collaboration},
pdftitle={Inclusive and differential measurements of the ttg cross section and the ttg/tt cross section ratio in proton-proton collisions at sqrt(s)=13 TeV},
pdfsubject={CMS},
pdfkeywords={CMS, ttbar, ttgamma, ratio, differential, charge asymmetry}}

\maketitle

\section{Introduction}

Precision measurements of top quark production can be used to test the standard model (SM) of particle physics. The associated production of a top quark-antiquark pair (\ttbar) with a photon (\PGg), is of particular interest as it is sensitive to new phenomena in the top quark electromagnetic coupling to the photon. Unlike other processes that offer sensitivity to new phenomena only at higher orders in perturbation theory~\cite{Baur:2004uw,Bouzas:2012av,Schulze:2016qas}, in \ttg production these effects are present already at leading order (LO) in perturbative quantum chromodynamics (QCD). New \PQt-\PGg interactions can be targeted in dedicated differential cross section measurements, for example, as a function of the photon momentum. Moreover, differential measurements provide useful information on the modelling of the \ttg production process.

\begin{figure}[!htp]
\centering
\includegraphics[width=0.31\textwidth]{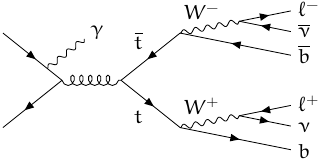}%
\hfill%
\includegraphics[width=0.31\textwidth]{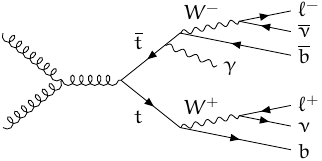}%
\hfill%
\includegraphics[width=0.31\textwidth]{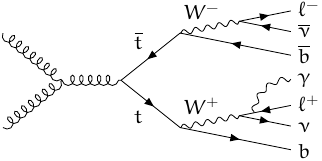}
\caption{Example Feynman diagrams for the production of \ttg, where both top quarks decay leptonically. The photon can be emitted from the initial state (left), from an off-shell top quark (centre), or from a top quark decay product (right).}
\label{feynman}
\end{figure}

In \ttg events, the photon can originate from initial-state radiation (ISR), intermediate-state particles, such as the top quarks, or final-state radiation (FSR) from the top quark decay products. Some examples of Feynman diagrams for \ttg production at LO in QCD are shown in Fig.~\ref{feynman}, with the top quarks decaying leptonically. Events consisting of a \ttbar pair with a photon emission during the hadronization stage are commonly treated as background. It has been shown in theoretical studies~\cite{Bevilacqua:2018woc,Bevilacqua:2019quz} that, in the narrow-width approximation, the cross section of the \ttg process with a high-momentum photon can be factorized into two parts: the first part describes the process where a photon is emitted from ISR or from an off-shell top quark (referred to as \ttg production), and the second part describes the process where a photon is emitted from an on-shell top quark or from the top quark decay products (referred to as \ttg decay). While the \ttg production process is the one offering sensitivity to the \PQt-\PGg coupling, and therefore the most interesting from a physics perspective, both processes have similar experimental signatures and thus the analysis reported in this paper aims to measure both. Single top quark production in association with a W boson and a photon (\twg) is another process that can be categorized in the same way into production and decay. In the following, it is treated as an irreducible background.

The interference between diagrams where the photon is emitted from one of the initial-state quarks and diagrams where it is radiated from one of the top quarks introduces a charge asymmetry in \ttg production at LO in QCD\@. This asymmetry induces an anisotropy in the rapidity distributions of the top quark and antiquark, where the top quark tends to scatter more centrally than the antiquark. A similar effect, but with gluon emission and in the opposite direction, is also present in \ttbar production, but the latter takes place only at next-to-LO (NLO) in QCD\@. The charge asymmetry in \ttg production can be more pronounced than that of \ttbar production~\cite{Bergner:2018lgm,Aguilar-Saavedra:2014vta}, partly because the contribution of quark-initiated production in the \ttg process is enhanced by about 60\% (at LO) when compared to \ttbar~\cite{Stremmer:2023kcd}.

Regardless of the photon origin, top quarks in \ttg events are mainly produced via QCD interactions, through similar diagrams as the standard \ttbar production at the CERN LHC\@.
The ratio of the cross section of \ttg to that of \ttbar, \ratio, benefits from the cancellation of several systematic uncertainties and yields an observable that can probe theory predictions more precisely. Differential measurements of \ratio bring additional insight into the \ttg process~\cite{Bevilacqua:2018dny}, as new physics effects can depend on the phase space. Such measurements can be used to place strong constraints on new \PQt-\PGg couplings in the context of SM effective field theory~\cite{Aguilar-Saavedra:2018ksv,Schulze:2016qas}.

The \ttg process was observed for the first time by the CDF Collaboration at the Fermilab Tevatron~\cite{cdf}. It was later observed at the LHC by the ATLAS Collaboration at $\sqrt{s}=7\TeV$~\cite{atlas7tev}, and by both the ATLAS and CMS Collaborations at $8\TeV$~\cite{atlas8tev,cms8tev}. At $\sqrt{s}=13\TeV$, the ATLAS and CMS Collaborations have each measured the inclusive \ttg cross section, as well as differential distributions of lepton and photon observables~\cite{CMS:2022lmh,CMS:2021klw, ATLAS:2018sos,ATLAS:2020yrp,ATLAS:2024hmk}. The ATLAS Collaboration has also performed a measurement of the charge asymmetry in \ttg events~\cite{ATLAS:2022wec}. To date, no differential measurements have been reported for \ratio, nor for \ttg as a function of observables related to the kinematic properties of the top quarks.

In this paper, we report the first differential measurements of the \ttg cross section as a function of top quark and \ttbar system observables, using proton-proton (\pp) collision data recorded with the CMS detector at $\sqrt{s}=13\TeV$ between 2016 and 2018, corresponding to an integrated luminosity of \il. Inclusive and differential \ratio measurements are also presented, and the top quark charge asymmetry is determined in \ttg events. The measurements are performed using events with two oppositely charged leptons and a photon. In this paper, leptons refer to electrons or muons, unless otherwise specified. Events with \PGt leptons are only considered if the \PGt leptons decay into electrons or muons passing the selection criteria. Throughout the paper, the designation ``top quark'' includes both the quark and the antiquark, unless otherwise indicated. These results are also tabulated in the HEPData record for this analysis~\cite{hepdata}.

The paper is organized as follows. In Section~\ref{sec:cms}, an overview of the CMS detector is given. Subsequently, in Section~\ref{sec:sim}, the data sets and Monte Carlo (MC) simulation samples used for the analysis are described. The criteria to reconstruct and select the physics objects and to select the events are discussed in Sections~\ref{sec:obj} and~\ref{sec:sel}, respectively. Section~\ref{sec:bkg} describes the methods used to estimate the background contributions, focusing on the backgrounds arising from misidentified photons and photons produced in hadron decays, using control samples in data. The systematic uncertainties are reported in Section~\ref{sec:sys}. The statistical analysis is introduced in Section~\ref{sec:stat}, while the results for the inclusive and differential cross section measurements are presented in Section~\ref{sec:meas}. The results for \ratio are reported in Section~\ref{sec:ratio}. Section~\ref{sec:charge_asym} presents the top quark charge asymmetry measurement in \ttg events, and a summary of the paper is given in Section~\ref{sec:sum}.

\section{The CMS experiment}
\label{sec:cms}

The central feature of the CMS apparatus is a superconducting solenoid of 6\unit{m} internal diameter, providing a magnetic field of 3.8\unit{T}. Within the solenoid volume are a silicon pixel and strip tracker, a lead tungstate crystal electromagnetic calorimeter (ECAL), and a brass and scintillator hadron calorimeter (HCAL), each composed of a barrel and two endcap sections. The ECAL consists of 75\,848 lead tungstate crystals, which provide coverage in absolute pseudorapidity $\abs{\eta} < 1.48$ in the barrel region (EB) and $1.48 < \abs{\eta} < 3.0$ in the two endcap regions (EE).
Preshower detectors consisting of two planes of silicon sensors interleaved with a total of three radiation lengths of lead are located in front of each EE detector.

Forward calorimeters extend the pseuddorapidity coverage provided by the barrel and endcap detectors. Muons are measured in gas-ionization detectors embedded in the steel flux-return yoke outside the solenoid, using three technologies: drift tubes, cathode strip chambers, and resistive plate chambers. A more detailed description of the CMS detector, together with a definition of the coordinate system used and the relevant kinematic variables, can be found in Refs.~\cite{CMS:2008xjf,CMS:2023gfb}.

Events of interest are selected using a two-tiered trigger system. The first level (L1), composed of custom hardware processors, uses information from the calorimeters and muon detectors to select events at a rate of around 100\unit{kHz} within a fixed latency of about 4\mus~\cite{CMS:2020cmk}. The second level, known as the high-level trigger (HLT), consists of a farm of processors running a version of the full event reconstruction software optimized for fast processing, and reduces the event rate to a few kHZ before data storage~\cite{CMS:2016ngn,CMS:TRG-19-001}.

\section{Data and simulated samples}
\label{sec:sim}

Events from \pp collision data are selected using a combination of single- and double-lepton triggers. The online selection requirements on leptons depend on the trigger type and vary according to the data-taking conditions. The threshold on transverse momentum \pt ranges from 27 (24)\GeV to 35 (27)\GeV for electrons (muons) in single-lepton triggers. In dilepton triggers, the \pt of the leading (subleading) lepton must exceed the threshold of 23 (12) and 17 (8)\GeV in the dielectron and dimuon paths, respectively. In the electron-muon paths, the leading lepton must have \pt above 23\GeV, while the subleading lepton must have \pt above 12\GeV if it is an electron, or above 8\GeV if it is a muon.

Monte Carlo simulations are used to estimate the contributions from both signal and background processes. Since the CMS detector configuration has undergone some changes between 2016 and 2018, samples are simulated for different data-taking periods separately. The \ttg production process is modelled as a $2{\to}3$ process ($\pp\to\ttg$) with two on-shell top quarks using the \MGvATNLO (v2.9.18)~\cite{Alwall:2014hca} MC generator, at NLO accuracy in QCD, with up to one additional jet at LO\@. Only photons with $\pt>10\GeV$ and $\abs{\eta}<2.6$ are accepted. They must have an angular separation from other generated particles of $\DR>0.05$, where $\DR=\sqrt{\smash[b]{(\Delta\eta)^2+(\Delta\phi)^2}}$, and $\phi$ is azimuthal angle, as defined in Ref.~\cite{Frixione:1998jh}, with $\varepsilon_{\PGg}=1$, and $n=1$. Top quark decays are simulated via \textsc{madspin}~\cite{Artoisenet:2012st}. A similar procedure is followed to simulate \twg production events at NLO accuracy in QCD, except that only one on-shell top quark must be present at matrix-element (ME) level, and no additional jets are simulated at ME level. The \twg and \ttg samples contain some overlap, which is removed using the diagram removal (DR) method~\cite{Frixione:2019fxg}.

To model the \ttg decay process, a sample simulating the $2{\to}2{\to}7$ ($\pp\to\ttbar\to\Pell^{+}\Pell^{-}\PGn\PAGn\bbbar\PGg$) process at LO accuracy in QCD is produced with \MGvATNLO, and the cross section is scaled to the NLO value of 2220.37\fb using a $K$-factor of 1.48, computed in Ref.~\cite{CMS:2021klw}. Simulated events in this sample contain a photon with $\pt>10\GeV$ and $\abs{\eta}<5$, isolated from other particles within a fixed cone of radius $R=0.1$. An alternative model for the \ttg decay process is provided by a \ttbar sample produced with the \POWHEG (v2.0) MC event generator~\cite{Nason:2004rx,Frixione:2007vw,Alioli:2010xd} at NLO accuracy in QCD and filtered for events with a ``signal'' photon, generated in the parton shower (PS), emitted from particles involved in the decay of one of the top quarks, before hadronization. ``Signal'' photons are defined as photons with generated $\pt>20\GeV$, isolated from nearby particles, and not emitted in hadron decays. The sample produced with \MGvATNLO is taken as the reference, and is designated as the ``nominal sample'' throughout the paper, but all the differential results are also compared with the \POWHEG model, designated as the ``alternative sample''.

The \ttbar process with no additional photons is simulated using the same sample produced with \POWHEG, mentioned above, but vetoing any ``signal'' photons in the event. For this sample, the inclusive \ttbar cross section (before any photon veto) is scaled to the most precise theoretical prediction at next-to-NLO (NNLO) accuracy in QCD, $832\pm42\unit{pb}$, which is calculated using the \textsc{top++} (v2.0) program~\cite{Czakon:2011xx}. The calculation also includes the resummation of next-to-next-to-leading-logarithm (NNLL) soft-gluon terms~\cite{Beneke:2011mq,Cacciari:2011hy,Czakon:2012zr,Czakon:2012pz,Czakon:2013goa}. The \ttbar and \ttg decay simulated events are reweighted in order to account for a known mismodelling of the top quark \pt, $\pt(\PQt)$, in the simulation at NLO, due to missing higher order corrections. This procedure is also referred to as ``NNLO QCD reweighting''~\cite{Czakon:2017wor}, and is performed using \textsc{fastnlo} tables~\cite{Czakon:2017dip,Czakon:2015owf,Czakon:2016dgf} to compute the $\pt(\PQt/\PAQt)$ distribution in fixed-order QCD at NNLO\@.

The \twg decay process is modelled with a \tw sample also produced with the \POWHEG event generator and filtered in the same way as the \ttbar sample. The \tw process without additional photons is obtained using the same \tw sample, but vetoing events with ``signal'' photons. Other processes involving top quarks are produced with \MGvATNLO at LO or at NLO accuracy in QCD, depending on the process.

The \MGvATNLO event generator is used for the remaining background processes, except for diboson processes with massive vector bosons ($\PZ\PZ$, $\PW\PZ$, and $\PW\PW$) processes, which are generated at LO accuracy in QCD with \PYTHIA (v8.240)~\cite{Sjostrand:2014zea}. The production of Drell--Yan events with up to four additional jets (\DY) is modelled at LO for dilepton invariant masses of $10<\mll<50\GeV$ and at NLO for $\mll>50\GeV$. The associated production of a \PZ (\PW) boson and a photon, with up to one additional jet, \Zg (\Wg), as well as triboson processes are generated at NLO\@. The overlap between \Zg and \DY, and \Wg and \Wj, is removed, by requiring that the events in the \DY and \Wj samples have no generated photons within the phase spaces generated in the \Zg and \Wg samples, respectively.

For all simulated samples involving top quarks, the top quark mass is set to 172.5\GeV~\cite{CMS:2015lbj}. All processes are interfaced with \PYTHIA using the CP5 tune~\cite{Skands:2014pea,Khachatryan:2015pea,CMS:GEN-17-001} to model the hadronization, PS, and underlying event. The NNPDF parton distribution functions (PDFs) version 3.1~\cite{Ball:2017nwa} at NNLO are used. The MLM~\cite{Alwall:2007fs} (FxFx~\cite{Frederix:2012ps}) matching scheme is used at LO (NLO) to account for the double-counting of jets from the ME calculations and the PS\@. Generated events are passed through the full description of the CMS detector, implemented in \GEANTfour~\cite{Agostinelli:2002hh}. Multiple simultaneous \pp collisions occur within the same bunch crossing (about 27--38 on average in 2016--2018). To model the effect of pileup caused by these additional collisions, as well as collisions from bunch crossings nearby in time, minimum bias interactions are simulated and superimposed on the hard-scattering events. Simulated events are then reweighted to reproduce the distribution of the number of interactions in each bunch crossing~\cite{Sirunyan:2018nqx} observed in data.

\section{Object reconstruction and selection}
\label{sec:obj}

The signature of the \ttg signal consists of a high-\pt photon, together with the decay products of \ttbar in the dilepton final state, \ie, two oppositely-charged leptons (\Pe or \PGm), two jets originating from \PQb quarks, and a \pt imbalance due to the presence of neutrinos.

Events are required to contain a primary vertex (PV), taken to be the vertex in which the \pt sum of the associated physics objects is the largest, as described in Section~9.4.1 of Ref.~\cite{CMS-TDR-15-02}. The particle-flow (PF) algorithm~\cite{Sirunyan:2017ulk} aims to reconstruct and identify each individual particle in an event, with an optimized combination of information from the various elements of the CMS detector.

The energy of electrons is evaluated from a combination of the electron momentum at the PV as determined by the tracker, the energy of the corresponding ECAL cluster, and the energy sum of all bremsstrahlung photons spatially compatible with originating from the electron track. The momentum resolution for electrons with $\pt\approx45\GeV$ from $\PZ\to\EE$ decays ranges from 1.6 to 5.0\%, depending on the electron $\eta$~\cite{Electrons:2021,CMS-DP-2020-021}. Electrons are selected with $\pt>25\GeV$ if they are the leading lepton in the event, and with $\pt>20\GeV$ otherwise. They are required to lie within $\abs{\eta}<2.5$, excluding candidates that are reconstructed in the ECAL superclusters (SC) located in the transition region between the barrel and endcap, $1.44<\abs{\eta_{\mathrm{SC}}}<1.57$. Electron identification requirements are based on the shower shape, the track-cluster matching, the hadronic over electromagnetic energy ratio, and the incompatibility with originating from a photon conversion. The isolation variable \Irel is calculated by summing the transverse energy deposited by other particles in a cone of radius $\DR=0.3$ centered around the lepton, normalized to the lepton \pt. The contribution of charged particles from pileup is suppressed by requiring the charged particles to be associated with the PV\@. An average pileup energy is subtracted from the total energy of neutral particles and photons within the isolation cone. To select electron candidates, \pt- and $\eta$-dependent maximum values are set on \Irel in the range of 5--10\%. A set of identification and isolation criteria are chosen such that genuine electrons are selected with an efficiency of about 70\%~\cite{Electrons:2021}.

Muons are measured in the range $\abs{\eta}<2.4$, and their momentum is obtained from the curvature of the corresponding track. Matching muons to tracks measured in the silicon tracker results in a relative \pt resolution, for muons with \pt up to 100\GeV, of 1\% in the barrel and 3\% in the endcaps~\cite{CMS:2018rym}. Muons are selected with $\pt>25$ (15)\GeV, if they are the leading (subleading) lepton in the event. Identification requirements are based on the quality of the geometrical matching between the measurements of the tracker and the muon system. The identification efficiency varies between 95 and 99\%, depending on $\eta$, where the data and simulation agree within 1--3\%~\cite{CMS:2018rym}. The misidentification rate for pions and kaons was found to be below 0.3\%. Muons are further required to be isolated with $\Irel<0.15$ where the isolation cone is of radius $\DR=0.4$.

Photons are selected with $\abs{\eta}<2.5$, excluding the ECAL transition region, and with $\pt>20\GeV$, where the energy of the photon is obtained from the ECAL measurement~\cite{ECALpaper}. Photon identification is based on the shower shape information, the hadronic to electromagnetic energy ratio, and a \pt- and $\eta$-dependent isolation from nearby charged and neutral reconstructed particles. A ``medium'' identification working point that is 80\% efficient for selecting genuine photons is used~\cite{Electrons:2021,CMS-DP-2017-004,CMS-DP-2020-037}. Photon candidates are rejected if a track in the silicon tracker is found to be compatible with the photon cluster (pixel seed veto), thus reducing the number of electrons misidentified as photons. Finally, photons must be well separated from all selected leptons in the event, fulfilling the criterion $\DRgl>0.4$.

Jets are reconstructed by clustering the PF candidates using the anti-\kt algorithm~\cite{antikt,Cacciari:2011ma} with a distance parameter of 0.4. The energy of neutral hadrons is obtained from the corresponding corrected ECAL and HCAL energies. The energy of charged hadrons is determined from a combination of their momentum measured in the tracker and the matching ECAL and HCAL energy deposits, corrected for the response function of the calorimeters to hadronic showers. Contamination from pileup and electronic noise is subtracted using the charged-hadron subtraction method~\cite{Sirunyan:2017ulk}. An $\eta$-dependent uncertainty is added to the jet energy in simulated events to better describe the data resolution~\cite{CMS-PAS-JME-16-003,CMSJetPaper}. Jets are selected with $\pt>30\GeV$ and $\abs{\eta}<2.4$, and are required to be away from leptons by $\DR(\Pell,\text{jet})>0.4$, and photons by $\DR(\PGg,\text{jet})>0.4$.

The \textsc{DeepJet} \PQb tagging algorithm~\cite{CMS:2017wtu, Bols:2020bkb, CMS:DP-2023-005} is used to identify jets originating from \PQb quarks in \ttbar decays. The medium working point of the algorithm is chosen, corresponding to an efficiency of 80\% and a misidentification rate of 1\% for light-quark and gluon jets.

The missing transverse momentum vector \ptvecmiss is computed as the negative vector sum of the transverse momenta of all the PF candidates in an event, and its magnitude is denoted as \ptmiss~\cite{CMS:2019ctu}, and it is a measure of the momentum imbalance in an event. This quantity is used to reconstruct the \ttbar system in the events, as described in Section~\ref{sec:sel}. The \ptvecmiss is modified to account for corrections to the energy scale of the reconstructed jets in the event. This observable is sensitive to pileup interactions, inefficiencies in reconstruction, energy calibrations, and detector effects.

Differences between data and simulation arising in lepton reconstruction, identification, and triggering efficiencies, the energy scale and resolution of jets, and the response of the \textsc{DeepJet} algorithm are accounted for through corrections applied to the simulated samples. These corrections are typically at the level of a few percent~\cite{CMS:2016lmd,CMS:2019ctu,CMS:2017wtu} and are measured using a variety of SM processes, such as $\PZ\to\EE$, $\PZ\to\MM$, \ttbar, and $\PGg{+}\text{jets}$ production.

\section{Event reconstruction and selection}
\label{sec:sel}

Following the HLT selection of events, a variety of offline requirements are made. Each event must have at least two charged leptons, and the two highest \pt (leading) leptons are required to have opposite charge and an invariant mass \mll greater than 40\GeV. Events where the two leading leptons are of the same flavour are rejected if \mll is compatible, within a $\pm$15\GeV window, with the mass of the \PZ boson, \mZ. Events are then required to have exactly one photon satisfying the criteria defined in Section~\ref{sec:obj}. For same-flavour lepton events, the invariant mass of the photon and two leading leptons, \mllg, must not be compatible with originating from a \PZ boson, $\mZ\pm15\GeV$. This is to reject the background from the $\PZ\to\Pell\Pell\PGg$ process. At least two jets must be present in the event, where at least one is identified as a \PQb jet. No explicit requirement on \ptmiss is applied. The \ttg events with only one generated lepton that pass the selection are considered as background in the analysis and grouped with the ``Others'' category in the figures, since their contribution is rather small.

\begin{figure}[!ht]
\centering
\includegraphics[width=0.495\textwidth]{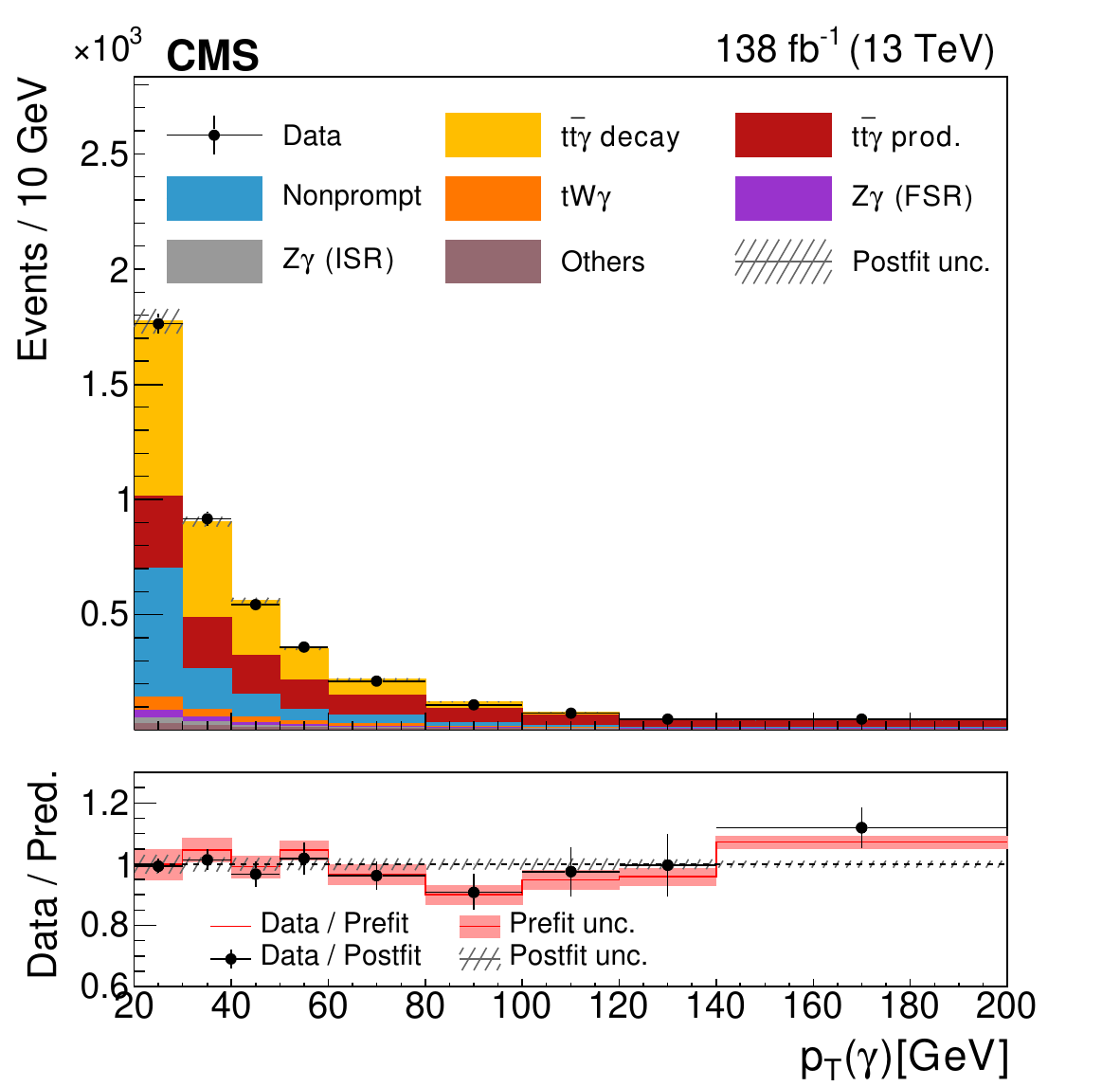}%
\hfill%
\includegraphics[width=0.495\textwidth]{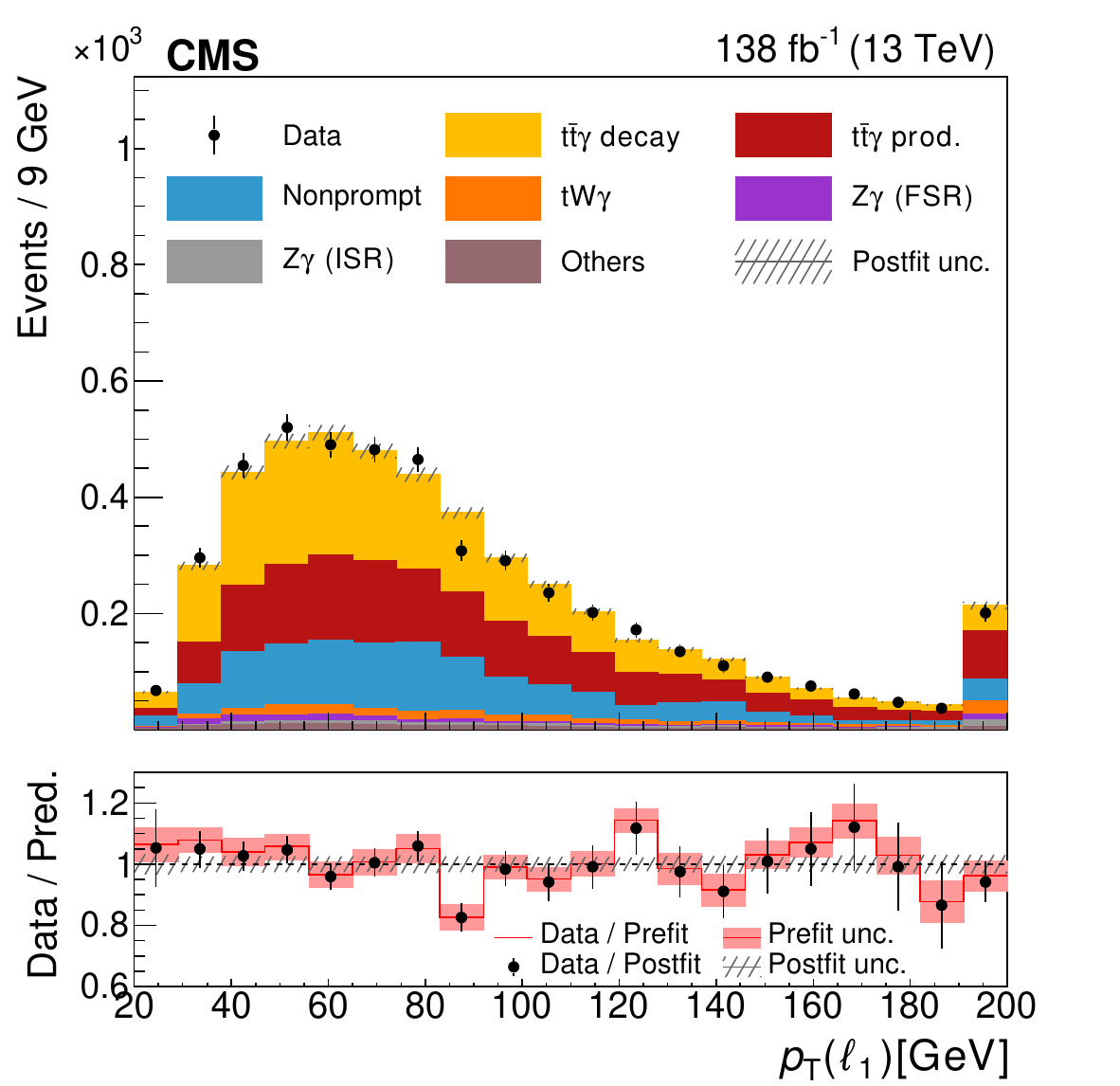} \\
\includegraphics[width=0.495\textwidth]{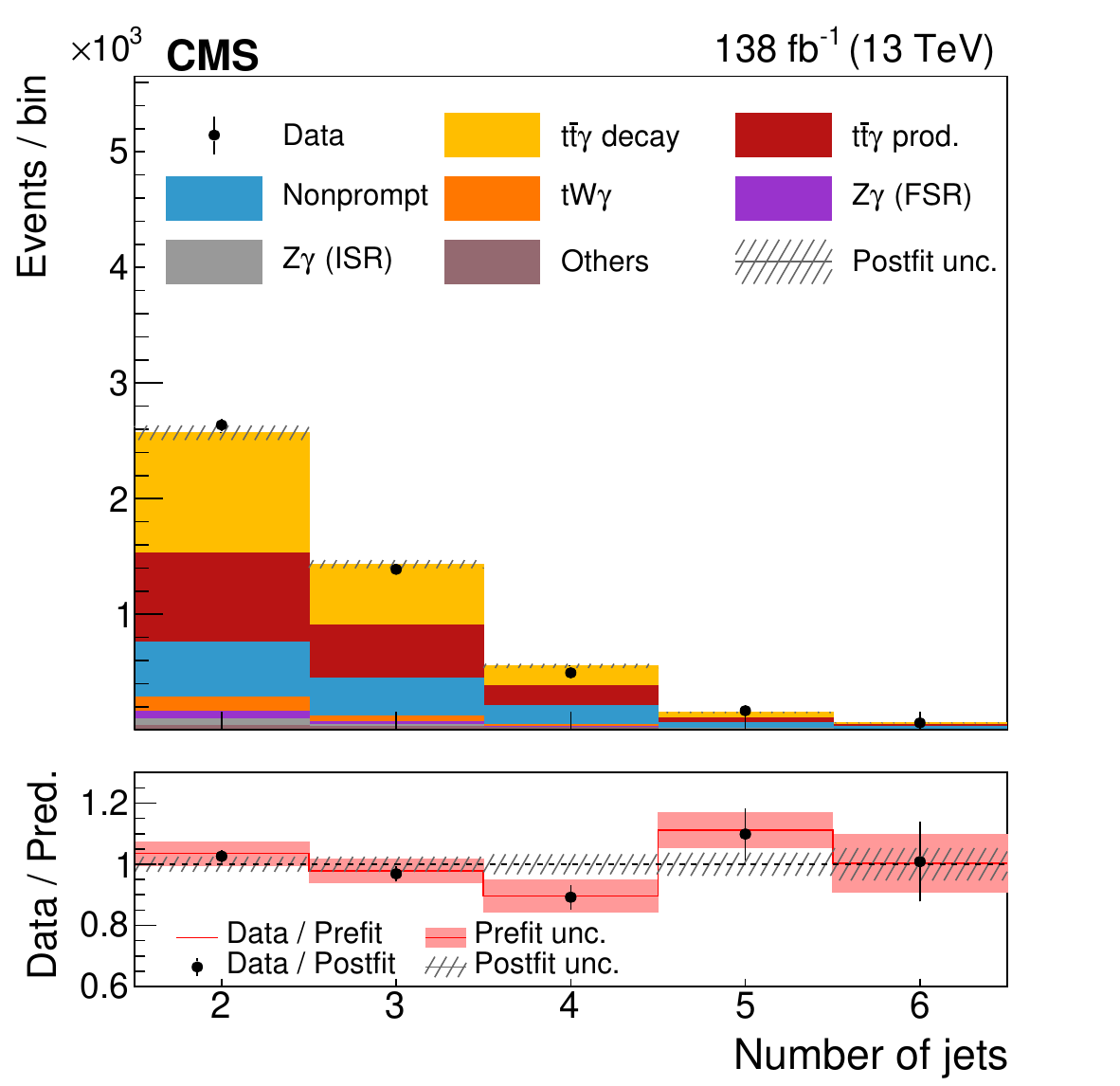}%
\hfill%
\includegraphics[width=0.495\textwidth]{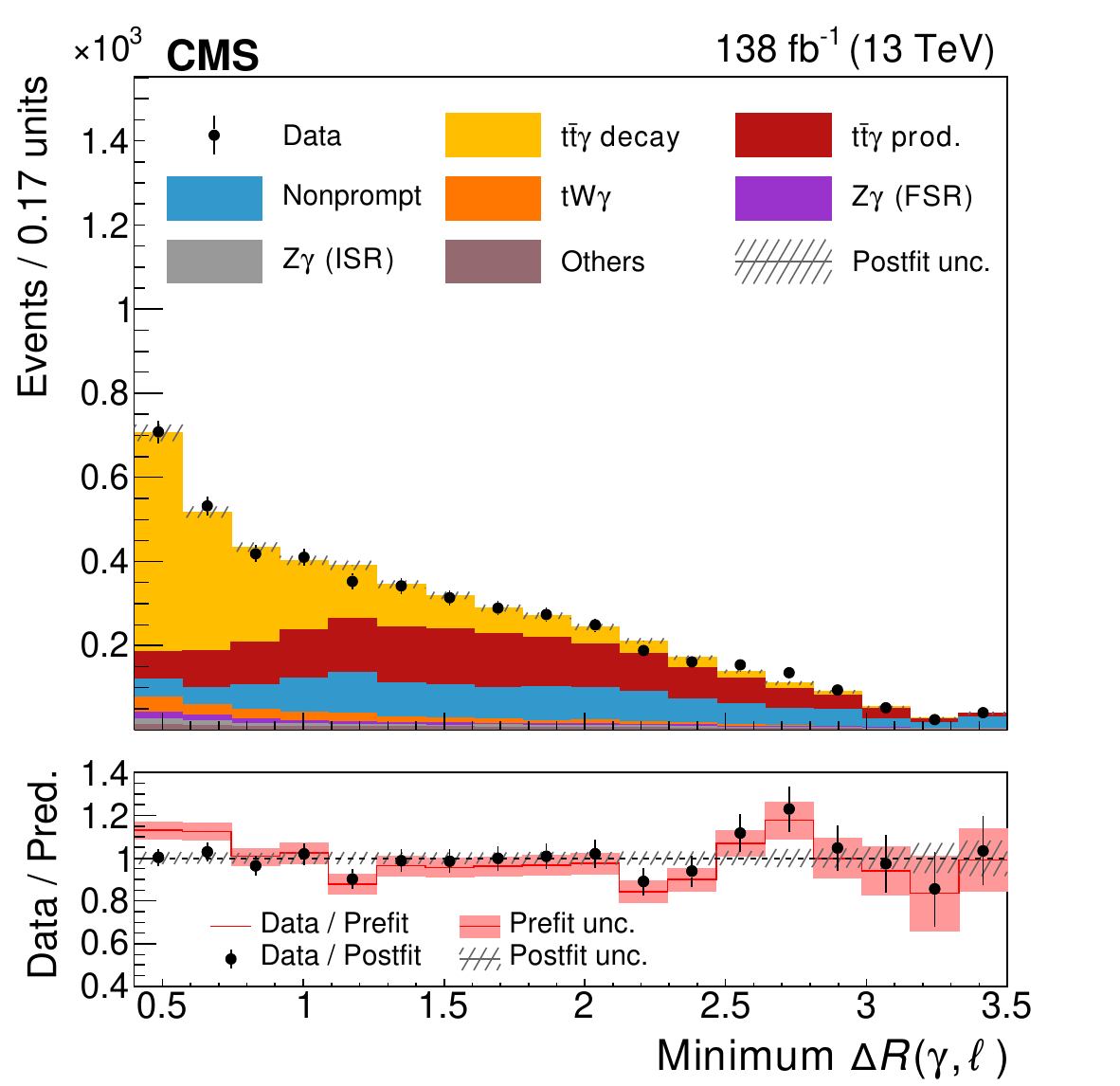}
\caption{Distributions after the event selection for:\ the \pt of the photon (upper left), the \pt of the lepton with the highest \pt (upper right), the number of jets (lower left), and the \DRglmin (lower right). The data and their statistical uncertainties are indicated by black points and error bars, respectively. The postfit prediction for the \ttg process is shown in red and yellow, after the fit to the production component of \ttg, described in Section~\ref{sec:measurement_inclusive}. The hatched area indicates the total uncertainty in the prediction. The lower panels show the ratio of the data to the sum of the postfit predictions (points) and the ratio of the data to the sum of the prefit predictions (red line). The black (red) hatched areas represent the postfit (prefit) uncertainties. The last bin includes all events above the plotted range.}
\label{fig:basic}
\end{figure}

Figure~\ref{fig:basic} shows comparisons between the data and the predictions, for several distributions, after the fit to the data to measure the production component of \ttg, described in Section~\ref{sec:measurement_inclusive}. The predictions for signal and most background processes are taken from MC simulations, as described above. The contribution labelled as ``\ttg prod.'' corresponds to the \ttg production process, while the one labelled as ``\ttg decay'' corresponds to the \ttg decay process, as defined in Section~\ref{sec:sim}. Background events containing nonprompt photons are estimated from the data, as described in Section~\ref{sec:bkg}, and labelled as ``Nonprompt''. The data and the estimated signal and background contributions after the fit are found to be consistent.

\begin{figure}[!ht]
\centering
\includegraphics[width=0.495\textwidth]{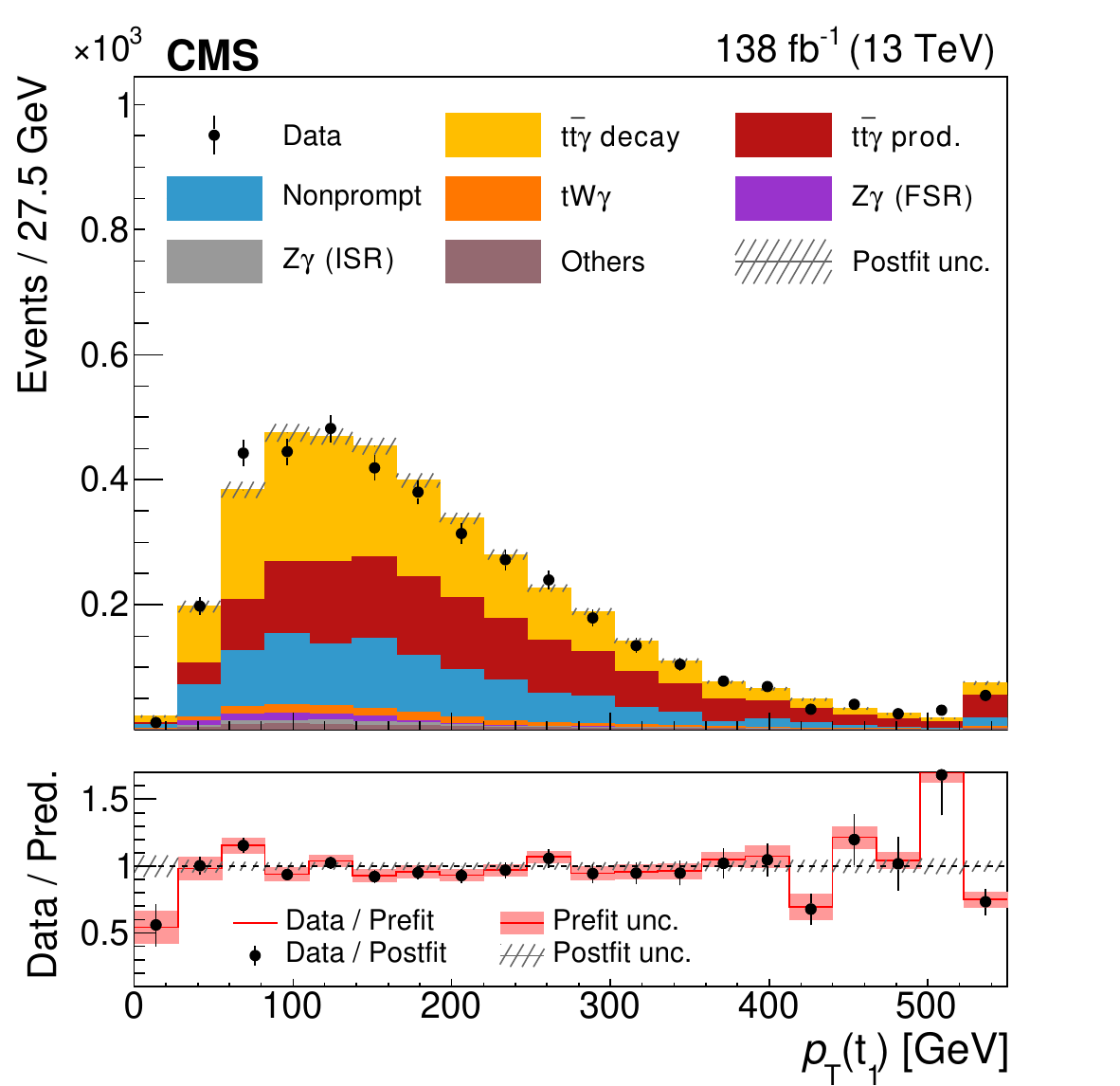}%
\hfill%
\includegraphics[width=0.495\textwidth]{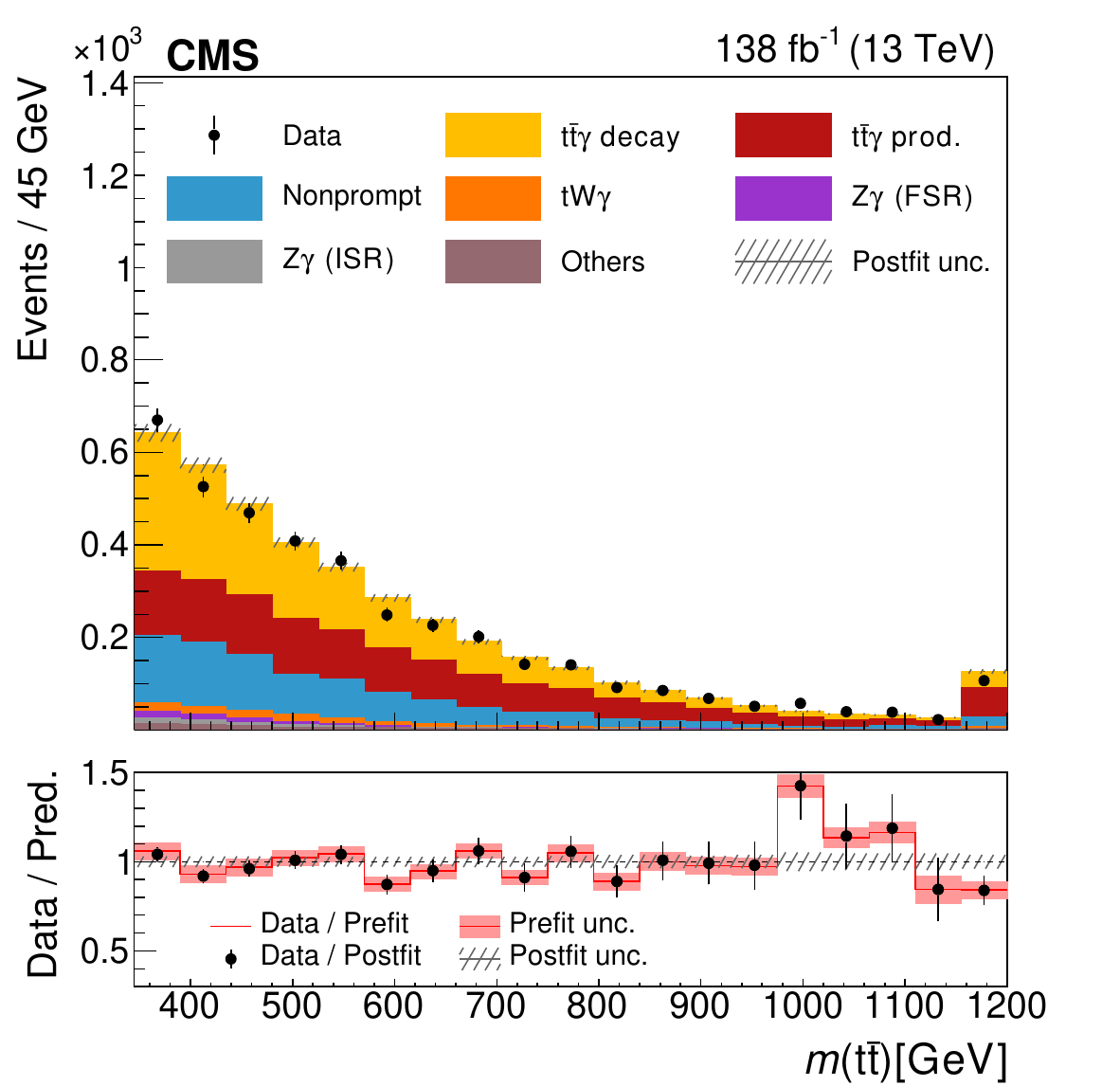} \\
\includegraphics[width=0.495\textwidth]{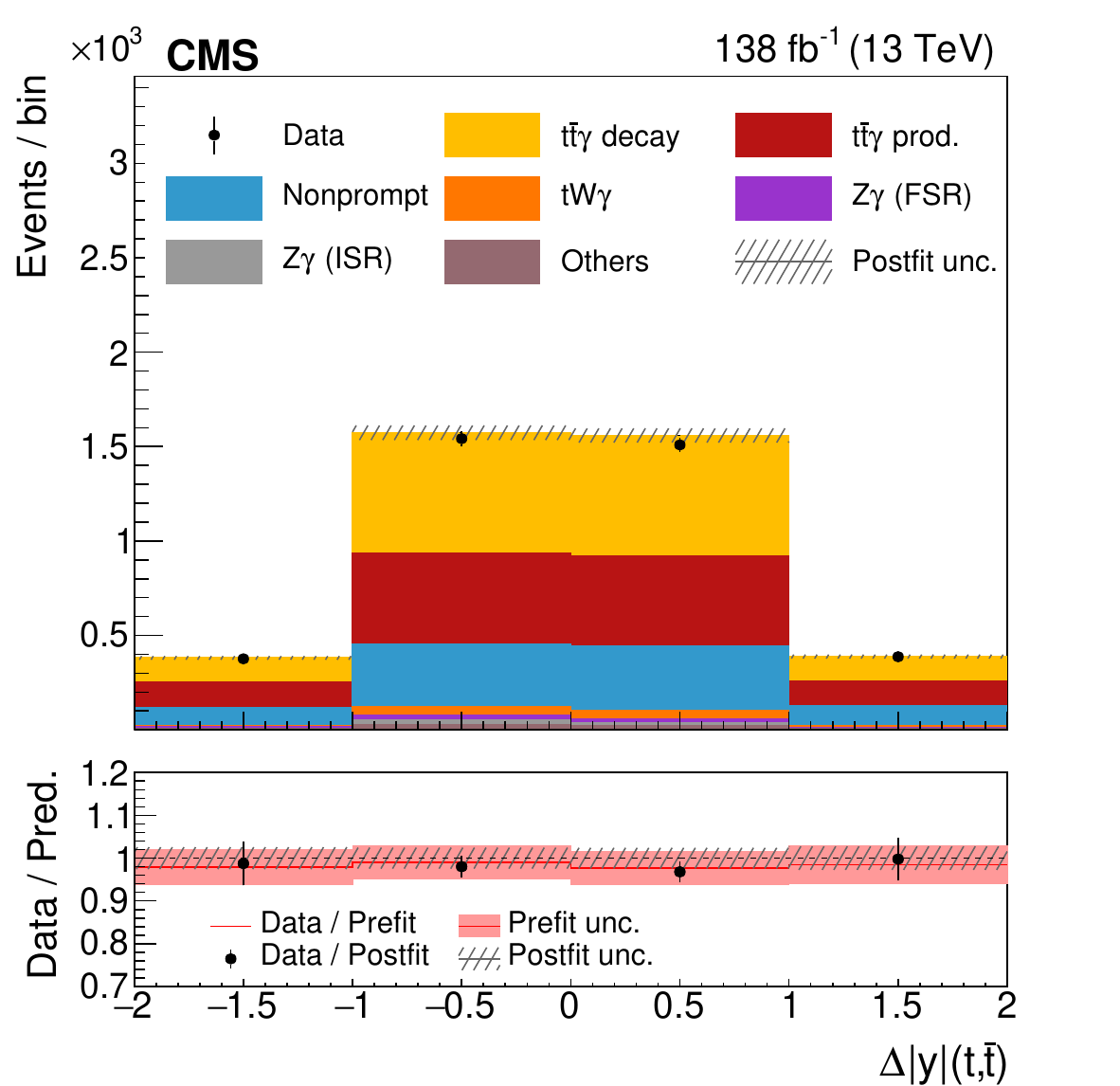}
\caption{The \toppt (upper left), \mttbar (upper right), and \dytops (lower) distributions in data and simulation, after the fit to the production component of \ttg, described in Section~\ref{sec:measurement_inclusive}. The hatched area indicates the total uncertainty in the prediction. The lower panels show the ratio of the data to the sum of the postfit predictions (points) and the ratio of the data to the sum of the prefit predictions (red line). The black (red) hatched areas represent the postfit (prefit) uncertainties. The last bin includes all events above the plotted range and, where applicable, the first bin includes all events below the plotted range.}
\label{fig:topdistributions}
\end{figure}

For the selected events, the kinematic properties of the top quarks and the \ttbar system are reconstructed. The four-momenta of the top quark and antiquark in each event are estimated using a kinematic reconstruction algorithm, extensively described in Ref.~\cite{CMS:2024irj}. The main challenge is to accurately estimate the four-momenta of the neutrinos that collectively give rise to the \ptvecmiss of the event. The reconstructed jets and leptons in each event are assigned to the \PQb quarks and leptons from the top quark decays, in all possible combinations. These combinations are inserted into a set of energy-momentum conservation equations. Constraints in those equations are the mass of the \PW boson, $m_{\PW}=80.4\GeV$~\cite{Tanabashi:2018oca}, and the mass of the top (anti)quark, $\mtop=172.5\GeV$. The assumed value of $m_{\PW}$ is varied according to a simulated Breit--Wigner distribution, with a width of 2.1\GeV~\cite{Tanabashi:2018oca}. It is also assumed that the \ptvecmiss originates only from the neutrinos. To account for the detector effects, the measured energy and the direction of the reconstructed jets and leptons are randomly smeared according to their simulated resolutions. The combination providing a neutrino solution that minimizes the invariant mass of the \ttbar system is chosen. More details can be found in Ref.~\cite{CMS:2019nrx}. The reconstruction efficiency in \ttbar, defined as the fraction of events where a solution is found, is about 90\% in both data and simulation. Events where no real solution can be found for the neutrino momenta are discarded. The presence of the photon is not considered explicitly in the reconstruction, but this is shown to have a negligible effect on the event reconstruction efficiency. The reconstruction efficiency was checked as a function of photon kinematic observables, and minimal dependence was observed. In \ttg events with at least two leptons, the reconstruction efficiency ranges from more than 95\% to about 80\% for the leading lepton \pt between 25 and 400\GeV. Figure~\ref{fig:topdistributions} shows the \pt distribution of the reconstructed leading top quark (\toppt), the invariant mass distribution of the reconstructed \ttbar system (\mttbar), and the difference between the absolute rapidities of the reconstructed top quark and antiquark (\dytops), in data and simulation, after the fit.

\section{Background determination}
\label{sec:bkg}

A variety of background processes have a similar signature to the signal, and events from these processes can pass the event selection. They can be categorized into processes with prompt or nonprompt photons, depending on the nature of the reconstructed photon. Each reconstructed photon is matched to the nearest generator-level particle within a cone of $\DR<0.3$, and with a \pt that agrees within 50\% with that of the photon. The photon is labelled as prompt if the generator-level particle is a photon radiated from either a lepton, a quark, or a boson; whereas it is considered nonprompt if the generator-level particle is a photon originating from hadronic sources, \eg a $\PGpz\to\PGg\PGg$ decay, if it is not a photon, \eg an electron that is misidentified as a photon, or if no match is found (for instance, photons from pileup interactions).

\begin{figure}[!htp]
\centering
\includegraphics[width=0.49\textwidth]{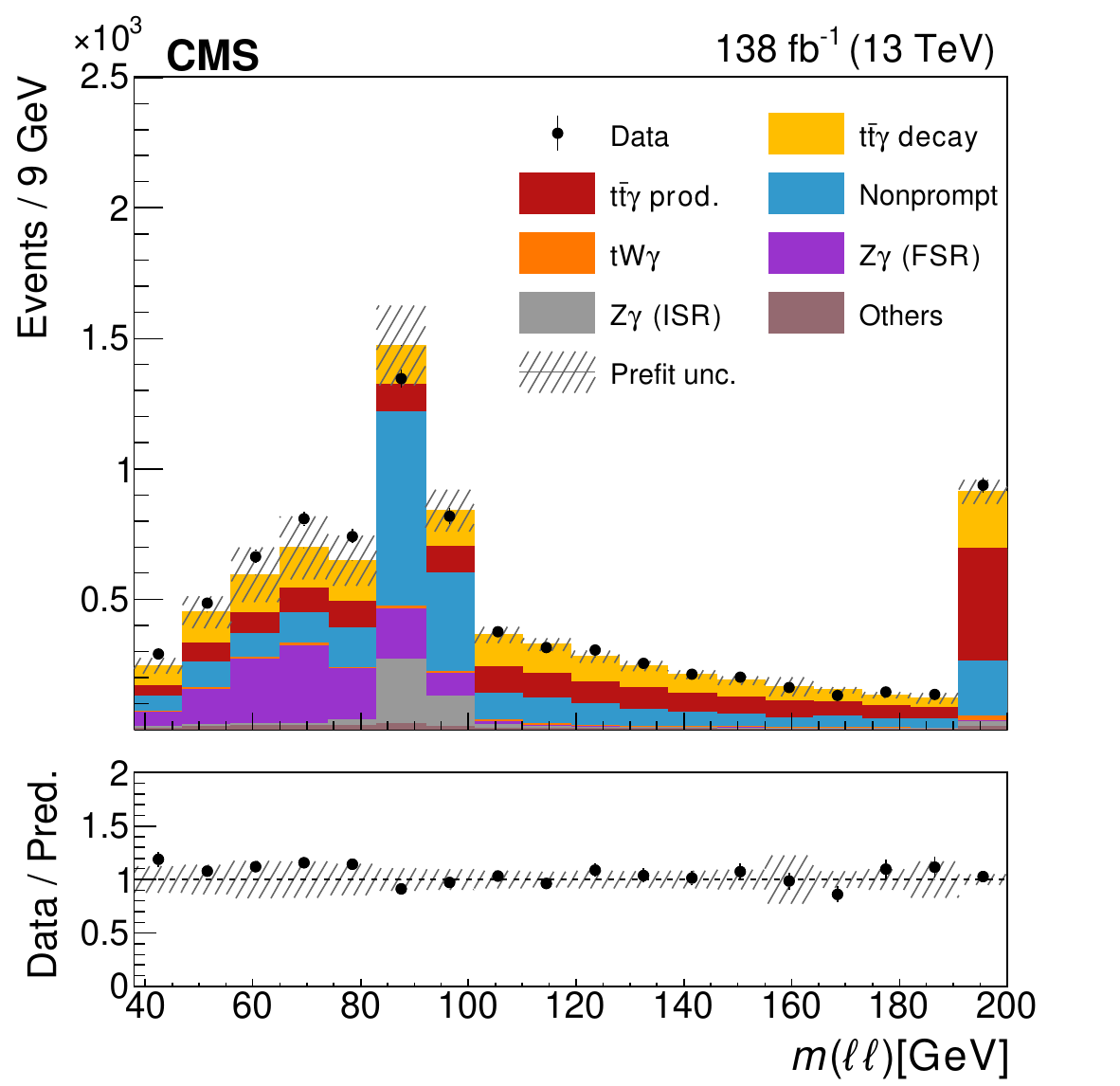}%
\hfill%
\includegraphics[width=0.49\textwidth]{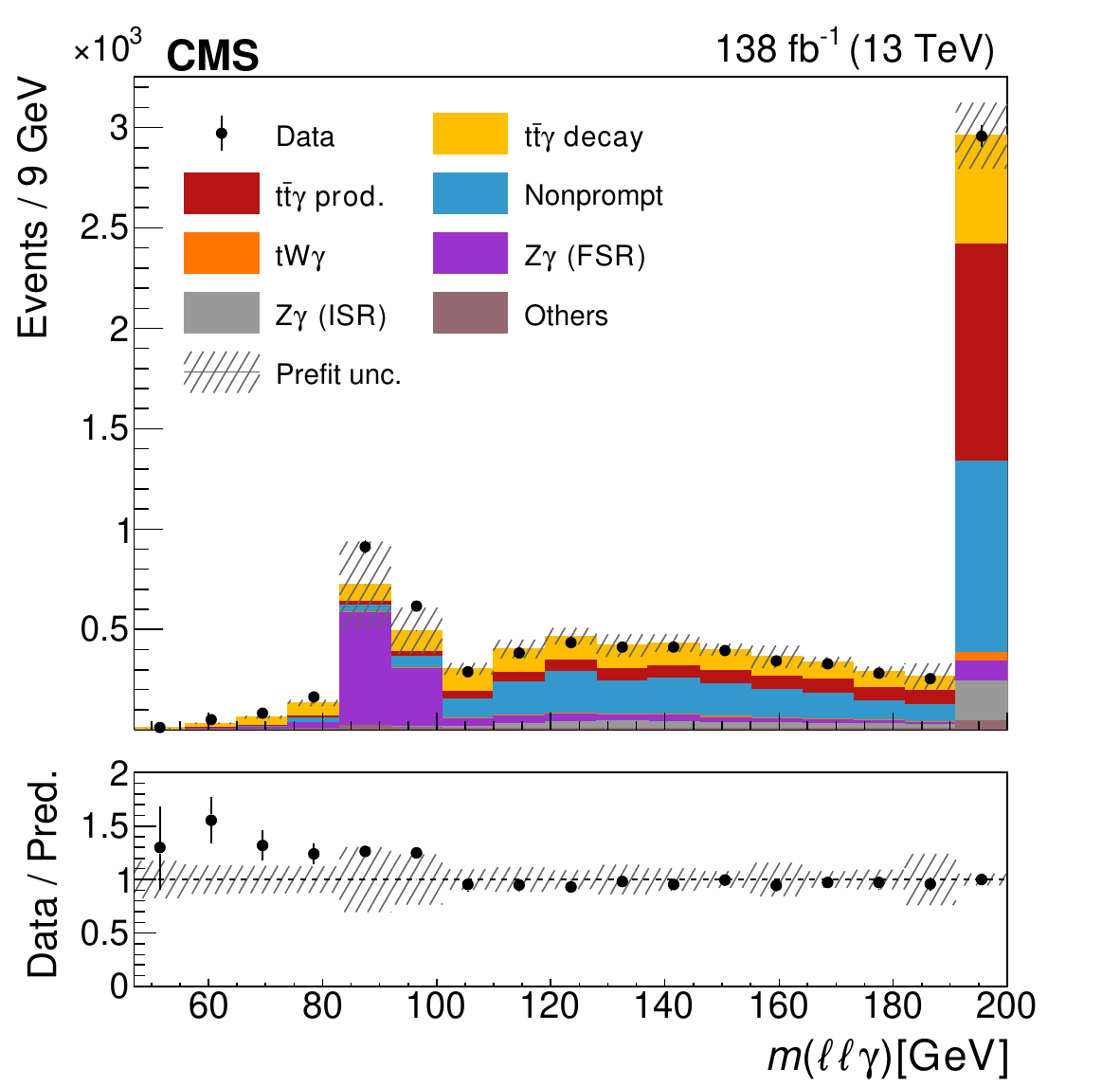}
\caption{Distributions of the invariant mass of the two leptons (left) and two leptons + photon (right) system, for events with two same-flavour leptons, as estimated by the simulation before the fit. These distributions are shown after requiring that the events pass the full event selection, but excluding the requirement that both invariant masses are reconstructed with a value not compatible with \mZ, within 15\GeV. The data and their statistical uncertainties are indicated by the black points and error bars, respectively. The hatched area indicates the total prefit uncertainty in the prediction. The last bin includes all events above the plotted range.}
\label{fig:zgamma}
\end{figure}

Among the background sources with prompt photons, the highest contribution arises from the \Zg process, where the \PZ boson decays into pairs of leptons. The \Zg contribution is split into \Zg (FSR) and \Zg (ISR) components, based on generator-level information related to the photon origin. The contribution from this process is estimated from simulation, and is constrained using a dedicated control region (CR). The CR is defined by selecting same-flavour lepton events with the same requirements as the signal region (SR), but with \mllg enforced to be compatible with \mZ within a $\pm$15\GeV window. Because of this requirement on the invariant mass, the photon in the CR mostly originates from FSR\@. Figure~\ref{fig:zgamma} shows the distributions of \mll and \mllg after the full event selection except for the requirements on these two variables, for events with two same-flavour leptons. The \Zg CR is included in the fit to the data, and the rates of the \Zg (FSR) and \Zg (ISR) processes are controlled by constrained nuisance parameters. Different nuisance parameters are defined for events with different jet multiplicities, to account for an observed mismodelling of this observable in the CR\@. All other backgrounds with prompt photons are estimated from simulation.

\begin{figure}[!htp]
\centering
\includegraphics[width=0.9\textwidth]{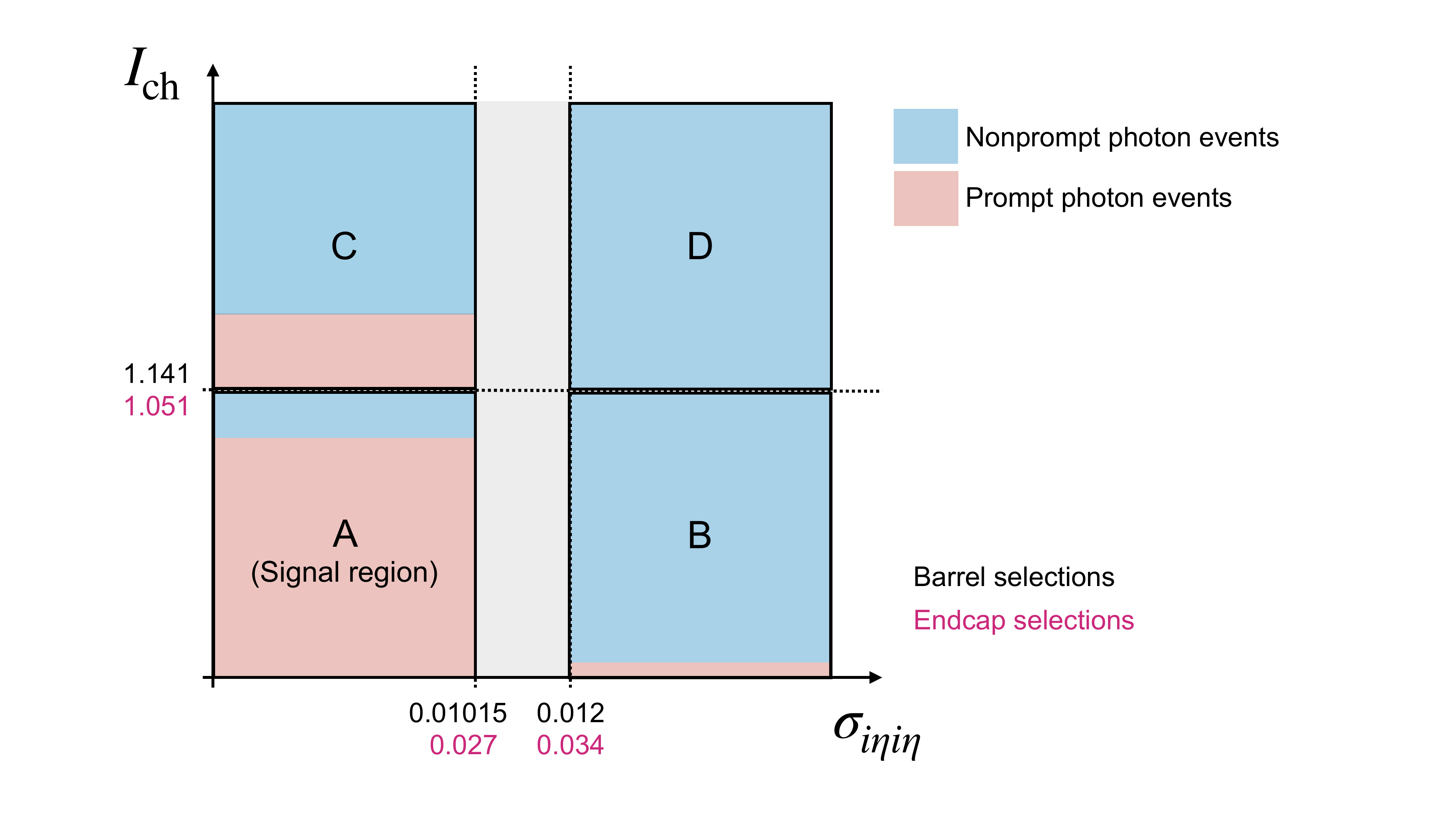}
\caption{Schematic representation of the regions used to estimate the contribution from events with a nonprompt photon in the SR\@. The blue (light red) areas represent the fraction of events with a nonprompt (prompt) photon in each region. The area shaded in grey represents a gap between the regions, and events falling in that gap are excluded. The numbers in black on the axes represent the selections applied to separate the regions for events with a photon reconstructed in the ECAL barrel, while those in pink represent the selections applied to events with a photon reconstructed in the endcaps.}
\label{fig:ABCD}
\end{figure}

Due to its potentially unreliable description in simulation, the contribution arising from processes with nonprompt photons is estimated from data using the so-called $\Ra\Rb\Rc\Rd$ method, with a similar strategy to that of Ref.~\cite{CMS:2022lmh}. Four exclusive regions (\Ra through \Rd) are defined, based on two variables that provide good discrimination between prompt and nonprompt photons, namely the photon PF charged isolation (\chiso) and the width of the electromagnetic shower measured in the ECAL (\sieie)~\cite{Electrons:2021}, as illustrated in Fig.~\ref{fig:ABCD}. The \chiso is obtained by summing the \pt of charged hadrons inside an isolation cone of $\DR<0.3$ with respect to the photon direction. These variables present a low correlation between them, below 2 (7)\% for photons reconstructed in the barrel (endcap) of the CMS detector. All regions are required to contain events with exactly one photon passing the identification criteria detailed above, but with looser requirements on the identification algorithm, in particular no requirements on \chiso and \sieie. Additionally, events in all regions are required to pass all criteria of the signal event selection except for those on the number of jets. Two regions (\Rc and \Rd) form the measurement region, and the ratio between the nonprompt photon contributions in those regions is used to measure the probability that a nonprompt photon also passes the identification criteria and therefore ends up in the final selection, referred to as the misidentification rate (\RmisID). In these regions, events are required to have at least one jet, or have two leptons of different flavours ($\Pe\PGm$), to suppress the contribution from the \DY process. The remaining two regions (\Ra and \Rb) form the application region, with the same requirements on jet multiplicity as those described for the signal event selection, where the \RmisID is applied in the form of a weight to photon events in region \Rb in order to predict the number of nonprompt photon events in the SR (\Ra).

In more detail, the measurement region contains events with one non-isolated photon, defined by requiring 1.141 (1.051)${}\leq\chiso<15$ for barrel (endcap) photons, and is further split into \Rc with the same \sieie requirement as the SR, \ie $\sieie<0.010$ (0.027) for barrel (endcap) photons; and a sideband \Rd defined by inverting the \sieie requirement, \ie $\sieie>0.012$ (0.034) for barrel (endcap) photons. With this definition, the events in \Rd contain mostly nonprompt photons (about 99.5\%). The misidentification rate is then measured by computing the ratio of data events in \Rc to those in \Rd in bins of \pt and $\eta$ of the photon, since low-\pt and high-$\abs{\eta}$ photons are more likely to be nonprompt. The small contribution from prompt photons in these regions is subtracted using simulated events. The application region contains events with isolated photons, \ie with the requirement $\chiso<1.141$ (1.051), and is further split into a region \Ra that corresponds to the SR, and a sideband \Rb defined by inverting the \sieie requirement. The gap between the two thresholds of the \sieie requirement ensures a higher contribution of nonprompt photons in both sideband regions. Figure~\ref{fig:ABCD} shows a scheme of the \Ra, \Rb, \Rc, and \Rd regions, illustrating the purity of each region in events with a nonprompt photon, computed from simulation. The contribution of nonprompt photons in \Ra is estimated by computing $\Na=\RmisID\,\Nb$. The \RmisID values are computed separately for each data-taking period, to account for differences in the detector configuration.

The method described above relies on the condition that the variables chosen to define the four regions are fully independent of one another. In this analysis, it has been found in simulation that this assumption does not always hold true, especially for endcap photons. The physical reason lies in the fact that nonprompt photons tend to have both high values of \sieie and to be poorly isolated from charged particles around them, for instance if they originate from jets misreconstructed as photons. To address this issue, correction factors $k_{\text{MC}}$ are calculated based on \ttbar simulated events using the same \pt and $\eta$ binning as the \RmisID. The correction factors are computed as the double ratio $(\Na^{\text{MC}}/\Nb^{\text{MC}})/(\Nc^{\text{MC}}/\Nd^{\text{MC}}) $, where $\Nx^{\text{MC}}$ is the number of nonprompt photon events in each region \Rx in \ttbar simulation. The correction factors are close to 1 for photons with low \pt and reconstructed in the barrel (the majority of events), but go up to 3 for some poorly populated bins in the higher \pt and/or higher rapidity regions. The number of nonprompt \PGg events in the SR is computed in each photon \pt and $\abs{\eta}$ bin, indexed by $i$ and $j$, respectively, by:
\begin{equation}
    \Na(\text{nonprompt }\PGg) = \sum_{i,j}\Big(\Nb^{ij}(\text{data})-\Nb^{ij}(\text{prompt MC})\Big)\,\RmisID^{ij}\,k_{\text{MC}}^{ij}.
    \label{eq:RmisID}
\end{equation}
The most significant contribution of nonprompt photon events in the SR arises from the \ttbar process, where an extra jet is misreconstructed as a photon. Therefore, in order to assess the consistency of this method, the \RmisID is also computed on simulated \ttbar events, and subsequently applied to simulated \ttbar events with nonprompt photons in \Rb, to estimate the number of nonprompt photon events in \Ra. This is then compared to the distribution of simulated \ttbar events with nonprompt photons in \Ra, which serves as a ``closure test'' to ensure the consistency of the method. Good agreement is observed in the normalized distributions, not only of the photon \pt and $\eta$, where these factors are derived, but of all the studied quantities. However, residual normalization differences were identified and are accounted for by applying a normalization uncertainty of 20\% to the nonprompt photon background, corresponding to the largest relative differences observed between the distributions, in specific bins. In addition, an extra shape uncertainty is included in the fits to data, to capture slight trends observed in the closure tests for the specific variables that are included in the fits, that may not be adequately addressed by the normalization uncertainty.

\section{Systematic uncertainties}
\label{sec:sys}

Several sources of systematic uncertainties are considered, which affect the statistical interpretation of the results through their modification of the normalization and/or the shape of the templates for the signal and background processes. All sources of systematic uncertainty considered in this analysis are summarized in Table~\ref{tab:systematics}, and are described below, grouped by experimental and theoretical uncertainties. Unless otherwise specified, the uncertainties are applied to all processes except the nonprompt photon background. The different components were studied one by one to evaluate whether they were dependent on the data-taking period and signal/background process, and incorporated accordingly as a correlated or uncorrelated nuisance parameter.

\begin{table}[!th]
\centering
\topcaption{%
    Summary of the systematic uncertainty sources in the inclusive and differential \ttg cross section, \ttg/\ttbar ratio, and charge asymmetry measurements.
    The first column lists the source of the uncertainty, while the second (third) column indicates the treatment of correlations of the uncertainties between different data-taking periods (processes), where \fullcorr means fully correlated, \partcorr means partially correlated (\ie contains sub-sources that are either fully correlated or uncorrelated), \nocorr means uncorrelated, and {\NA} means not applicable.
}
\renewcommand{\arraystretch}{1.1}
\begin{tabular}{clcc}
    & & \multicolumn{2}{c}{Correlation} \\
    & Source & Period & Process \\
    \hline
    \multirow{12}{*}{\rotatebox{90}{Experimental}}
    & Integrated luminosity                         & \partcorr & \fullcorr  \\
    & Pileup reweighting                            & \fullcorr & \fullcorr   \\
    & L1 prefiring                                  & \fullcorr & \fullcorr   \\
    & Electron/muon reconstruction and identification    & \fullcorr & \fullcorr   \\
    & Photon reconstruction and identification      & \partcorr & \fullcorr   \\
    & \PQb tagging                                  & \partcorr & \fullcorr   \\
    & JEC                                           & \partcorr & \fullcorr   \\
    & JER                                           & \nocorr   & \fullcorr   \\
    & Unclustered energy                         & \fullcorr & \fullcorr   \\
    & Trigger efficiencies                          & \nocorr   & \fullcorr   \\
    & Limited size of simulated samples                & \nocorr & \partcorr \\
    & Nonprompt photon estimation                   & \partcorr & \NA   \\[\cmsTabSkip]
    \multirow{8}{*}{\rotatebox{90}{Theoretical}}
    & \muR and \muF scales                        & \fullcorr & \partcorr   \\
    & PDF+\alpS                         & \fullcorr & \fullcorr   \\
    & PS scales: ISR and FSR                   & \fullcorr & \partcorr     \\
    & ME-PS matching (\hdamp)           & \fullcorr & \NA   \\
    & NNLO QCD reweighting              & \fullcorr & \NA   \\
    & Background normalization          & \fullcorr & \NA         \\
    & \Zg mormalization depending on jet multiplicity   & \fullcorr & \NA         \\
    & \ttg production/decay fraction   & \fullcorr & \NA         \\
\end{tabular}
\label{tab:systematics}
\end{table}

\begin{itemize}
\item \textit{Integrated luminosity}: The uncertainty in the integrated luminosity from the considered data-taking periods is propagated as a systematic uncertainty in the yields of all processes except the nonprompt photon contribution, estimated to be between 1.2 and 2.5\%~\cite{lumipaper,CMS:LUM-17-004,CMS:LUM-18-002}. This uncertainty affects only the normalization of the processes and is partially correlated across data-taking periods.
\item \textit{Pileup reweighting}: The uncertainty in the pileup is taken into account by using two sets of alternative event weights derived with a variation of 4.6\% in the total inelastic cross section~\cite{Sirunyan:2018nqx}. This uncertainty affects both the shape and normalization of the predictions.
\item \textit{L1 trigger inefficiency}: In 2016 and 2017 data, a problem in L1 triggers related to the ECAL and the muon systems during a portion of the data-taking period led to the selection of data events from the previous bunch crossing~\cite{ECALpaper}, and consequently to a loss of signal efficiency. To account for this, a correction factor is applied to simulated events, and the corresponding uncertainties are assigned, affecting both the shape and normalization of the predictions.
\item \textit{Lepton selection}: The uncertainties associated to the lepton reconstruction, identification, and isolation affect both the shape and normalization of the processes~\cite{Electrons:2021,CMS:2018rym}. Separate nuisance parameters are included for electrons and muons.
\item \textit{Photon selection}: The uncertainties in the photon identification and pixel seed veto are considered, which affect both the shape and normalization of the processes~\cite{Electrons:2021}.
\item \textit{\PQb tagging}: Uncertainties in the \PQb tagging efficiency corrections are considered, which affect both the shape and normalization of the predictions. These factors are derived from two different regions: a heavy-flavour one, \ie a QCD-enriched region, and a light-flavour one, \ie a \Zj-enriched region~\cite{CMS:DP-2023-005, CMS:2017wtu}.
\item \textit{Jet energy scale and resolution}: Uncertainties in the determination of the jet energy scale (JES) are taken into account by shifting the jet momenta in the simulation up and down, separately for several sources of uncertainty such as the overall energy scale, differences in flavour response, and residual differences between energy scale measurements. These corrections to the JES are denoted jet energy corrections (JEC). The JEC uncertainties are also propagated to the \ptmiss. These uncertainties affect both the shape and normalization of the predictions. Some of the sources are treated as uncorrelated per data-taking period, while others are correlated for all periods, depending on whether the underlying effect depends on the different detector configurations across periods. Uncertainties in the jet energy resolution (JER) are evaluated by increasing or decreasing the variation of jet energies between the reconstructed and particle levels, or by randomly modifying the measured jet energy by a small amount in case no matching particle-level jet could be found~\cite{CMS:2016lmd}.
\item \textit{Unclustered energy}: An uncertainty in \ptmiss is derived by varying the energies of reconstructed particles not clustered into jets within their respective resolutions~\cite{CMS:2019ctu}, affecting both the shape and normalization of the predictions.
\item \textit{Trigger efficiencies}: Uncertainties in the trigger efficiency corrections are considered, affecting both the shape and normalization of the predictions.
{\tolerance=800
\item \textit{Limited size of MC simulation samples}: The uncertainty originating from a finite number of simulated events is accounted for using the Beeston--Barlow ``lite'' method~\cite{Barlow:1993dm}.\par}
\item \textit{Nonprompt photon background}: Shape and normalization uncertainties are applied to account for the residual mismatches in the nonprompt photon estimation. These are built by comparing the distributions obtained with the $\Ra\Rb\Rc\Rd$ method with those obtained directly from the simulation, as detailed in Section~\ref{sec:bkg}, and are considered as partly correlated between data-taking periods.
\end{itemize}
Uncertainties arising from generator parameters and settings affect both the shape and normalization of the templates and are incorporated into the statistical analysis. To avoid double counting of pure normalization uncertainties, the templates containing the "up" and "down" variations of these parameters are normalized to the same cross section as the nominal ones, in the full phase space before any event selection, for all processes. These uncertainties are listed below:
\begin{itemize}
\item \textit{Renormalization and factorization scales}: The uncertainties in the choice of renormalization and factorization scales (\muF and \muR) are considered for the \ttg, \ttbar, and \Zg processes. These uncertainties are estimated by varying each scale up and down independently, by a factor of two, and are treated as correlated between the \ttg production and \ttbar processes, and uncorrelated between those and the \ttg decay and \Zg processes. The choice to correlate or uncorrelate each process was based on how similar the simulated processes are in terms of their production mechanisms and order of perturbation theory at which they are generated.
\item \textit{PDF and \alpS}: The impact of the uncertainties related to the PDFs and the variation of the strong coupling constant \alpS for each simulated background and signal process is obtained using the corresponding variations in the NNPDF sets~\cite{Ball:2017nwa,Ball:2014uwa}. In particular, one nuisance parameter per eigenvector is included.
\item \textit{PS scales}: Uncertainties in the PS simulation are accounted for by varying \alpS independently for ISR and FSR by a factor of $1/2$ and 2, for all processes. The ISR uncertainties are treated as uncorrelated between double- and single-top quark processes, while the FSR uncertainties are treated as correlated for all processes involving top quarks. Both uncertainties are treated as uncorrelated between processes involving and not involving top quarks.
\item \textit{ME-PS matching (\hdamp)}: In the \POWHEG generator, the scale that separates the phase space of the first QCD emission into soft and hard parts is controlled by the \hdamp parameter.
The nominal value in the CP5 tune~\cite{CMS:GEN-17-001} is $\hdamp=1.379\,\mtop$ and the uncertainties are estimated with varied values of $2.305\,\mtop$ and $0.874\,\mtop$, for the \ttbar simulation sample only.
\item \textit{NNLO QCD reweighting}: The \ttbar simulation sample is reweighted such that the top quark \pt spectrum matches that from a fixed-order ME calculation at NNLO in QCD, as described in Section~\ref{sec:sim}. The difference between the results obtained with this sample and with the unweighted version is taken as a systematic uncertainty.
\end{itemize}
Additionally, theoretical uncertainties in the total cross sections of the background processes are considered as systematic uncertainties in the yield predictions. In particular, 20\% is estimated for \twg and 30\% for the remaining backgrounds, as used in Ref.~\cite{CMS:2022lmh}. For the \Zg process, a large uncertainty is assigned as a function of the number of jets, ranging from 50 to 150\%, as it was observed that the modelling of this process is significantly worse for larger jet multiplicities~\cite{CMS:2021gme}. These large uncertainties are constrained using the data, by including the \Zg CR in all the fits. An additional uncertainty of 20\% is applied to the ratio between the \ttg production and decay components; this uncertainty does not modify the total cross section, but only the relative contributions of the two processes. The choice of 20\% is based on the observed differences between the \ttg sample produced with \MGvATNLO and the alternative sample produced with \POWHEG, as described in Section~\ref{sec:sim}, but it was checked that the results are stable with respect to this choice. The impact of all systematic uncertainties on the results are discussed in Sections~\ref{sec:meas}, \ref{sec:ratio}, and \ref{sec:charge_asym}.

\section{Statistical analysis}
\label{sec:stat}

All results are extracted via a maximum likelihood fit to data, in the \ttg SR and \Zg CR simultaneously, where systematic uncertainties are accounted for as nuisance parameters. The likelihood function is built from the product of Poisson probabilities $\mathcal{P}$ over $N$ bins, calculated as functions of the number of data events $n_i$ and the expected number of signal and background events $\nu_i$ in a given bin $i$. This is multiplied by an auxiliary function $\pi$, which controls the constraint of the nuisance parameter $\theta_k$, given a set of nuisance parameters \boldtheta of length $M$.

The likelihood function reads:
\begin{equation}
    L\big(\mathbf{n}\big|\mu,\boldtheta\big) = \prod_{i=1}^N\mathcal{P}\big(n_i\big|\nu_i(\mu,\boldtheta)\big)\prod_{k=1}^M\pi(\theta_k),
\end{equation}
where the vector $\mathbf{n}=n_{1},\dots,n_{N}$ represents the observed number of events in each bin, \boldtheta represents the set of nuisance parameters, and $\mu$ are the parameters of interest (POIs) to be measured. In the inclusive measurement of the cross section, the POI is a single parameter, \ie the \ttg cross section modifier $\mu_{\ttg}=\sigma_{\ttg}^{\text{obs}}/\sigma_{\ttg}^{\text{exp}}$, with $\sigma_{\ttg}^{\text{obs}}$ and $\sigma_{\ttg}^{\text{exp}}$ being the observed and expected \ttg cross sections, respectively. The fits are implemented with \textsc{combine}~\cite{combine_tool}.

In differential measurements, where cross sections are measured in bins of a chosen observable, a set of POIs is used instead, with one POI per bin. Each observable of interest $x$ is divided into $N$ bins at the generator level. A likelihood-based unfolding technique is used to correct for the finite resolution and limited acceptance of the detector, which cause the measured distribution to be distorted and the values of $x$ to deviate from the ones at the generator level. In practice, we perform a fit to extract a cross section modifier $\mu_l$ for each generator-level bin $l$, and multiply it by $\sigma_l^{\text{exp}}$ taken from the predictions, which results in the measured value of $\sigma_l$. The binning is chosen for each observable such that the purity (fraction of reconstructed events observed in the reconstruction-level bin $x_i^{\text{reco}}$ that were also generated in the generator-level bin $x_i$) and stability (fraction of reconstructed events generated in the generator-level bin $x_i$ that are also observed in the reconstruction-level bin $x_i^{\text{reco}}$) are above 50\%, and the condition number for the response matrices comparing the generator- and reconstruction-level quantities are reasonably small. The condition number is a measure of the stability of the unfolding problem. Typically, numbers of the order of 10 or below guarantee that the unfolding can be performed without applying any regularization technique~\cite{Blobel:2203257}.

{\tolerance=800
For the inclusive and differential measurements of \ratio, in addition to the \ttg SR and the \Zg CR, two separate regions enriched in \ttbar and \DY events are defined. These regions are constructed similarly to the \ttg SR and the \Zg CR, respectively, except that events are required to have no associated reconstructed photons. We refer to the \ttg and \ttbar SRs with the subscripts ``$\ttbar,1\PGg$'' and ``$\ttbar,0\PGg$'', respectively. We perform a simultaneous fit to the \ttbar cross section modifier $\mu_{\ttbar}$ and to the \ratio modifier $\mu_{\ratio} = \ratio/\ratioexp$ in the four regions, where:
\begin{equation}
    \ratio = \frac{\sigma_{\ttbar,1\PGg}}{\sigma_{\ttbar,0\PGg}+\sigma_{\ttbar,1\PGg}},
    \label{eq:ratio}
\end{equation}}
and \ratioexp is the expected value of \ratio, computed using the simulated samples.

The top quark charge asymmetry at the LHC can be defined as:
\begin{equation}
    A_C = \frac{\sigma(\Delta\abs{y}>0)-\sigma(\Delta\abs{y}<0)}{\sigma(\Delta\abs{y}>0)+\sigma(\Delta\abs{y}<0)},
\end{equation}
where $\Delta\abs{y}=\abs{y(\PQt)}-\abs{y(\bar{\PQt})}$ is the difference in absolute rapidity between the top quark and antiquark, and $\sigma$ is the associated cross section.
For the measurement of this quantity, events are divided in two categories, according to the observable $\Delta\abs{y}$ being positive or negative, and the charge asymmetry is extracted directly from a maximum likelihood fit to a distribution of this observable in the \ttg SR, as shown in Fig.~\ref{fig:topdistributions}, and in the \Zg CR\@. The two signal categories are used to build two separate signal templates, and the cross section modifiers for one of the signals is reparametrized to be a function of the charge asymmetry. This way, the charge asymmetry is extracted directly, with all the uncertainties and respective correlations being taken into account in the fit.

\section{Cross section measurements}
\label{sec:meas}

\subsection{Cross section definition}
\label{crosssection_def}

The \ttg cross section is measured in two different phase spaces: a fiducial phase space defined at the particle level after hadronization, and a broader phase space defined at the parton level. Both phase spaces are defined in terms of the generator-level objects. For the parton-level phase space, events are required to have two leptons in the final state, originating from top quark decays, and exactly one final-state photon with \pt greater than 20\GeV and $\abs{\eta}<2.5$, which is isolated from all other final-state particles with \pt larger than 5\GeV (leptons, other photons, and hadrons), excluding neutrinos, by $\DR>0.1$. These photons must not originate from the hadronization process. No further requirements are imposed on any other objects.

Concerning the particle-level object definition~\cite{Collaboration:2267573}, leptons are required to originate from the prompt decays of the \PW boson (or from the decay of a \PGt lepton which in turn originates from the \PW boson) and satisfy $\pt>15\GeV$ and $\abs{\eta}<2.5$. The momenta of photons within a $\DR=0.1$ cone around leptons are added to the respective leptons. Photons are required to not originate from the hadronization process, have \pt greater than 20\GeV, $\abs{\eta}<2.5$, and that the sum of \pt of all particles surrounding the photon within $\DR=0.4$ is less than 50\% of the photon \pt. Photons separated from leptons by $\DR<0.4$ are excluded. Particle-level jets are clustered with the anti-\kt algorithm with a distance parameter of 0.4, and are required to satisfy $\pt>30\GeV$ and $\abs{\eta}<2.4$, and to be separated from leptons by $\DR>0.4$. Particle-level \PQb jets are identified as jets that have a \PQb quark matched to them through a hadron ghost-matching technique~\cite{Cacciari:2008gn}, where for each \PQb hadron, an additional collinear four-vector of infinitesimal magnitude is included in the jet clustering, and each jet that includes such a ``ghost'' is identified as a \PQb jet. The fiducial phase space contains signal events with at least two leptons with opposite-sign charges (OS), exactly one photon, and at least two jets, of which at least one must be a \PQb jet. Additionally, a requirement on the invariant mass of the dilepton system, $\mll>30\GeV$ is applied. The fiducial phase space definition is summarized in Table~\ref{tab:fiducial}. This selection reduces the total predicted \ttg fiducial cross section (including the production and decay components) when compared to the parton level phase space, from $1118\pm194\fb$ to $126\pm19\fb$, using the nominal prediction defined in Section~\ref{sec:sim}. The uncertainties include the choice of \muR and \muF, PS, and the PDF uncertainties.

\begin{table}[!ht]
\centering
\topcaption{Definition of the fiducial phase space.}
\label{tab:fiducial}
\renewcommand{\arraystretch}{1.1}
\begin{tabular}{lcccc}
    Criteria & Lepton & Photon & Jet & \PQb jet \\
    \hline
    \pt [{\GeVns}] & $>$15 & $>$20 & $>$30 & $>$30 \\
    $\abs{\eta}$ & $<$2.5 & $<$2.5 & $<$2.4 & $<$2.4 \\
    Number & 2 & 1 & $\geq$2 & $\geq$1 \\
    Origin  & Prompt & Not from hadrons & \NA & \NA \\
    Others & OS, $\mll>30\GeV$ & $\DRgl>0.4$ & $\DR(\PGg/\Pell,\text{jet})>0.4$ & \NA \\
\end{tabular}
\end{table}

\subsection{Fixed-order predictions}
\label{sec:fixedorder}

All results are compared to the predictions from simulation. In addition, the particle-level results are compared to a set of fixed-order calculations. To obtain them, NLO QCD corrections to the $\pp\to\ttg$ process in the dilepton decay channel are computed, taking into account higher-order effects in both the \ttbar production and top quark decays, using the narrow-width-approximation (NWA)~\cite{Bevilacqua:2019quz,Stremmer:2024ecl}. In this approach, all unstable particles are treated as on shell and NLO spin correlation effects are preserved throughout the calculation. In addition, the effects of photon emission from the charged top quark decay products (part of the \ttg decay process) are consistently included. The same applies to gluon radiation encountered in the NLO QCD corrections.

The five-flavour scheme is considered, and non-diagonal elements of the Cabibbo--Kobayashi--Maskawa matrix are neglected. The calculation is performed within the \textsc{Helac-NLO} MC framework~\cite{Bevilacqua:2011xh}, which is interfaced to the \textsc{recola}~\cite{Actis:2016mpe,Actis:2012qn} programme for the calculation of tree-level and one-loop matrix elements, together with the \textsc{collier}~\cite{Denner:2016kdg} library for the numerical evaluation of one-loop scalar and tensor integrals. The calculation of the real emission part is performed with the Nagy--Soper subtraction scheme~\cite{Bevilacqua:2013iha} with its extension to the NWA~\cite{Bevilacqua:2022ozv} and a phase-space restriction on the subtraction terms~\cite{Czakon:2015cla}. In addition, the parameter in the phase-space restriction is varied to cross check the computation of the real correction part. The numerical input parameters and the photon isolation prescription for this calculation are adapted from Ref.~\cite{Stremmer:2024zhd} to match the fiducial phase space of the analysis. Specifically, we employ a hybrid approach where we first use a rather inclusive smooth-photon isolation condition, as introduced in Ref.~\cite{Frixione:1998jh}, with $\varepsilon_{\PGg}=1$, $n=1$, and $R=0.05$. Afterwards, a fixed-cone isolation condition is applied where the event is rejected unless the scalar sum of \pt of all partons in the final-state within a cone of radius $R=0.4$ centered around the photon candidate is smaller than half of the photon \pt. Events are required to have at least two jets and at least one jet originating from a \PQb quark, matching the fiducial phase space selections of the analysis; however, the jets obtained from the calculation are before hadronization, and therefore do not correspond exactly to the particle-level jets defined in Section~\ref{crosssection_def}. This effect is not explicitly corrected for in the unfolding procedure, but is expected to be small.

\subsection{Inclusive measurements}
\label{sec:measurement_inclusive}

We extract the inclusive \ttg cross section in the particle-level fiducial phase space by performing a maximum likelihood fit using two distributions: the \DR between the photon and the closest lepton in the SR, shown in Fig.~\ref{fig:basic} (lower right), and the number of jets in the \Zg CR\@. The variable in the SR is chosen because it provides good separation between the \ttg production and decay components. The measured cross section modifier is $\mu_{\ttg}=1.09\pm0.18$. The uncertainty in this value and other cross section modifiers throughout the paper includes both the uncertainty in the measurement and the one on the theoretical cross section used to normalize it. The resulting fiducial cross section is $\sigma_{\ttg}=137\pm3\stat\pm7\syst\fb$. This value is consistent with the prediction of $126\pm19\fb$ obtained from simulation, where the uncertainty includes the choice of \muR and \muF, PS, and the PDF uncertainties. The leading sources of systematic uncertainty in the measurement are those related to the normalization of the nonprompt photon background and the photon identification. Other experimental sources such as luminosity, \PQb tagging, and the normalization of the \twg and \Zg backgrounds also play an important role, as well as the uncertainty in the fraction between the \ttg production and decay components. The fit was also performed using as input alternative distributions, such as the photon or the leading lepton \pt (\llpt), and the results were found to be consistent, with the same or larger uncertainties.

The \ttg production component is more sensitive to potential new physics effects than the decay component. Therefore, its cross section is also extracted in a separate fit, where the \ttg decay component is treated as background and accounted for as a free parameter of the fit. In this case, we obtain a cross section modifier of \smash[b]{$\mu_{\ttg}^{\text{prod}}=0.98\pm0.12$}, corresponding to a fiducial cross section of \smash[b]{$\sigma_{\ttg}^{\text{prod}}=56\pm2\stat\pm4\syst\fb$}, consistent with the expectation from the nominal simulation, $57\pm5\fb$. The fitted \ttg decay normalization factor is \smash[b]{$\mu_{\ttg}^{\text{decay}}=1.21\pm0.28$}, which has a small (5\%) anticorrelation with the \ttg production cross section. This normalization factor can be translated into a measured cross section of \smash[b]{$\sigma_{\ttg}^{\text{decay}}=84\pm5\fb$}, in agreement with the nominal simulation prediction of $70\pm16\fb$ for the \ttg decay component. The leading sources of systematic uncertainty are the same as those for the measurement of the combined production+decay \ttg cross section.

\begin{figure}[!b]
\centering
\includegraphics[width=\textwidth]{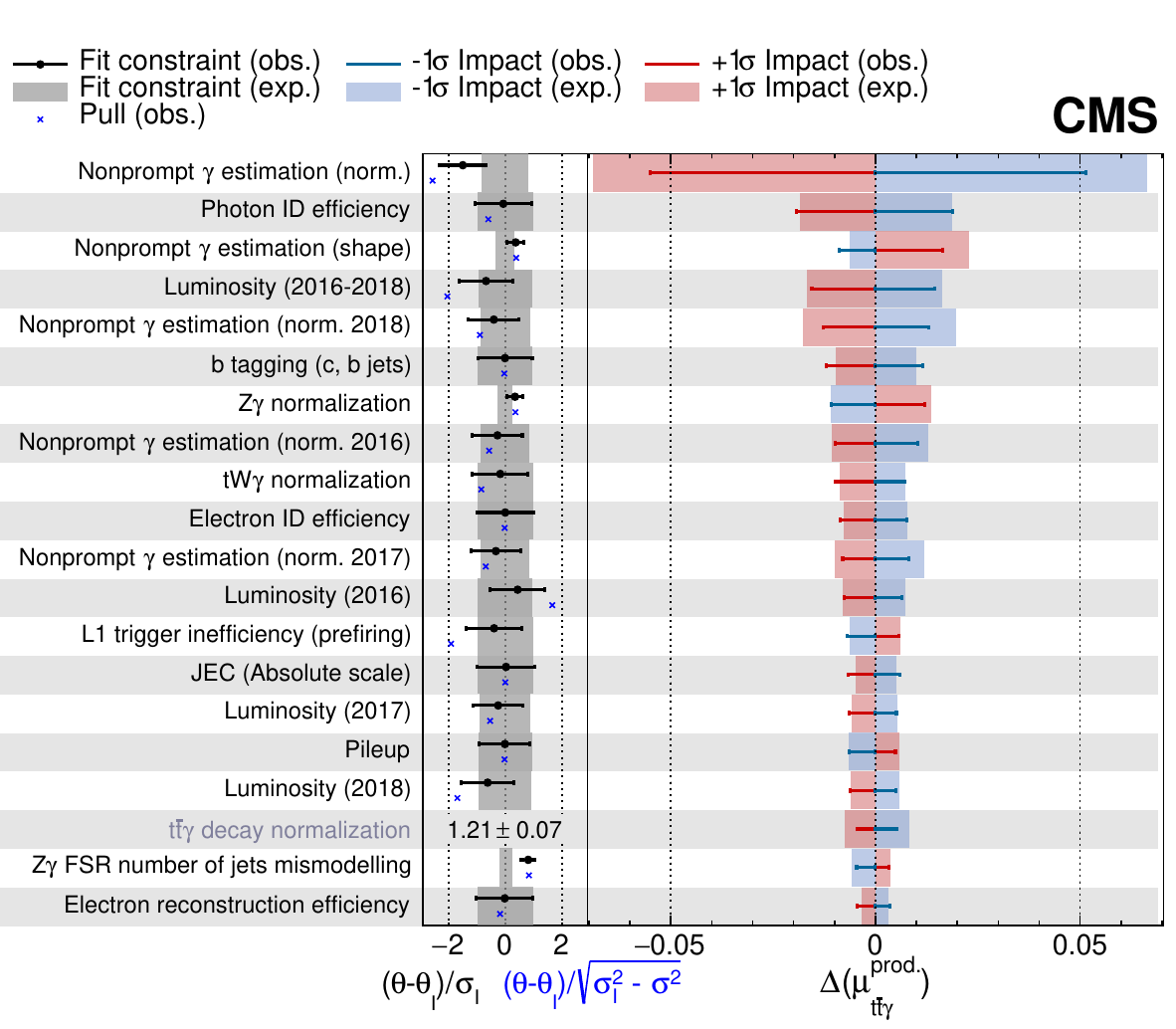}
\caption{Impacts $\Delta(\mu_{\ttg}^{\text{prod}})$ (right column) and fit pulls and constraints (middle column) of the twenty most important nuisance parameters (listed in the left column) in the fit to extract the cross section of the \ttg production component. In the middle column, two different pull definitions are presented: the difference between the postfit value of each nuisance parameter $\theta$ and its prefit value $\theta_I$, normalized either by the initial size of the uncertainty $\sigma_I$ (in black and gray) or by $\sqrt{{\sigma_I}^2-\sigma^2}$ (in blue) \cite{CDF:AN5776}, where $\sigma$ is the postfit size of the uncertainty. For the first definition, the uncertainty on this quantity is also reported and calculated as $\sigma/\sigma_I$, effectively giving the constraints on the nuisance parameters. The contraints are shown as gray horizontal bands (black horizontal lines) for the expected (observed) fit. In the impacts column, the red and blue horizontal bands (lines) show the expected (observed) impact of each nuisance parameter on the cross section modifier, defined as the change in $\mu_{\ttg}^{\text{prod}}$ when the nuisance parameter is shifted by one standard deviation from its fitted value.}
\label{fig:impacts_inclusive}
\end{figure}

An assessment of the quality of the fit to the production component of \ttg is shown in Fig.~\ref{fig:impacts_inclusive} through the postfit deviations of nuisance parameters from their prefit values (``pulls'') as well as their postfit reduction in uncertainty (``constraints''). The nuisance parameters are ranked by the impact that shifting their fitted values would have on the central value of the cross section modifier. In the figure, if the data-taking period is specified alongside the uncertainty name, it indicates that this component of the uncertainty is uncorrelated between periods. Similarly, if "norm." or "shape" is specified alongside the uncertainty name, it indicates an uncertainty modifying only the normalization or only the shape of the template histograms, respectively. The normalization of the \ttg decay component is an unconstrained nuisance in the fit, and its value and uncertainty after the fit are shown in the respective row, instead of the constraint. Only the uncertainty on the measurement is quoted in the figure, not the uncertainty on the theoretical prediction used to normalize the \ttg decay component. A goodness of fit test is performed using the saturated model~\cite{Gross:2010qma}, and the $p$-value is found to be 0.71, indicating a good agreement between the data and the postfit predictions.

\subsection{Differential measurements}

Additionally, we report the measurement of the differential \ttg cross section as a function of several observables of interest at parton and particle level. At the parton level, the cross section is measured as a function of \toppt, the \DR between the photon and the closest top quark (\DRgt), the \DR between the photon and the \ttbar system (\DRgtt), and \mttbar. The observables chosen for the measurements at the particle level are \llpt, \gammapt, and the $\Delta\phi$ between the two charged leptons (\Dphill). Observables such as the \pt of particles are sensitive to new physics in the tails of the distributions. Angular variables, on the other hand, are sensitive to the spin correlations of the top quark pair, to the coupling between the top quark and the photon, and to the modelling of the origin of the photon. Moreover, angular observables can be used to probe the charge-parity ($CP$) structure of the SM~\cite{Bevilacqua:2018dny}. Measurements at the parton level are performed in a broader phase space, whereas those at the particle level are performed on particle-level objects in a fiducial phase space, as described in Section~\ref{crosssection_def}.

In the unfolding process, the condition numbers of the response matrices relating reconstructed and parton level quantities were computed for all variables, and the largest number was found to be 11.4, for \toppt, well within the range of stability. Each differential measurement is the result of a fit to two distributions: that of the variable of interest in the SR and the number of jets in the \Zg CR\@.

\begin{figure}[!htp]
\centering
\includegraphics[width=0.495\textwidth]{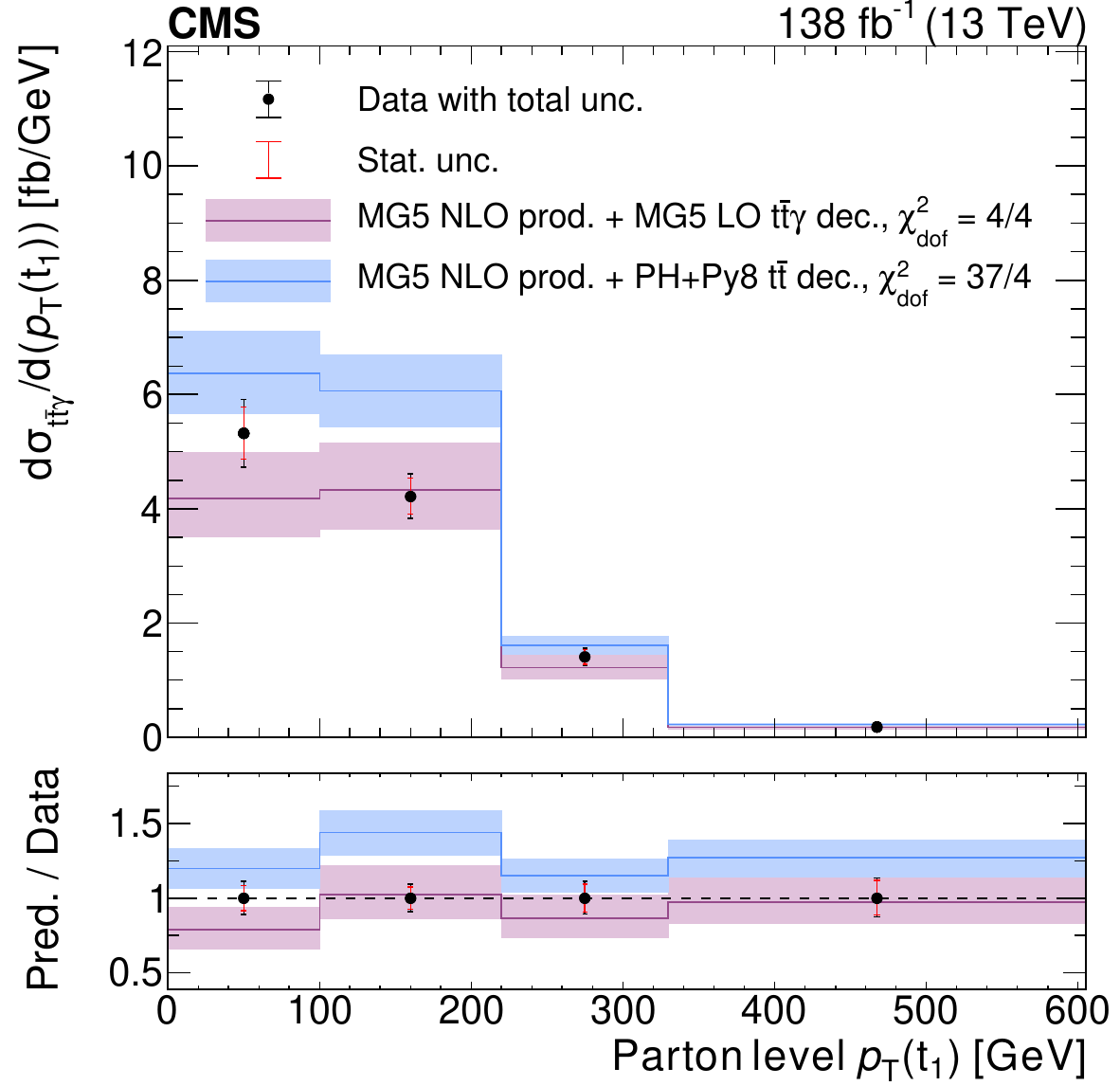}%
\hfill%
\includegraphics[width=0.495\textwidth]{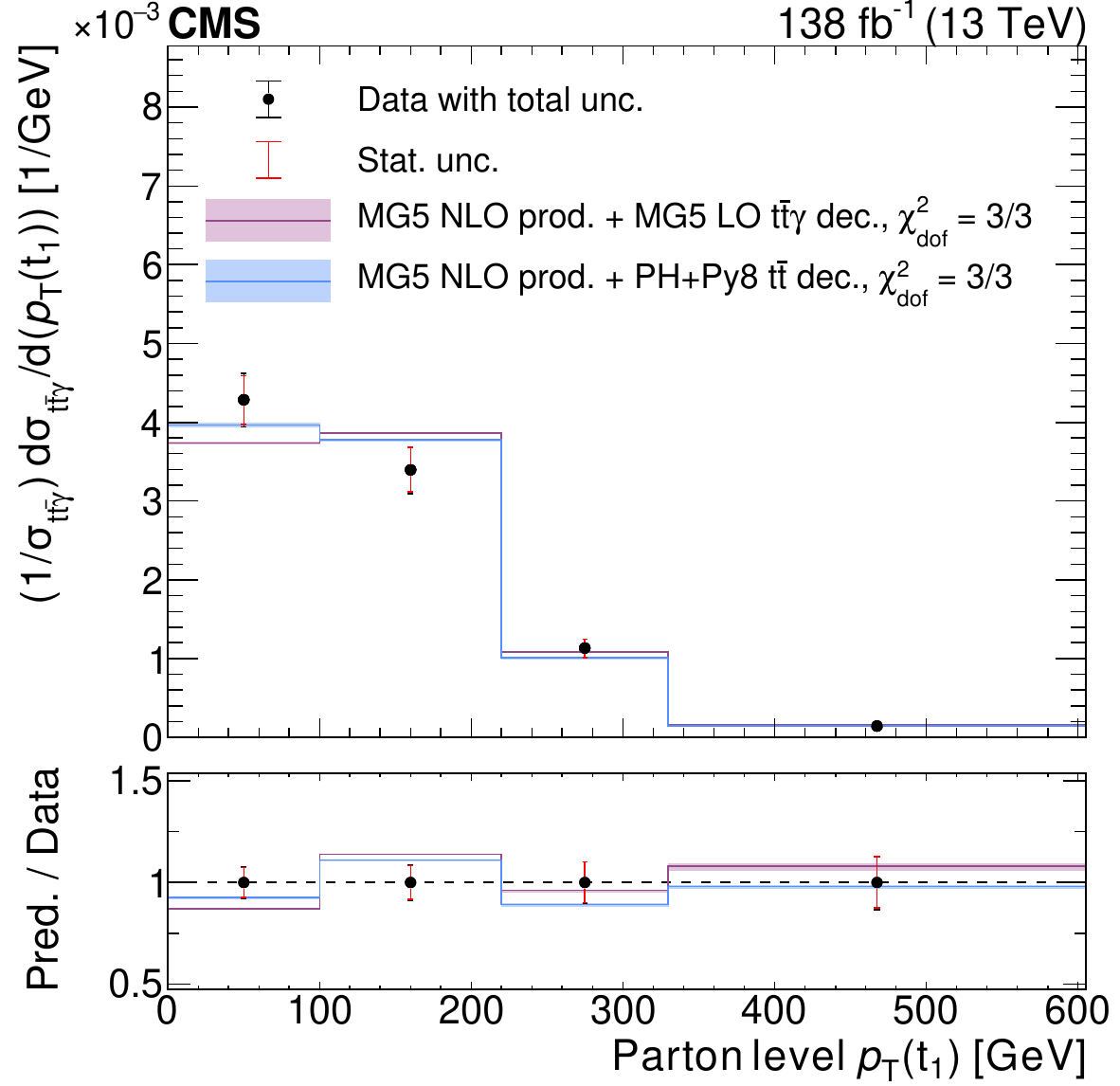} \\
\includegraphics[width=0.495\textwidth]{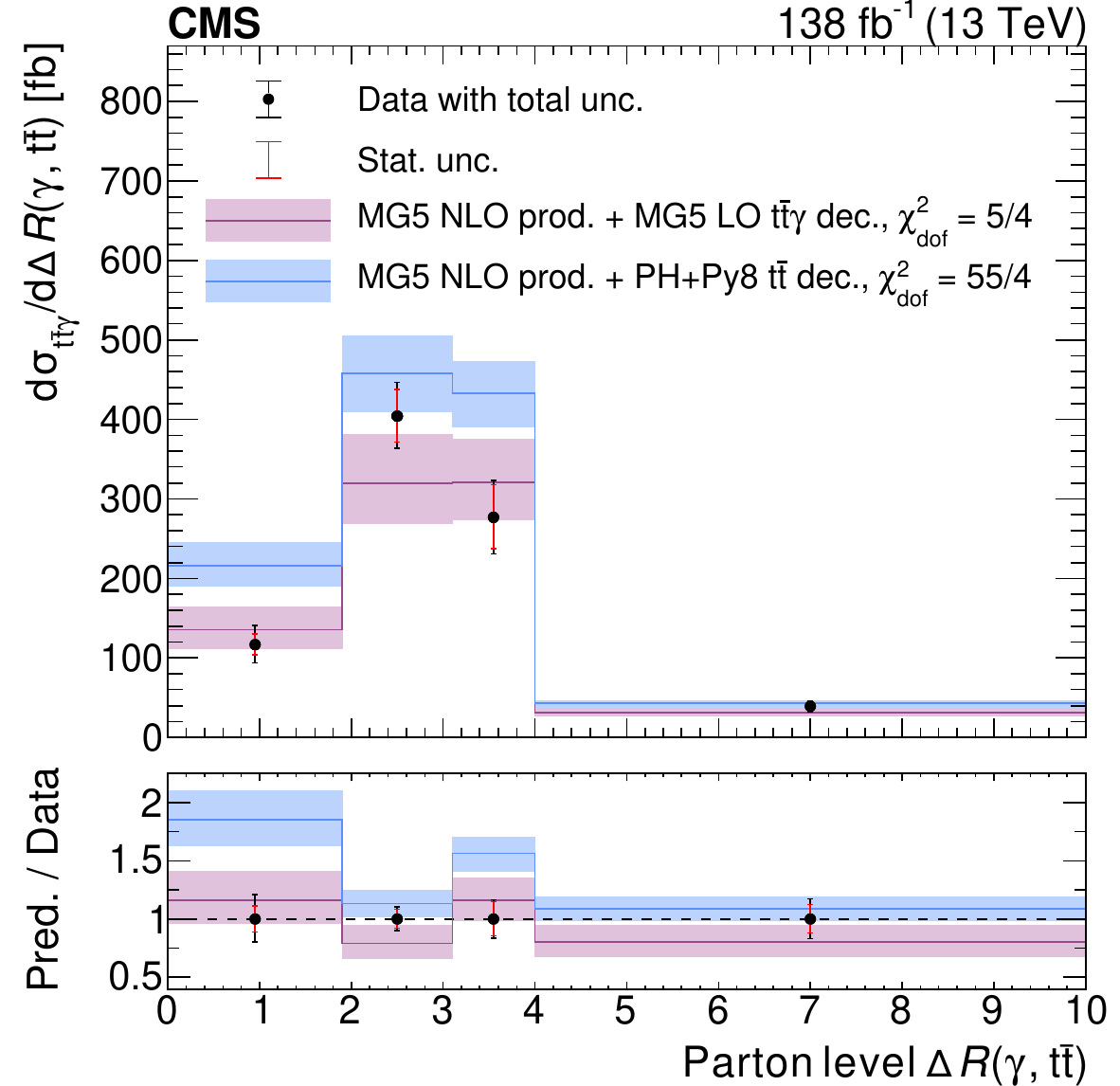}%
\hfill%
\includegraphics[width=0.495\textwidth]{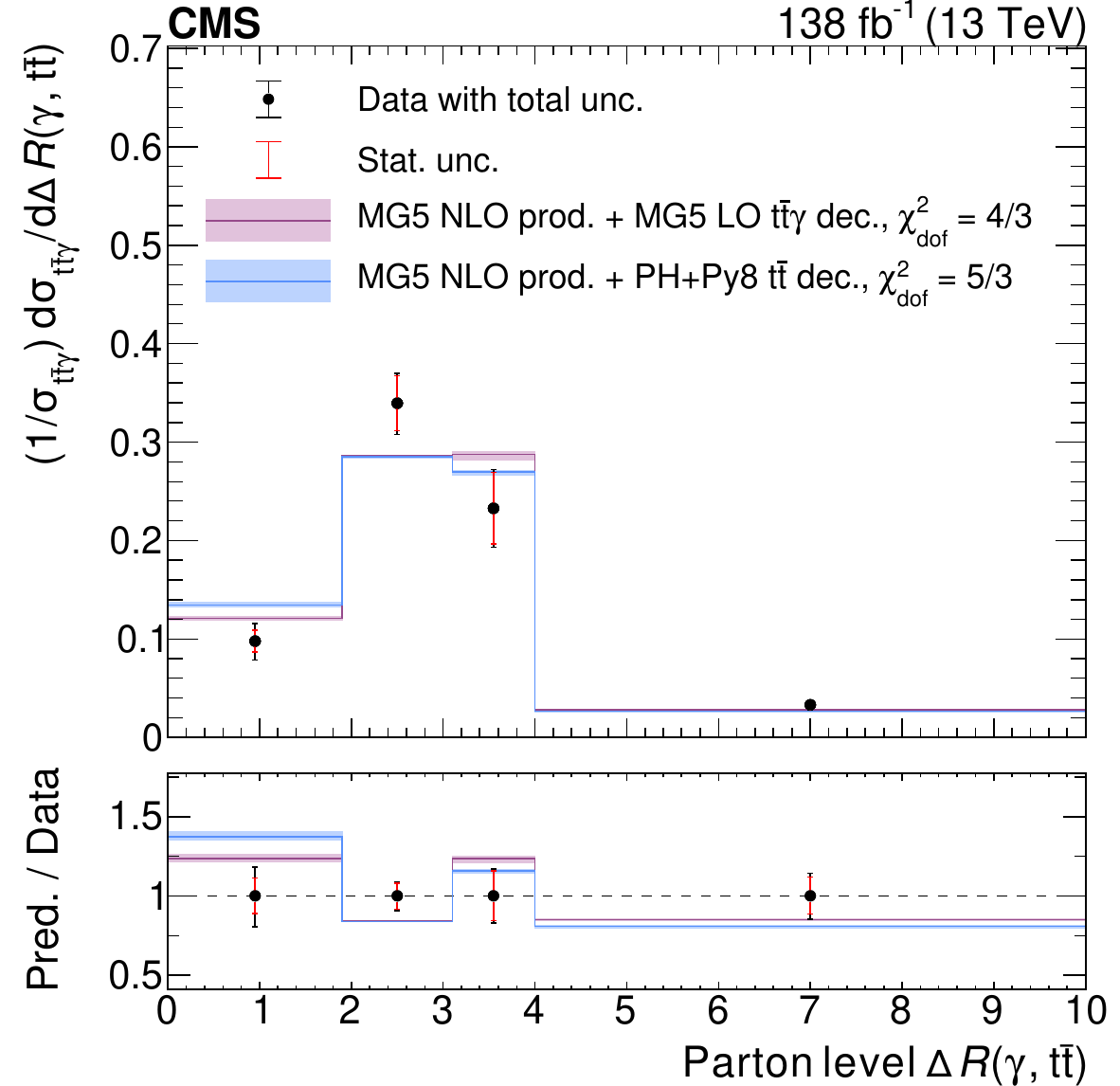}
\caption{Absolute (left) and normalized (right) differential \ttg cross sections at the parton level as a function of \toppt (upper) and \DRgtt (lower). The purple (blue) lines show the predictions from the nominal (alternative) simulation, and the lighter purple (blue) shaded areas represent the theoretical uncertainties in the predictions. In the legends, ``MG5'' refers to \MGvATNLO, while ``PH+Py8'' refers to \POWHEG and \PYTHIA. The theoretical uncertainties include the choice of \muR and \muF and PDFs, including \alpS variations. The black points represent the measured values, with the total uncertainty, while the red error bar shows the results considering only the statistical uncertainty.}
\label{fig:diff1}
\end{figure}

\begin{figure}[!htp]
\centering
\includegraphics[width=0.495\textwidth]{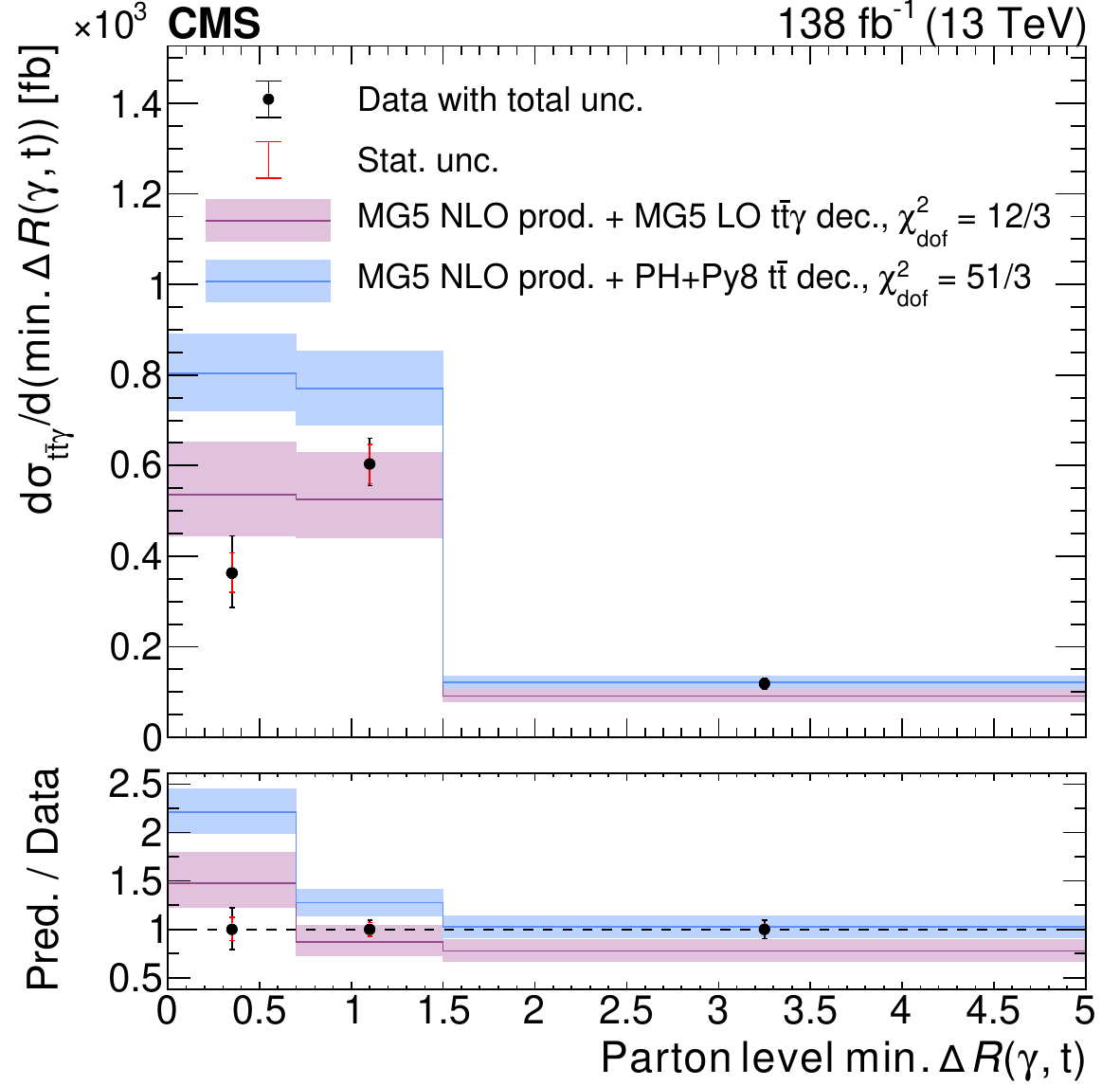}%
\hfill%
\includegraphics[width=0.495\textwidth]{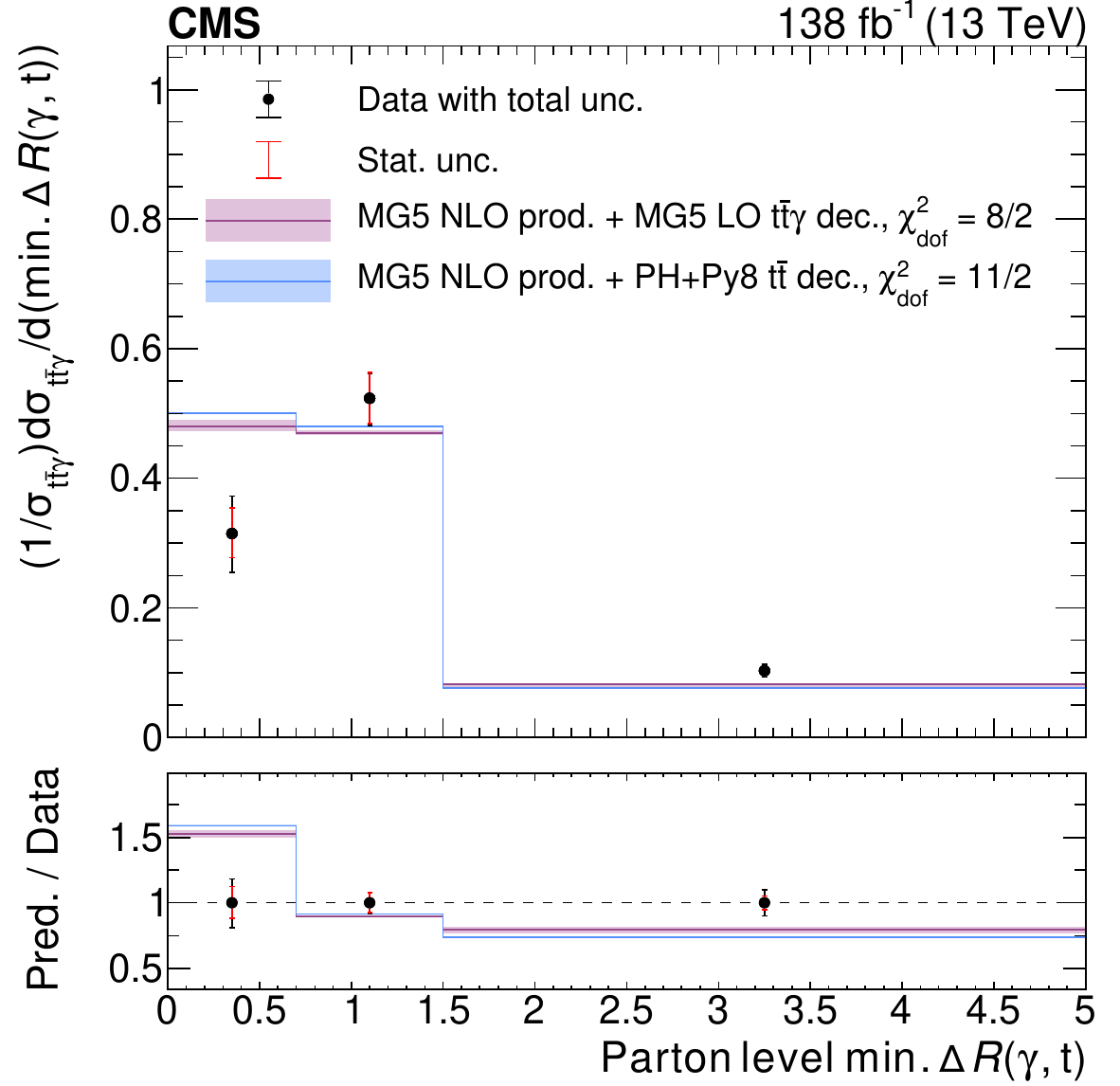} \\
\includegraphics[width=0.495\textwidth]{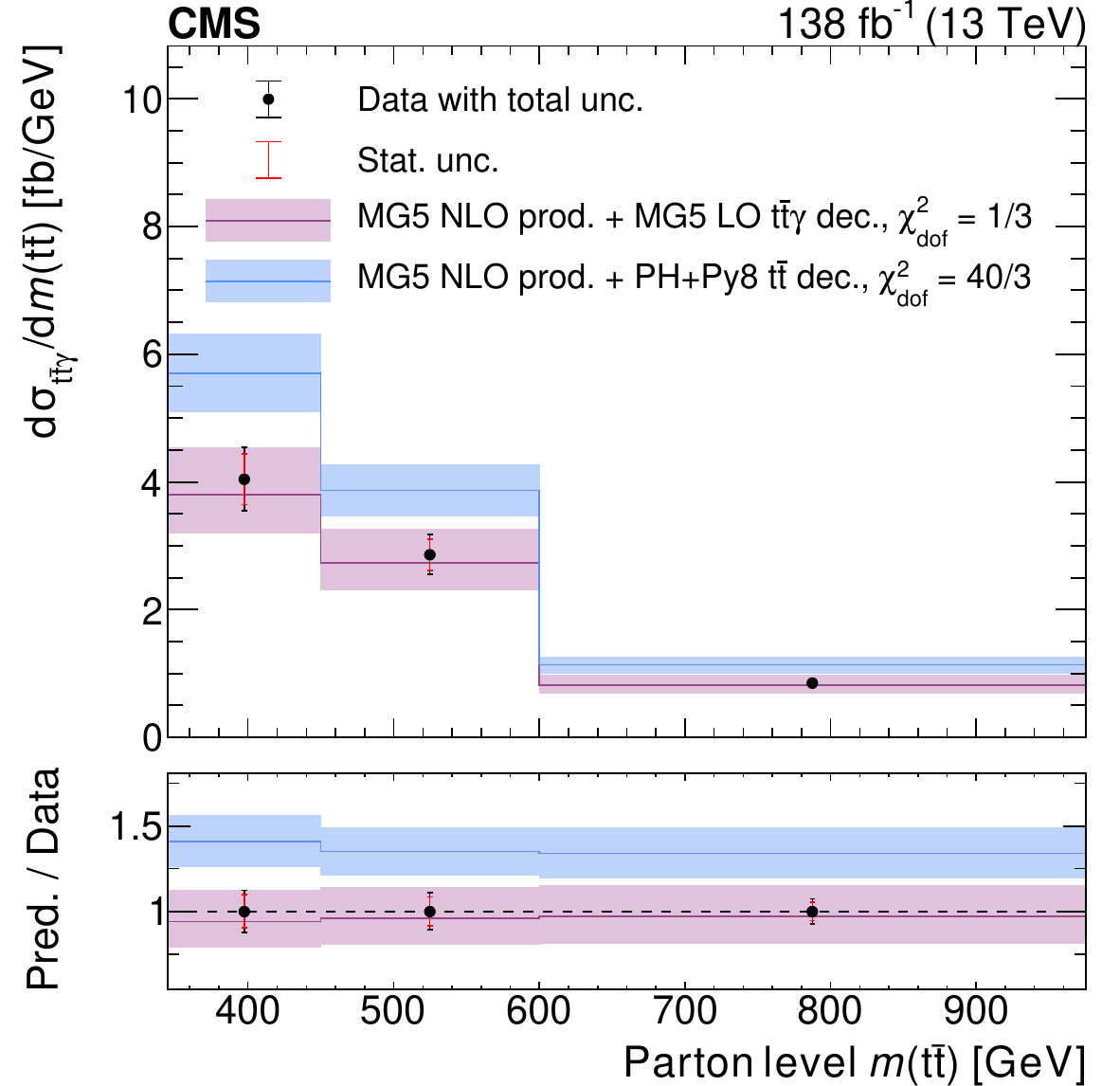}%
\hfill%
\includegraphics[width=0.495\textwidth]{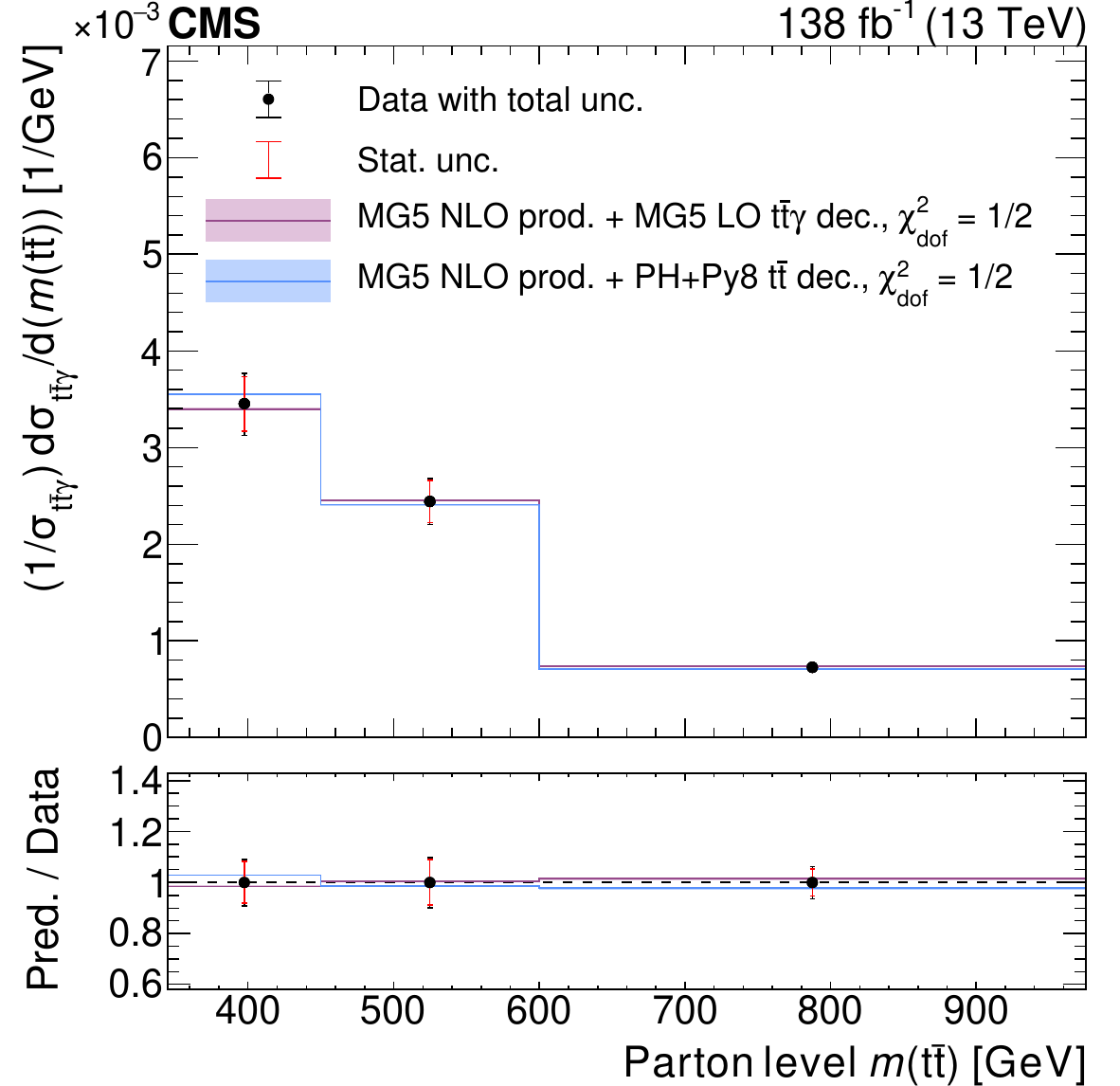}
\caption{Absolute (left) and normalized (right) differential \ttg cross sections at the parton level as a function of the \DRgt (upper) and \mttbar (lower). The purple (blue) lines show the predictions from the nominal (alternative) simulation, and the lighter purple (blue) shaded areas represent the theoretical uncertainties in the predictions. In the legends, ``MG5'' refers to \MGvATNLO, while ``PH+Py8'' refers to \POWHEG and \PYTHIA. The theoretical uncertainties include the choice of \muR and \muF and PDFs, including \alpS variations. The black points represent the measured values, with the total uncertainty, while the red error bar shows the results considering only the statistical uncertainty.}
\label{fig:diff2}
\end{figure}

\begin{figure}[!htp]
\centering
\includegraphics[width=0.495\textwidth]{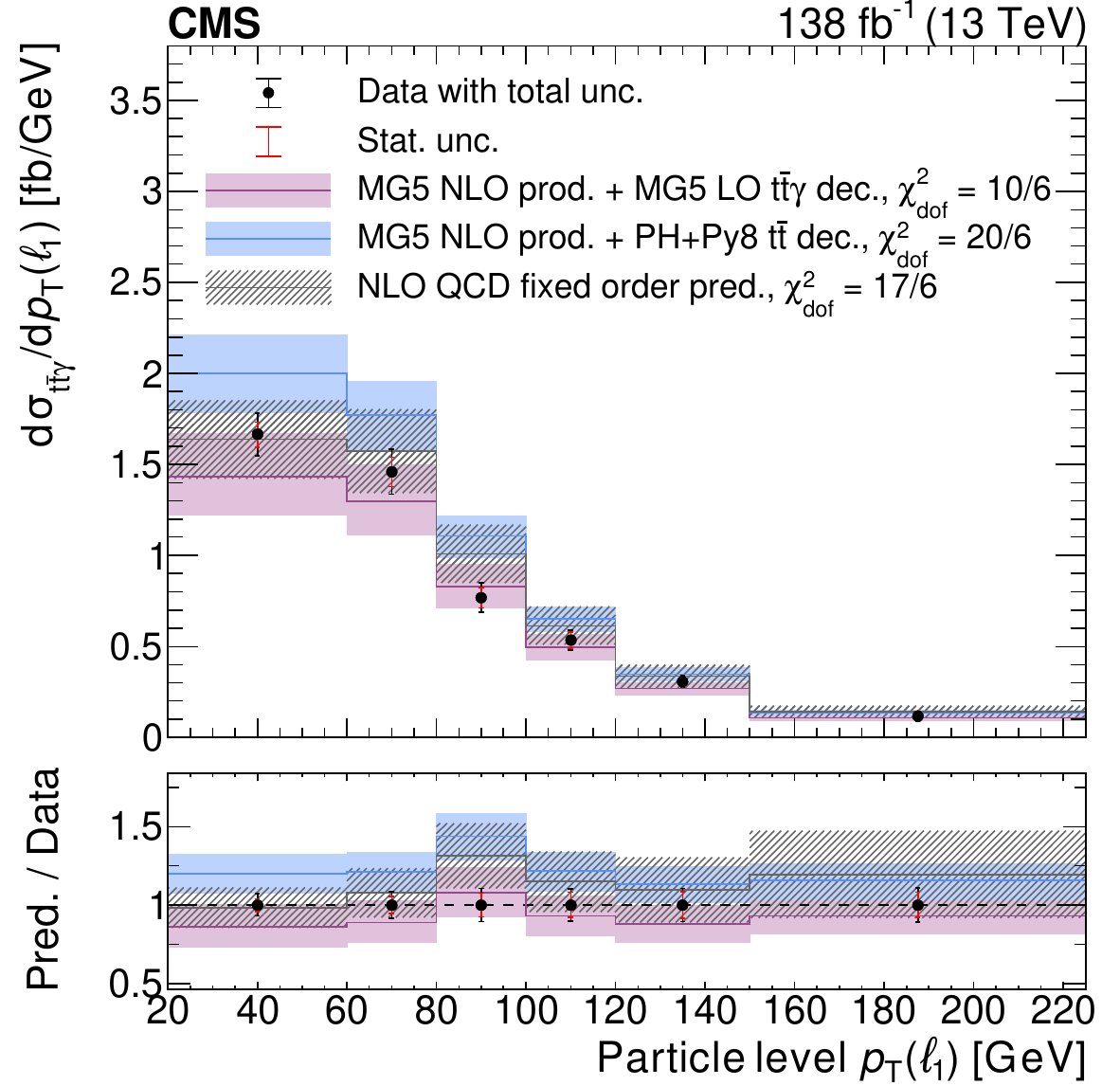}%
\hfill%
\includegraphics[width=0.495\textwidth]{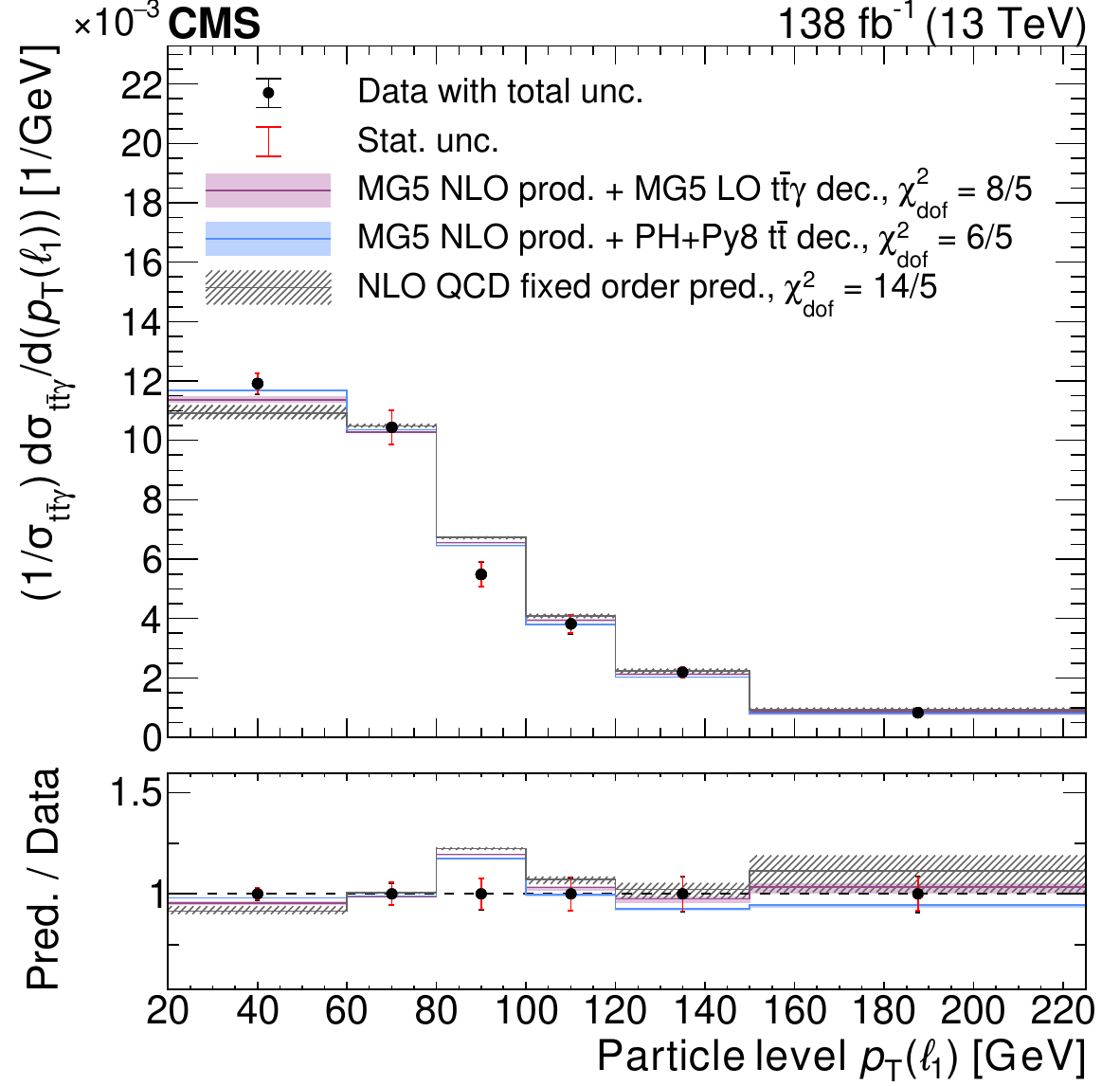} \\
\includegraphics[width=0.495\textwidth]{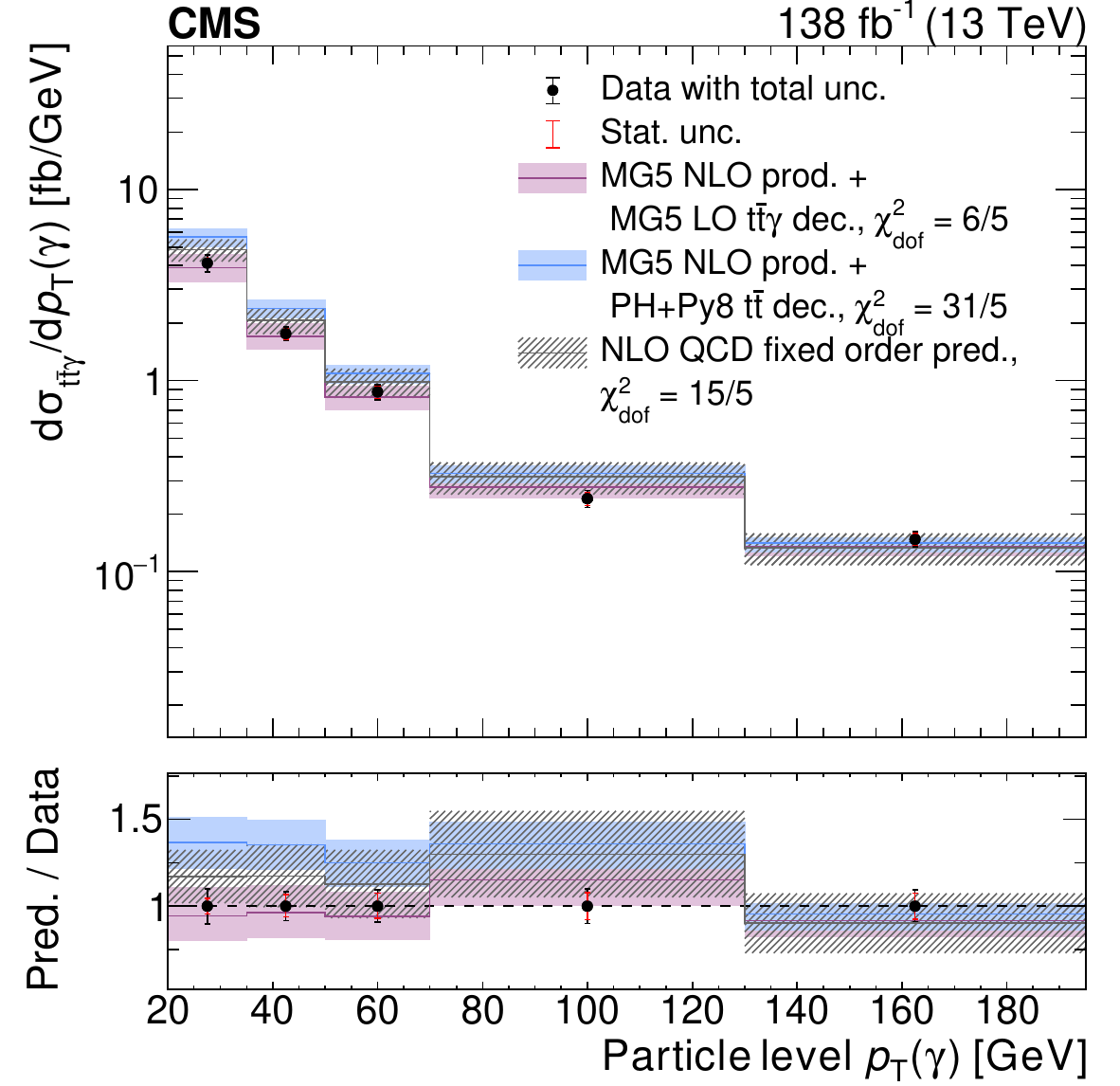}%
\hfill%
\includegraphics[width=0.495\textwidth]{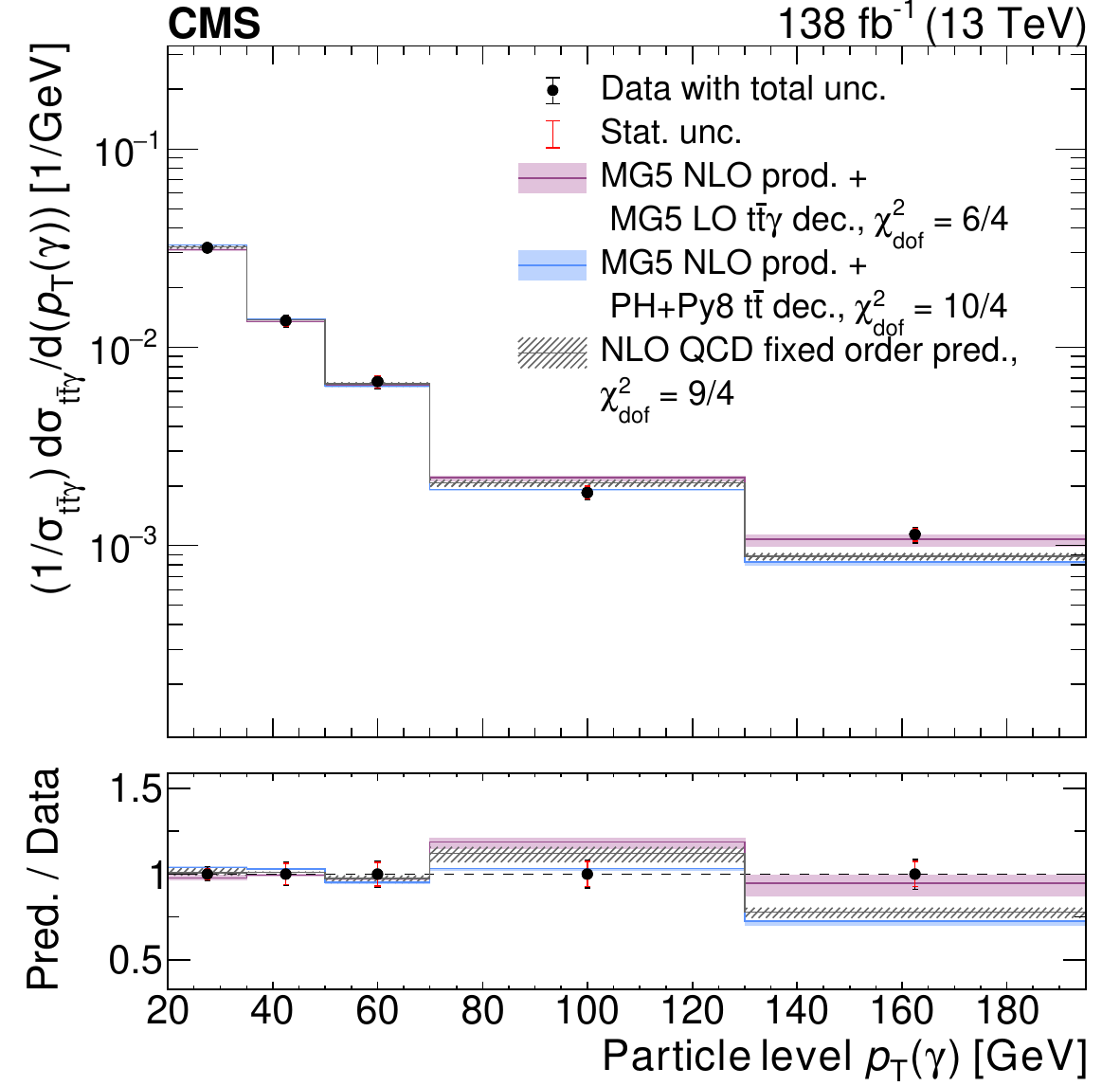}
\caption{Absolute (left) and normalized (right) differential \ttg cross sections at the particle level as a function of \llpt (upper) and \gammapt (lower). The purple (blue) lines show the predictions from the nominal (alternative) simulation, and the lighter purple (blue) shaded areas represent the theoretical uncertainty in the predictions. The gray lines and bands represent the fixed-order prediction and their respective uncertainty. In the legends, ``MG5'' refers to \MGvATNLO, while ``PH+Py8'' refers to \POWHEG and \PYTHIA. The theoretical uncertainty includes the choice of \muR and \muF and PDFs, including \alpS variations. The black points represent the measured values, with the total uncertainty, while the red error bar shows the results considering only the statistical uncertainty.}
\label{fig:diff3}
\end{figure}

\begin{figure}[!ht]
\centering
\includegraphics[width=0.495\textwidth]{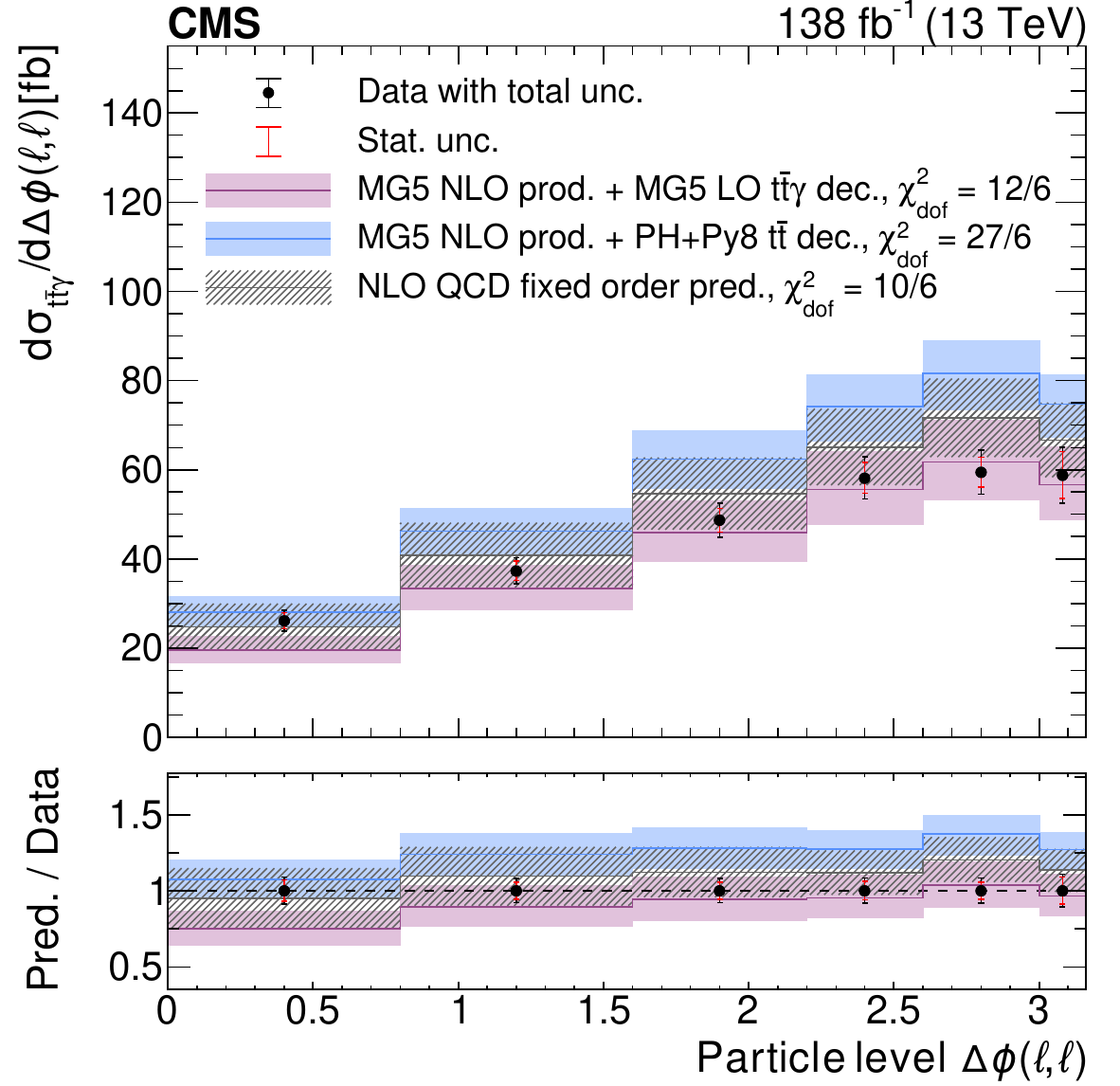}%
\hfill%
\includegraphics[width=0.495\textwidth]{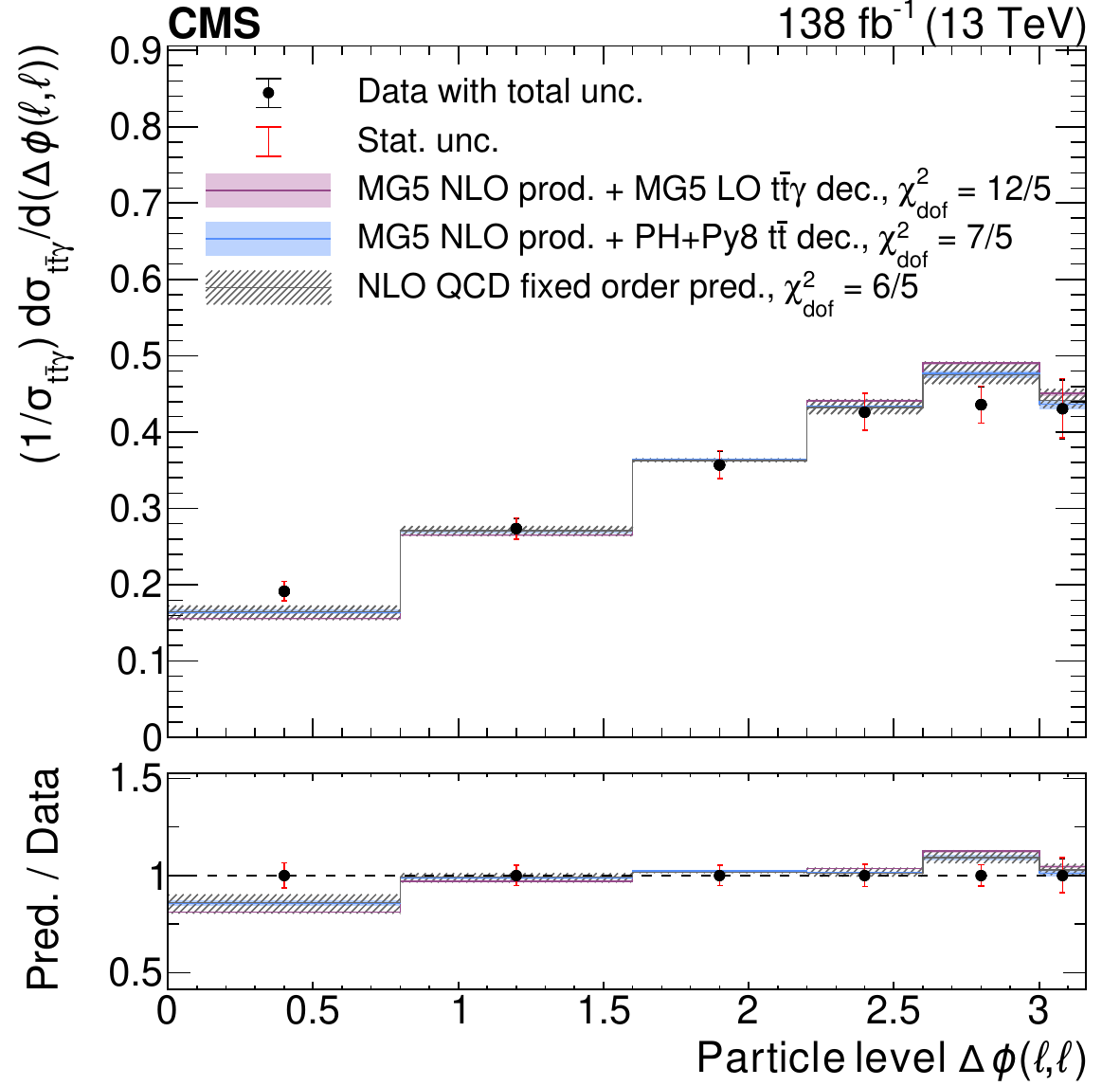}
\caption{Absolute (left) and normalized (right) differential \ttg cross sections at the particle level as a function of the \Dphill. The purple (blue) lines show the predictions from the nominal (alternative) simulation, and the lighter purple (blue) shaded areas represent the theoretical uncertainty in the predictions. The gray lines and bands represent the fixed-order prediction and their respective uncertainty. In the legends, ``MG5'' refers to \MGvATNLO, while ``PH+Py8'' refers to \POWHEG and \PYTHIA. The theoretical uncertainty include the choice of \muR and \muF and PDFs, including \alpS variations. The black points represent the measured values, with the total uncertainty, while the red error bar shows the results considering only the statistical uncertainty.}
\label{fig:diff4}
\end{figure}

The absolute and normalized differential cross sections at the parton level are displayed in Figs.~\ref{fig:diff1} and~\ref{fig:diff2}, while those at the particle level are shown in Figs.~\ref{fig:diff3} and~\ref{fig:diff4}. The purple and blue lines represent the predictions using the two modelling options described in Section~\ref{sec:sim}, while the hashed bands show the theoretical uncertainty in those predictions, coming from the choice of \muR and \muF, and the choice of PDFs including \alpS variations. For particle-level measurements, the results are also compared to the fixed-order prediction introduced in Section~\ref{sec:fixedorder}, shown in gray. The measurements are limited by the statistical uncertainties in most bins, with the exception of the most populated bins in some distributions. The leading systematic uncertainties affecting the unfolded data are those in the photon and electron identification, as well as the nonprompt photon contribution. The normalizations of the \twg process and other backgrounds also play a significant role.

In terms of normalization, good agreement of the data with the nominal simulation is observed, while the alternative simulation model systematically overpredicts the data. The normalized cross sections show, however, that the shape of the data is well described by both models for the invariant masses and \pt observables, while for angular observables some trends are observed. The latter are especially sensitive to the modelling of the photon origin, showing that no model describes the photon emission perfectly. In general, it can be concluded that the alternative simulation model, which relies on the PS to model the photon emissions from the top quark decay products, does not describe the data well, while the nominal simulation, which relies on the ME calculations at LO, provides a better, though not ideal, description.

The fixed-order predictions are in good agreement with the data both in terms of shape and normalization, especially for the angular observables, where they are expected to be more accurate, providing a better description when compared to both simulation models.

\section{Inclusive and differential ratio measurements}
\label{sec:ratio}

\begin{figure}[!ht]
\centering
\includegraphics[width=0.495\textwidth]{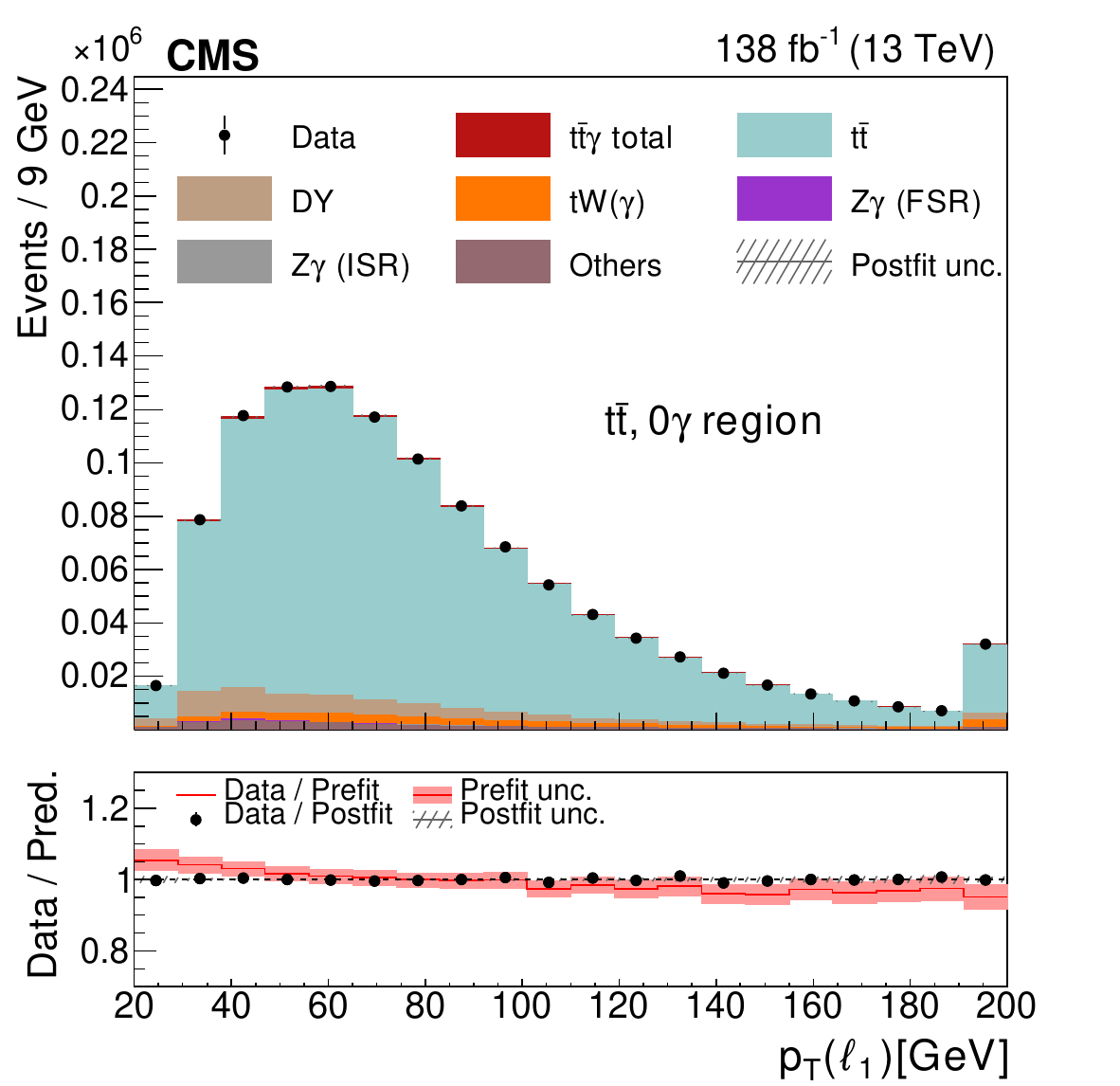}%
\hfill%
\includegraphics[width=0.495\textwidth]{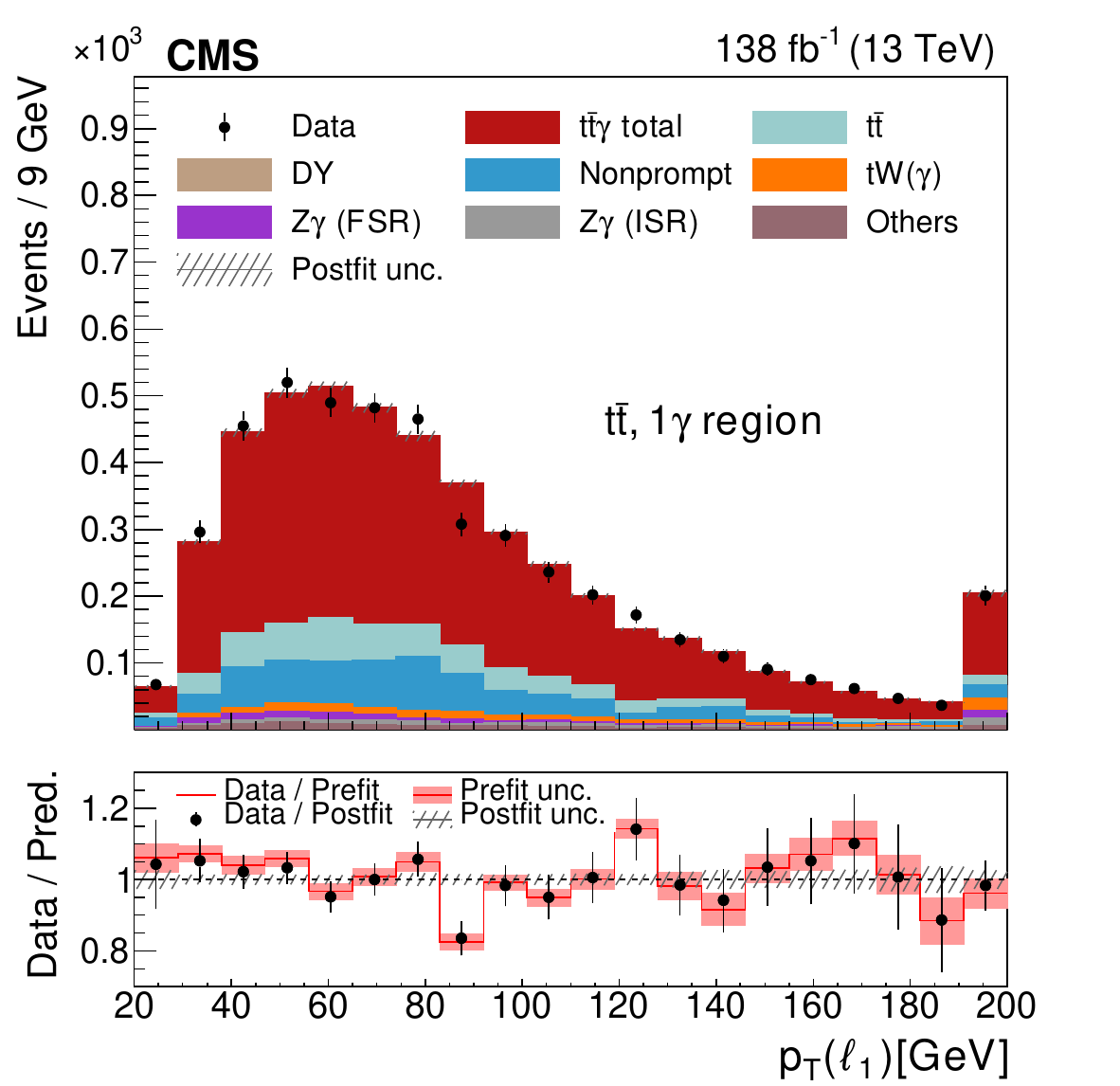} \\
\includegraphics[width=0.495\textwidth]{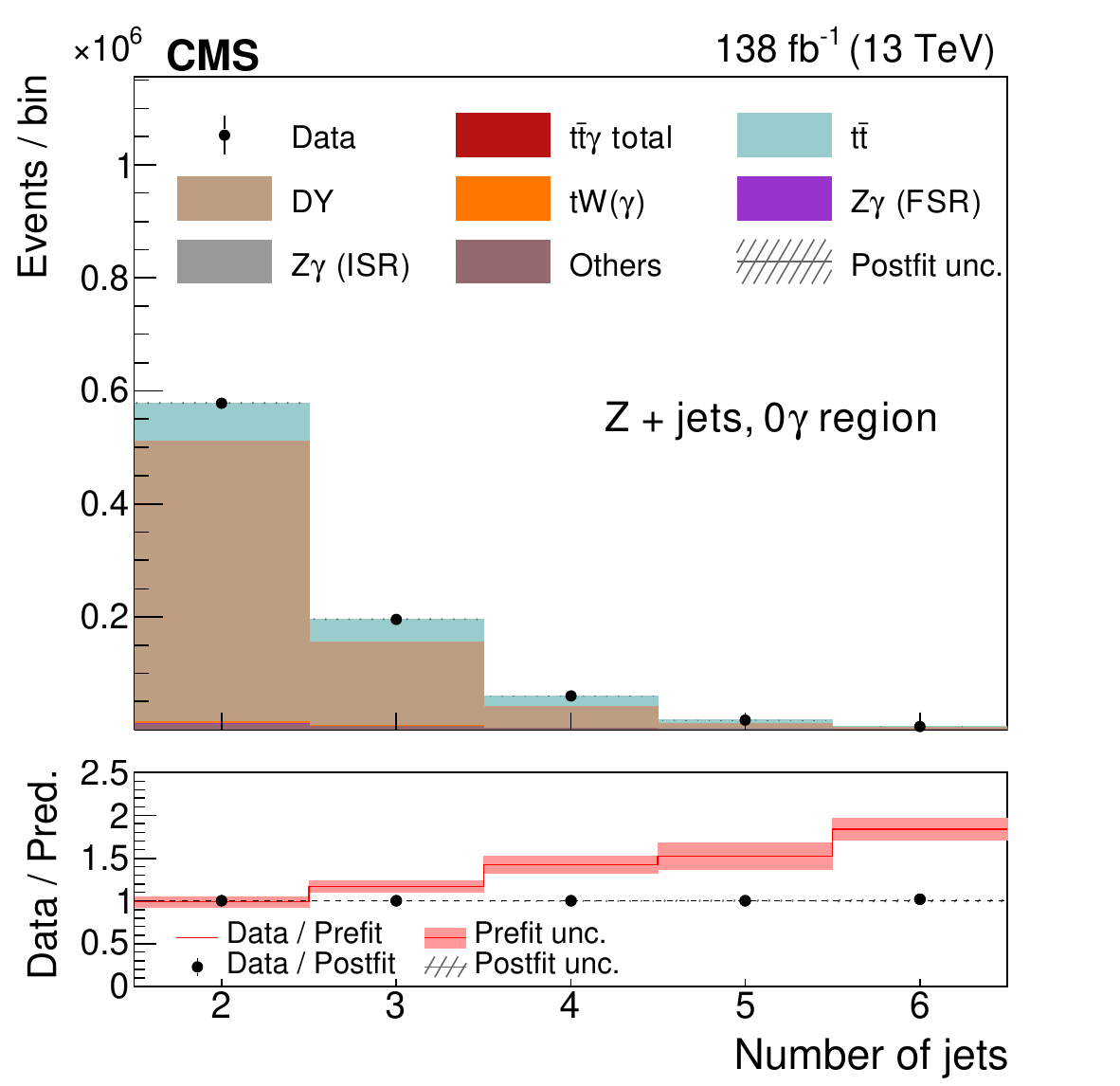}%
\hfill%
\includegraphics[width=0.495\textwidth]{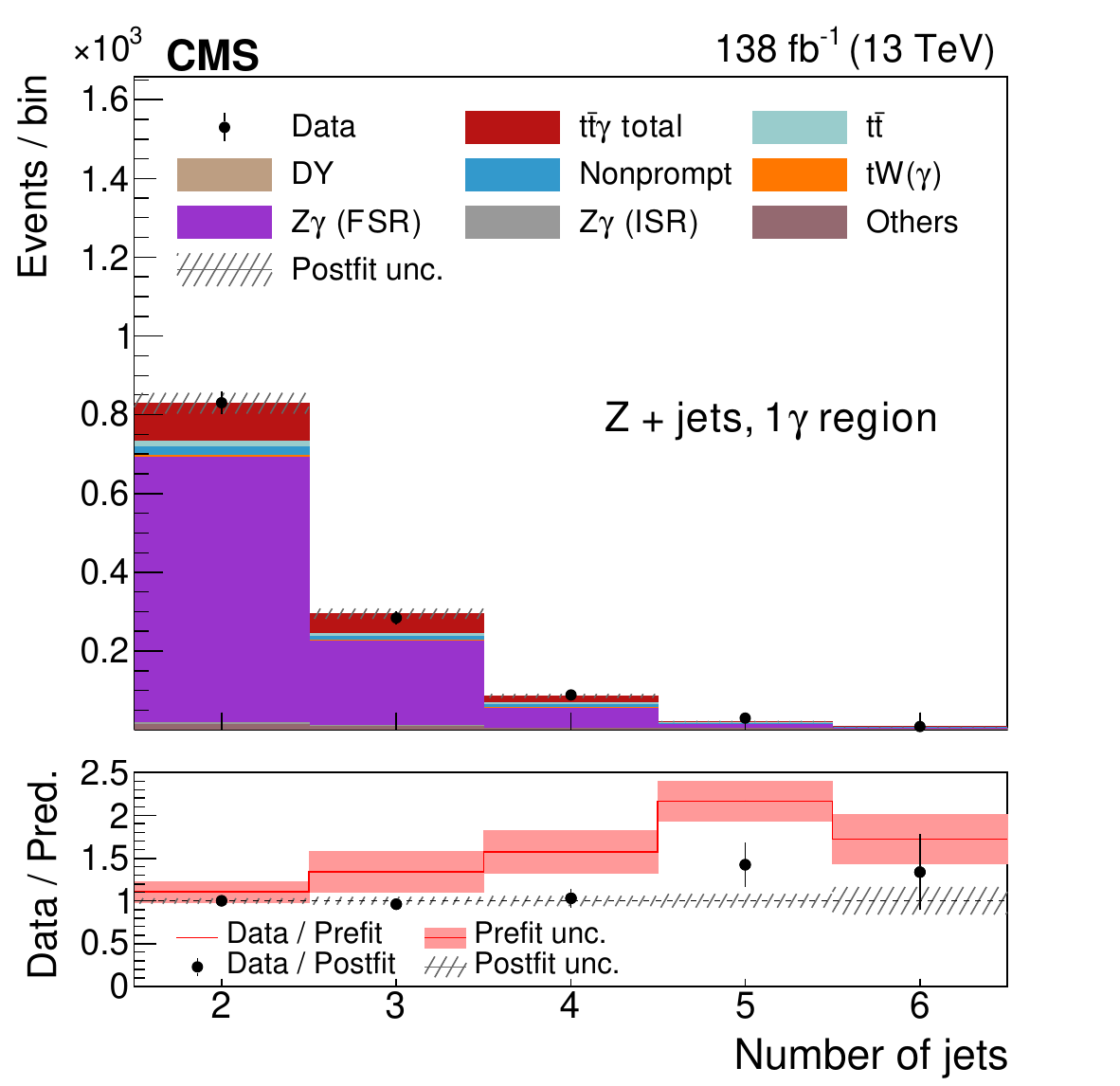}
\caption{Distribution of the \pt of the leading lepton for the ``$\ttbar, 0 \PGg$'' region (upper left) and the ``$\ttbar, 1 \PGg$'' SR (upper right), and the number of jets for the \DY (lower left) and \Zg (lower right) CRs after the fit. The hatched area indicates the total uncertainty in the prediction. The lower panels show the ratio of the data to the sum of the postfit predictions (points) and the ratio of the data to the sum of the prefit predictions (red line).}
\label{fig:inclratio}
\end{figure}

In order to extract the \ratio modifier $\mu_{\ratio} = \ratio/\ratioexp$ introduced in Eq.~\eqref{eq:ratio}, the distributions of the \pt of the leading lepton for the ``$\ttbar,1\PGg$'' region and the ``$\ttbar,0\PGg$'' region, and the number of jets distributions in the two CRs are fitted simultaneously. The CRs are the \Zg CR and a \DY CR, built similarly to the ``$\ttbar,0\PGg$'' region, but selecting events with same-flavour leptons and \mll within 15\GeV of \mZ. The fit is performed to four distributions: that of the \llpt in the ``$\ttbar,1\PGg$'' and the ``$\ttbar,0\PGg$'' regions, and the number of jets in the \Zg and \DY CRs. The distributions after the fit are shown in Fig.~\ref{fig:inclratio}. In this section, ``\ttg'' refers to the sum of the \ttg production and decay processes. This measurement is performed in the extended phase space used for the parton-level measurements but without the photon requirement, where the predicted value of the ratio is $0.0127\pm0.0008$, computed using the nominal simulation samples. The uncertainty on this value comes from variations in the ME \muR and \muF, and PDFs, including \alpS variations, all treated as uncorrelated between the \ttbar and \ttg samples. In this measurement, the contribution from \ttbar with nonprompt photons in the ``$\ttbar,1\PGg$ ''region is estimated from simulation, unlike in the cross section measurements, where it is part of the nonprompt-photon background that is taken from data. This procedure ensures that correlations between the \ttg and \ttbar process without associated photons are properly accounted for. The residual nonprompt-photon background is then estimated from data after subtracting the \ttbar contribution obtained from simulation.

{\tolerance=800
We measure $\mu_{\ratio}=1.05\pm0.06$, corresponding to an inclusive ratio of $\ratio=0.0133\pm0.0002\stat\pm0.0005\syst$, compatible with the prediction. The \ttbar normalization factor is found to be $\mu_{\ttbar}=1.008\pm0.015$, consistent with the predictions at NNLO in QCD\@.
The systematic uncertainties with the largest impact in the \ratio measurement are those related to the photon identification and the estimation of the nonprompt photon background (with an impact of about 2\% each), followed by modelling uncertainties such as FSR and the \muF choice, and the normalization of the \Zg and other minor backgrounds (with impacts below 1\% each).
\par}

\begin{figure}[!ht]
\includegraphics[width=0.495\textwidth]{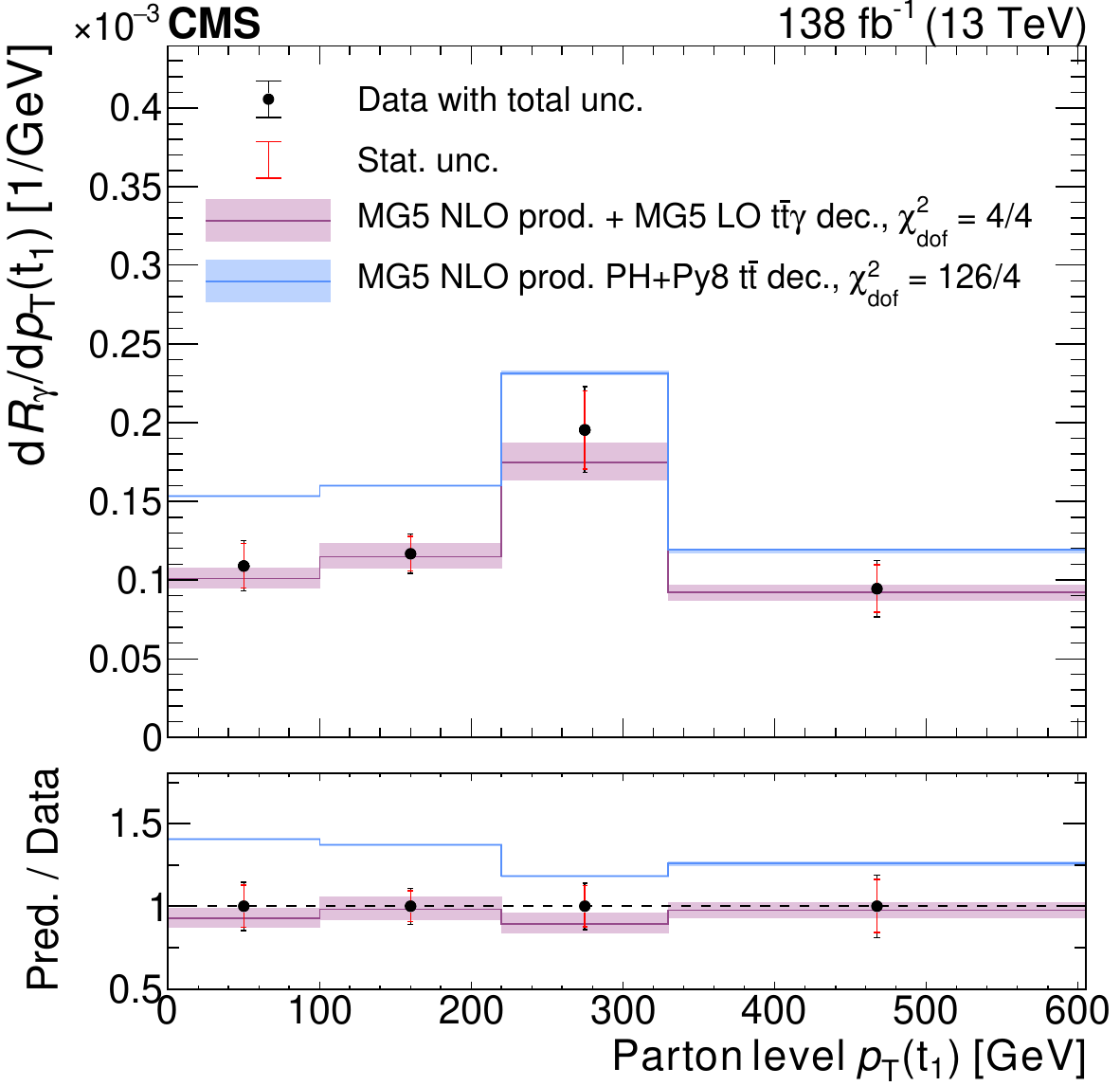}%
\hfill%
\includegraphics[width=0.495\textwidth]{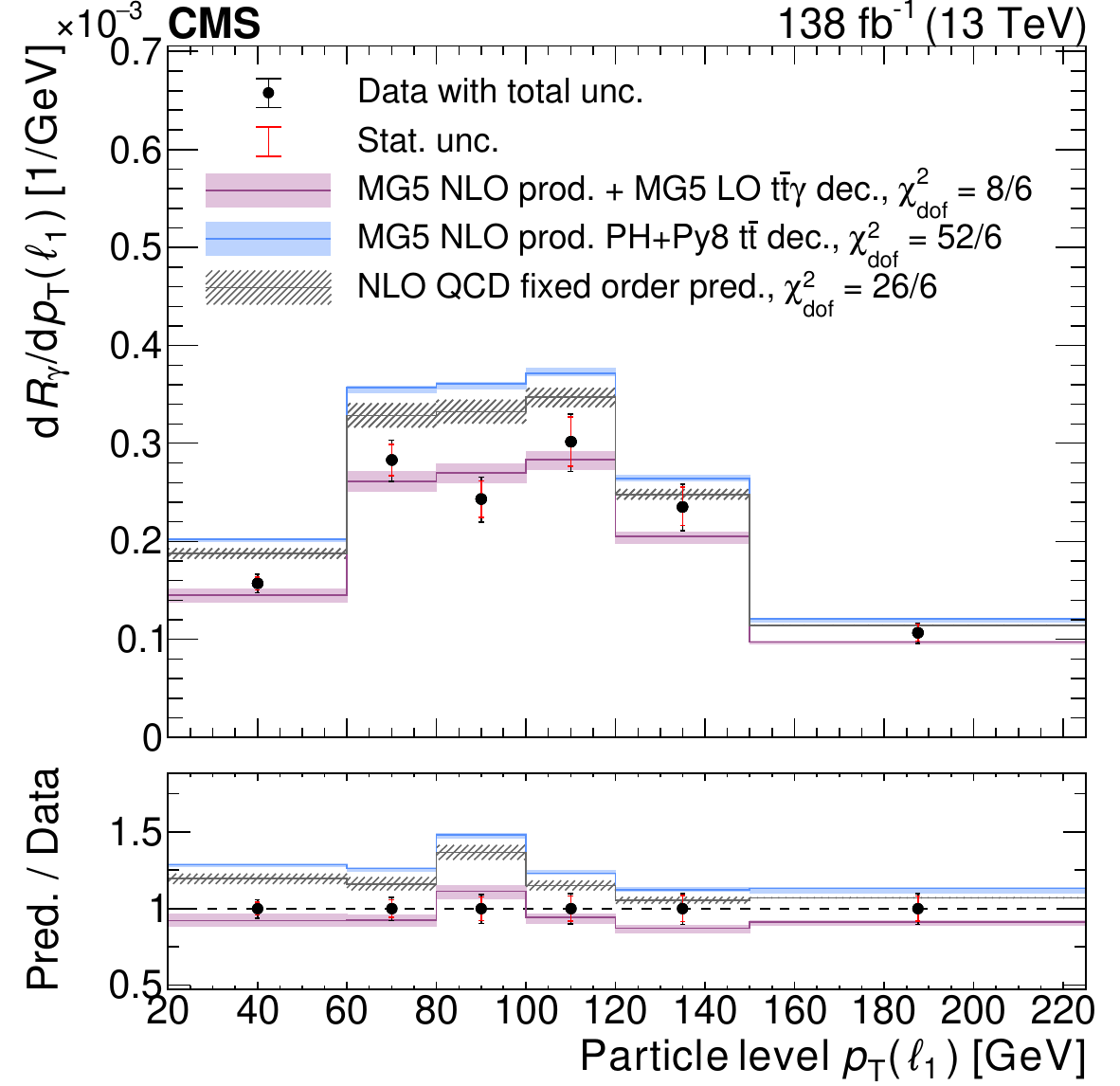}
\caption{Absolute differential measurements of \ratio as a function of \toppt at the parton level (left) and \llpt at the particle level (right). The purple (blue) lines show the predictions from the nominal (alternative) simulation, and the lighter purple (blue) shaded areas represent the theoretical uncertainties in the predictions. The gray lines and bands represent the fixed-order predictions and their respective uncertainties. In the legends, ``MG5'' refers to \MGvATNLO, while ``PH+Py8'' refers to \POWHEG and \PYTHIA. The theoretical uncertainties include the choice of \muR and \muF and PDFs, including \alpS variations. The black points represent the measured values, with the total uncertainty, while the red error bar shows the results considering only the statistical uncertainty.}
\label{fig:diffratio1}
\end{figure}

Similarly, we perform a differential measurement of \ratio as a function of \toppt, at the parton level, and of the \llpt, at the particle level. The absolute differential cross section ratios are displayed in Fig.~\ref{fig:diffratio1}. Each result is extracted from a fit to four distributions: that of the variable of interest in the ``$\ttbar,1\PGg$'' and the ``$\ttbar,0\PGg$'' regions, and the number of jets in the \Zg and \DY CRs. The results are in agreement with the nominal prediction and with the fixed-order calculation, while the alternative prediction overpredicts the data. These results are limited by the statistical uncertainty. The systematic uncertainties with the largest contribution are those in the photon identification efficiency and in the reweighting of the top quark \pt distribution to NNLO in QCD\@.

\section{Measurement of the top quark charge asymmetry in \texorpdfstring{\ttg}{ttg}}
\label{sec:charge_asym}

The top quark charge asymmetry in \ttg production is measured in the extended phase space at the parton level, through a fit to the difference in absolute rapidity between the top quark and antiquark in the SR, shown in Fig.~\ref{fig:topdistributions}, and the jet multiplicity in the \Zg CR\@. The fit procedure and parametrization are described in Section~\ref{sec:stat}. The measured value of the charge asymmetry is $A_C=(-1.2\pm4.1\stat\pm0.9\syst)\%$, consistent with the expectation of $(-0.4\pm0.1)\%$, obtained from the \MGvATNLO MC simulation of \ttg production, where the uncertainty accounts for the choice of \muR and \muF, and PDFs, including \alpS variations. The \ttg decay process is simulated without including any charge asymmetry. The precision of this measurement is heavily limited by the statistical uncertainty, both of the data and the simulated signal samples, and further investigations will be needed as new data become available, in order to have sensitivity to new physics scenarios that might alter the expected value.

\section{Summary}
\label{sec:sum}

A comprehensive study of the top quark pair (\ttbar) production in association with a photon (\PGg) at the LHC is presented, using data collected by the CMS experiment in 2016--2018 at a centre-of-mass energy of 13\TeV, and corresponding to an integrated luminosity of \il. Inclusive and differential measurements are performed in the dilepton decay channels, in a fiducial region at the particle level including events with photon transverse momentum larger than 20\GeV.

The inclusive fiducial cross section, defined at the particle level, for \ttg with a photon radiated at any stage of the process is $137\pm3\stat\pm7\syst\fb$, while the cross section for events with a photon radiated at the production stage of the process is $56\pm2\stat\pm4\syst\fb$. The measured cross sections agree with the predictions from the standard model (SM) for the combined \ttg process and for the \ttg process with photons from the production stage.

The cross section is also measured differentially, in bins of seven different observables, related to the kinematic properties and topology of the photon, the leptons, and the top quarks reconstructed in the event. Measurements are performed at the particle or parton level, in different phase spaces, depending on the observable. The predictions from simulation accurately describe the shape of the measured cross sections. The $\ttg / \ttbar$ cross section ratio is measured for the first time, inclusively and differentially. The inclusive ratio is found to be $0.0133\pm0.0002\stat\pm0.0005\syst$, in a phase space defined at the parton level, in agreement with the nominal predictions from simulation. The differential ratios are well described by the predictions, within the total uncertainty. The top quark charge asymmetry in \ttg events is also measured in a phase space defined at the parton level to be $(-1.2\pm4.1\stat\pm0.9\syst)\%$, compatible with both the SM prediction at next-to-leading order in quantum chromodynamics and with no asymmetry.

\begin{acknowledgments}
We congratulate our colleagues in the CERN accelerator departments for the excellent performance of the LHC and thank the technical and administrative staffs at CERN and at other CMS institutes for their contributions to the success of the CMS effort. In addition, we gratefully acknowledge the computing centres and personnel of the Worldwide LHC Computing Grid and other centres for delivering so effectively the computing infrastructure essential to our analyses. Finally, we acknowledge the enduring support for the construction and operation of the LHC, the CMS detector, and the supporting computing infrastructure provided by the following funding agencies: SC (Armenia), BMBWF and FWF (Austria); FNRS and FWO (Belgium); CNPq, CAPES, FAPERJ, FAPERGS, and FAPESP (Brazil); MES and BNSF (Bulgaria); CERN; CAS, MoST, and NSFC (China); MINCIENCIAS (Colombia); MSES and CSF (Croatia); RIF (Cyprus); SENESCYT (Ecuador); ERC PRG, TARISTU24-TK10 and MoER TK202 (Estonia); Academy of Finland, MEC, and HIP (Finland); CEA and CNRS/IN2P3 (France); SRNSF (Georgia); BMFTR, DFG, and HGF (Germany); GSRI (Greece); NKFIH (Hungary); DAE and DST (India); IPM (Iran); SFI (Ireland); INFN (Italy); MSIT and NRF (Republic of Korea); MES (Latvia); LMTLT (Lithuania); MOE and UM (Malaysia); BUAP, CINVESTAV, CONACYT, LNS, SEP, and UASLP-FAI (Mexico); MOS (Montenegro); MBIE (New Zealand); PAEC (Pakistan); MES, NSC, and NAWA (Poland); FCT (Portugal);  MESTD (Serbia); MICIU/AEI and PCTI (Spain); MOSTR (Sri Lanka); Swiss Funding Agencies (Switzerland); MST (Taipei); MHESI (Thailand); TUBITAK and TENMAK (T\"{u}rkiye); NASU (Ukraine); STFC (United Kingdom); DOE and NSF (USA).

\hyphenation{Rachada-pisek} Individuals have received support from the Marie-Curie programme and the European Research Council and Horizon 2020 Grant, contract Nos.\ 675440, 724704, 752730, 758316, 765710, 824093, 101115353, 101002207, 101001205, and COST Action CA16108 (European Union); the Leventis Foundation; the Alfred P.\ Sloan Foundation; the Alexander von Humboldt Foundation; the Science Committee, project no. 22rl-037 (Armenia); the Fonds pour la Formation \`a la Recherche dans l'Industrie et dans l'Agriculture (FRIA) and Fonds voor Wetenschappelijk Onderzoek contract No. 1228724N (Belgium); the Beijing Municipal Science \& Technology Commission, No. Z191100007219010, the Fundamental Research Funds for the Central Universities, the Ministry of Science and Technology of China under Grant No. 2023YFA1605804, the Natural Science Foundation of China under Grant No. 12061141002, 12535004, and USTC Research Funds of the Double First-Class Initiative No.\ YD2030002017 (China); the Ministry of Education, Youth and Sports (MEYS) of the Czech Republic; the Shota Rustaveli National Science Foundation, grant FR-22-985 (Georgia); the Deutsche Forschungsgemeinschaft (DFG), among others, under Germany's Excellence Strategy -- EXC 2121 ``Quantum Universe" -- 390833306, and under project number 400140256 - GRK2497; the Hellenic Foundation for Research and Innovation (HFRI), Project Number 2288 (Greece); the Hungarian Academy of Sciences, the New National Excellence Program - \'UNKP, the NKFIH research grants K 131991, K 133046, K 138136, K 143460, K 143477, K 146913, K 146914, K 147048, 2020-2.2.1-ED-2021-00181, TKP2021-NKTA-64, and 2021-4.1.2-NEMZ\_KI-2024-00036 (Hungary); the Council of Science and Industrial Research, India; ICSC -- National Research Centre for High Performance Computing, Big Data and Quantum Computing, FAIR -- Future Artificial Intelligence Research, and CUP I53D23001070006 (Mission 4 Component 1), funded by the NextGenerationEU program (Italy); the Latvian Council of Science; the Ministry of Education and Science, project no. 2022/WK/14, and the National Science Center, contracts Opus 2021/41/B/ST2/01369, 2021/43/B/ST2/01552, 2023/49/B/ST2/03273, and the NAWA contract BPN/PPO/2021/1/00011 (Poland); the Funda\c{c}\~ao para a Ci\^encia e a Tecnologia, grant CEECIND/01334/2018 (Portugal); the National Priorities Research Program by Qatar National Research Fund; MICIU/AEI/10.13039/501100011033, ERDF/EU, "European Union NextGenerationEU/PRTR", and Programa Severo Ochoa del Principado de Asturias (Spain); the Chulalongkorn Academic into Its 2nd Century Project Advancement Project, the National Science, Research and Innovation Fund program IND\_FF\_68\_369\_2300\_097, and the Program Management Unit for Human Resources \& Institutional Development, Research and Innovation, grant B39G680009 (Thailand); the Kavli Foundation; the Nvidia Corporation; the SuperMicro Corporation; the Welch Foundation, contract C-1845; and the Weston Havens Foundation (USA).
\end{acknowledgments}

\bibliography{auto_generated}

\providecommand{\href}[2]{#2}\begingroup\raggedright\begin{thebibliography}{10}%
\makeatletter
\providecommand{\hrefCMSnoop }[0]{\@secondoftwo}%
\makeatother
\providecommand{\doi}{\texttt{doi:}\begingroup \urlstyle{tt}\Url}

\bibitem{Baur:2004uw}
\hrefCMSnoop {}{U.~Baur, A.~Juste, L.~H. Orr, and D.~Rainwater, ``Probing
  electroweak top quark couplings at hadron colliders'',} \textit{ Phys. Rev.
  D} \textbf{ 71} (2005) 054013,
  \href{http://dx.doi.org/10.1103/PhysRevD.71.054013}{\doi{10.1103/PhysRevD.71.054013}},
  \href{http://www.arXiv.org/abs/hep-ph/0412021}{\texttt{arXiv:hep-ph/0412021}}.

\bibitem{Bouzas:2012av}
\hrefCMSnoop {}{A.~O. Bouzas and F.~Larios, ``Electromagnetic dipole moments of
  the top quark'',} \textit{ Phys. Rev. D} \textbf{ 87} (2013) 074015,
  \href{http://dx.doi.org/10.1103/PhysRevD.87.074015}{\doi{10.1103/PhysRevD.87.074015}},
  \href{http://www.arXiv.org/abs/1212.6575}{\texttt{arXiv:1212.6575}}.

\bibitem{Schulze:2016qas}
\hrefCMSnoop {}{M.~Schulze and Y.~Soreq, ``Pinning down electroweak dipole
  operators of the top quark'',} \textit{ Eur. Phys. J. C} \textbf{ 76} (2016)
  466,
  \href{http://dx.doi.org/10.1140/epjc/s10052-016-4263-x}{\doi{10.1140/epjc/s10052-016-4263-x}},
  \href{http://www.arXiv.org/abs/1603.08911}{\texttt{arXiv:1603.08911}}.

\bibitem{Bevilacqua:2018woc}
G.~Bevilacqua\hrefCMSnoop {}{ { et~al.}, ``Hard photons in hadroproduction of
  top quarks with realistic final states'',} \textit{ JHEP} \textbf{ 10} (2018)
  158,
  \href{http://dx.doi.org/10.1007/JHEP10(2018)158}{\doi{10.1007/JHEP10(2018)158}},
  \href{http://www.arXiv.org/abs/1803.09916}{\texttt{arXiv:1803.09916}}.

\bibitem{Bevilacqua:2019quz}
G.~Bevilacqua\hrefCMSnoop {}{ { et~al.}, ``Off-shell vs on-shell modelling of
  top quarks in photon associated production'',} \textit{ JHEP} \textbf{ 03}
  (2020) 154,
  \href{http://dx.doi.org/10.1007/JHEP03(2020)154}{\doi{10.1007/JHEP03(2020)154}},
  \href{http://www.arXiv.org/abs/1912.09999}{\texttt{arXiv:1912.09999}}.

\bibitem{Bergner:2018lgm}
\hrefCMSnoop {}{J.~Bergner and M.~Schulze, ``The top quark charge asymmetry in
  ${\ttbar\PGg}$ production at the {LHC}'',} \textit{ Eur. Phys. J. C} \textbf{
  79} (2019) 189,
  \href{http://dx.doi.org/10.1140/epjc/s10052-019-6707-6}{\doi{10.1140/epjc/s10052-019-6707-6}},
  \href{http://www.arXiv.org/abs/1812.10535}{\texttt{arXiv:1812.10535}}.

\bibitem{Aguilar-Saavedra:2014vta}
\hrefCMSnoop {}{J.~A. Aguilar-Saavedra, E.~{\'A}lvarez, A.~Juste, and F.~Rubbo,
  ``Shedding light on the \ttbar asymmetry: the photon handle'',} \textit{
  JHEP} \textbf{ 04} (2014) 188,
  \href{http://dx.doi.org/10.1007/JHEP04(2014)188}{\doi{10.1007/JHEP04(2014)188}},
  \href{http://www.arXiv.org/abs/1402.3598}{\texttt{arXiv:1402.3598}}.

\bibitem{Stremmer:2023kcd}
\hrefCMSnoop {}{D.~Stremmer and M.~Worek, ``Associated production of a
  top-quark pair with two isolated photons at the {LHC} through {NLO} in
  {QCD}'',} \textit{ JHEP} \textbf{ 08} (2023) 179,
  \href{http://dx.doi.org/10.1007/JHEP08(2023)179}{\doi{10.1007/JHEP08(2023)179}},
  \href{http://www.arXiv.org/abs/2306.16968}{\texttt{arXiv:2306.16968}}.

\bibitem{Bevilacqua:2018dny}
G.~Bevilacqua\hrefCMSnoop {}{ { et~al.}, ``Precise predictions for
  ${\ttbar\PGg}/\ttbar$ cross section ratios at the {LHC}'',} \textit{ JHEP}
  \textbf{ 01} (2019) 188,
  \href{http://dx.doi.org/10.1007/JHEP01(2019)188}{\doi{10.1007/JHEP01(2019)188}},
  \href{http://www.arXiv.org/abs/1809.08562}{\texttt{arXiv:1809.08562}}.

\bibitem{Aguilar-Saavedra:2018ksv}
\href {https://cds.cern.ch/record/2305783}{J.~A. Aguilar-Saavedra { et~al.},
  ``Interpreting top-quark {LHC} measurements in the standard-model effective
  field theory'',} LHC TOP Working Group Public Note CERN-LPCC-2018-01, 2018.
\newblock
  \href{http://www.arXiv.org/abs/1802.07237}{\texttt{arXiv:1802.07237}}.

\bibitem{cdf}
\hrefCMSnoop {}{{CDF} Collaboration, ``Evidence for ${\ttbar\PGg}$ production
  and measurement of $\sigma_{\ttbar\PGg}/\sigma_{\ttbar}$'',} \textit{ Phys.
  Rev. D} \textbf{ 84} (2011) 031104,
  \href{http://dx.doi.org/10.1103/PhysRevD.84.031104}{\doi{10.1103/PhysRevD.84.031104}},
  \href{http://www.arXiv.org/abs/1106.3970}{\texttt{arXiv:1106.3970}}.

\bibitem{atlas7tev}
\hrefCMSnoop {}{{ATLAS Collaboration}, ``Observation of top-quark pair
  production in association with a photon and measurement of the ${\ttbar\PGg}$
  production cross section in ${\Pp\Pp}$ collisions at $\sqrt{s}={7\TeV}$ using
  the {ATLAS} detector'',} \textit{ Phys. Rev. D} \textbf{ 91} (2015) 072007,
  \href{http://dx.doi.org/10.1103/PhysRevD.91.072007}{\doi{10.1103/PhysRevD.91.072007}},
  \href{http://www.arXiv.org/abs/1502.00586}{\texttt{arXiv:1502.00586}}.

\bibitem{atlas8tev}
\hrefCMSnoop {}{{ATLAS Collaboration}, ``Measurement of the ${\ttbar\PGg}$
  production cross section in proton-proton collisions at $\sqrt{s}={8\TeV}$
  with the {ATLAS} detector'',} \textit{ JHEP} \textbf{ 11} (2017) 086,
  \href{http://dx.doi.org/10.1007/JHEP11(2017)086}{\doi{10.1007/JHEP11(2017)086}},
  \href{http://www.arXiv.org/abs/1706.03046}{\texttt{arXiv:1706.03046}}.

\bibitem{cms8tev}
\hrefCMSnoop {}{{CMS Collaboration}, ``Measurement of the semileptonic
  {\ttbar}+{\PGg} production cross-section in ${\Pp\Pp}$ collisions at
  $\sqrt{s}={8\TeV}$'',} \textit{ JHEP} \textbf{ 10} (2017) 006,
  \href{http://dx.doi.org/10.1007/JHEP10(2017)006}{\doi{10.1007/JHEP10(2017)006}},
  \href{http://www.arXiv.org/abs/1706.08128}{\texttt{arXiv:1706.08128}}.

\bibitem{CMS:2022lmh}
\hrefCMSnoop {}{{CMS Collaboration}, ``Measurement of the inclusive and
  differential ${\ttbar\PGg}$ cross sections in the dilepton channel and
  effective field theory interpretation in proton-proton collisions at
  $\sqrt{s}={13\TeV}$'',} \textit{ JHEP} \textbf{ 05} (2022) 091,
  \href{http://dx.doi.org/10.1007/JHEP05(2022)091}{\doi{10.1007/JHEP05(2022)091}},
  \href{http://www.arXiv.org/abs/2201.07301}{\texttt{arXiv:2201.07301}}.

\bibitem{CMS:2021klw}
\hrefCMSnoop {}{{CMS Collaboration}, ``Measurement of the inclusive and
  differential ${\ttbar\PGg}$ cross sections in the single-lepton channel and
  {EFT} interpretation at $\sqrt{s}={13\TeV}$'',} \textit{ JHEP} \textbf{ 12}
  (2021) 180,
  \href{http://dx.doi.org/10.1007/JHEP12(2021)180}{\doi{10.1007/JHEP12(2021)180}},
  \href{http://www.arXiv.org/abs/2107.01508}{\texttt{arXiv:2107.01508}}.

\bibitem{ATLAS:2018sos}
\hrefCMSnoop {}{{ATLAS Collaboration}, ``Measurements of inclusive and
  differential fiducial cross-sections of ${\ttbar\PGg}$ production in leptonic
  final states at $\sqrt{s}={13\TeV}$ in {ATLAS}'',} \textit{ Eur. Phys. J. C}
  \textbf{ 79} (2019) 382,
  \href{http://dx.doi.org/10.1140/epjc/s10052-019-6849-6}{\doi{10.1140/epjc/s10052-019-6849-6}},
  \href{http://www.arXiv.org/abs/1812.01697}{\texttt{arXiv:1812.01697}}.

\bibitem{ATLAS:2020yrp}
\hrefCMSnoop {}{{ATLAS Collaboration}, ``Measurements of inclusive and
  differential cross-sections of combined ${\ttbar\PGg}$ and ${\PQt\PW\PGg}$
  production in the ${\Pe\PGm}$ channel at {13\TeV} with the {ATLAS}
  detector'',} \textit{ JHEP} \textbf{ 09} (2020) 049,
  \href{http://dx.doi.org/10.1007/JHEP09(2020)049}{\doi{10.1007/JHEP09(2020)049}},
  \href{http://www.arXiv.org/abs/2007.06946}{\texttt{arXiv:2007.06946}}.

\bibitem{ATLAS:2024hmk}
\hrefCMSnoop {}{{ATLAS Collaboration}, ``Measurements of inclusive and
  differential cross-sections of ${\ttbar\PGg}$ production in ${\Pp\Pp}$
  collisions at $\sqrt{s}={13\TeV}$ with the {ATLAS} detector'',} \textit{
  JHEP} \textbf{ 10} (2024) 191,
  \href{http://dx.doi.org/10.1007/JHEP10(2024)191}{\doi{10.1007/JHEP10(2024)191}},
  \href{http://www.arXiv.org/abs/2403.09452}{\texttt{arXiv:2403.09452}}.

\bibitem{ATLAS:2022wec}
\hrefCMSnoop {}{{ATLAS Collaboration}, ``Measurement of the charge asymmetry in
  top-quark pair production in association with a photon with the {ATLAS}
  experiment'',} \textit{ Phys. Lett. B} \textbf{ 843} (2023) 137848,
  \href{http://dx.doi.org/10.1016/j.physletb.2023.137848}{\doi{10.1016/j.physletb.2023.137848}},
  \href{http://www.arXiv.org/abs/2212.10552}{\texttt{arXiv:2212.10552}}.

\bibitem{hepdata}
\hrefCMSnoop {}{}{HEPData} record for this analysis, 2025.
\newblock
  \href{http://dx.doi.org/10.17182/hepdata.157848}{\doi{10.17182/hepdata.157848}}.

\bibitem{CMS:2008xjf}
\hrefCMSnoop {}{{CMS Collaboration}, ``The {CMS} experiment at the {CERN}
  {LHC}'',} \textit{ JINST} \textbf{ 3} (2008) S08004,
  \href{http://dx.doi.org/10.1088/1748-0221/3/08/S08004}{\doi{10.1088/1748-0221/3/08/S08004}}.

\bibitem{CMS:2023gfb}
\hrefCMSnoop {}{{CMS Collaboration}, ``Development of the {CMS} detector for
  the {CERN} {LHC} \mbox{Run 3}'',} \textit{ JINST} \textbf{ 19} (2024) P05064,
  \href{http://dx.doi.org/10.1088/1748-0221/19/05/P05064}{\doi{10.1088/1748-0221/19/05/P05064}},
  \href{http://www.arXiv.org/abs/2309.05466}{\texttt{arXiv:2309.05466}}.

\bibitem{CMS:2020cmk}
\hrefCMSnoop {}{{CMS Collaboration}, ``Performance of the {CMS} {\Lone} trigger
  in proton-proton collisions at $\sqrt{s}={13\TeV}$'',} \textit{ JINST}
  \textbf{ 15} (2020) P10017,
  \href{http://dx.doi.org/10.1088/1748-0221/15/10/P10017}{\doi{10.1088/1748-0221/15/10/P10017}},
  \href{http://www.arXiv.org/abs/2006.10165}{\texttt{arXiv:2006.10165}}.

\bibitem{CMS:2016ngn}
\hrefCMSnoop {}{{CMS Collaboration}, ``The {CMS} trigger system'',} \textit{
  JINST} \textbf{ 12} (2017) P01020,
  \href{http://dx.doi.org/10.1088/1748-0221/12/01/P01020}{\doi{10.1088/1748-0221/12/01/P01020}},
  \href{http://www.arXiv.org/abs/1609.02366}{\texttt{arXiv:1609.02366}}.

\bibitem{CMS:TRG-19-001}
\hrefCMSnoop {}{{CMS Collaboration}, ``Performance of the {CMS} high-level
  trigger during {LHC} \mbox{Run 2}'',} \textit{ JINST} \textbf{ 19} (2024)
  P11021,
  \href{http://dx.doi.org/10.1088/1748-0221/19/11/P11021}{\doi{10.1088/1748-0221/19/11/P11021}},
  \href{http://www.arXiv.org/abs/2410.17038}{\texttt{arXiv:2410.17038}}.

\bibitem{Alwall:2014hca}
J.~Alwall\hrefCMSnoop {}{ { et~al.}, ``The automated computation of tree-level
  and next-to-leading order differential cross sections, and their matching to
  parton shower simulations'',} \textit{ JHEP} \textbf{ 07} (2014) 079,
  \href{http://dx.doi.org/10.1007/JHEP07(2014)079}{\doi{10.1007/JHEP07(2014)079}},
  \href{http://www.arXiv.org/abs/1405.0301}{\texttt{arXiv:1405.0301}}.

\bibitem{Frixione:1998jh}
\hrefCMSnoop {}{S.~Frixione, ``Isolated photons in perturbative {QCD}'',}
  \textit{ Phys. Lett. B} \textbf{ 429} (1998) 369,
  \href{http://dx.doi.org/10.1016/S0370-2693(98)00454-7}{\doi{10.1016/S0370-2693(98)00454-7}},
  \href{http://www.arXiv.org/abs/hep-ph/9801442}{\texttt{arXiv:hep-ph/9801442}}.

\bibitem{Artoisenet:2012st}
\hrefCMSnoop {}{P.~Artoisenet, R.~Frederix, O.~Mattelaer, and R.~Rietkerk,
  ``Automatic spin-entangled decays of heavy resonances in {Monte Carlo}
  simulations'',} \textit{ JHEP} \textbf{ 03} (2013) 015,
  \href{http://dx.doi.org/10.1007/JHEP03(2013)015}{\doi{10.1007/JHEP03(2013)015}},
  \href{http://www.arXiv.org/abs/1212.3460}{\texttt{arXiv:1212.3460}}.

\bibitem{Frixione:2019fxg}
S.~Frixione\hrefCMSnoop {}{ { et~al.}, ``Automated simulations beyond the
  standard model: supersymmetry'',} \textit{ JHEP} \textbf{ 12} (2019) 008,
  \href{http://dx.doi.org/10.1007/JHEP12(2019)008}{\doi{10.1007/JHEP12(2019)008}},
  \href{http://www.arXiv.org/abs/1907.04898}{\texttt{arXiv:1907.04898}}.

\bibitem{Nason:2004rx}
\hrefCMSnoop {}{P.~Nason, ``A new method for combining {NLO} {QCD} with shower
  {Monte Carlo} algorithms'',} \textit{ JHEP} \textbf{ 11} (2004) 040,
  \href{http://dx.doi.org/10.1088/1126-6708/2004/11/040}{\doi{10.1088/1126-6708/2004/11/040}},
  \href{http://www.arXiv.org/abs/hep-ph/0409146}{\texttt{arXiv:hep-ph/0409146}}.

\bibitem{Frixione:2007vw}
\hrefCMSnoop {}{S.~Frixione, P.~Nason, and C.~Oleari, ``Matching {NLO} {QCD}
  computations with parton shower simulations: the {\POWHEG} method'',}
  \textit{ JHEP} \textbf{ 11} (2007) 070,
  \href{http://dx.doi.org/10.1088/1126-6708/2007/11/070}{\doi{10.1088/1126-6708/2007/11/070}},
  \href{http://www.arXiv.org/abs/0709.2092}{\texttt{arXiv:0709.2092}}.

\bibitem{Alioli:2010xd}
\hrefCMSnoop {}{S.~Alioli, P.~Nason, C.~Oleari, and E.~Re, ``A general
  framework for implementing {NLO} calculations in shower {Monte Carlo}
  programs: the {\POWHEG} \textsc{box}'',} \textit{ JHEP} \textbf{ 06} (2010)
  043,
  \href{http://dx.doi.org/10.1007/JHEP06(2010)043}{\doi{10.1007/JHEP06(2010)043}},
  \href{http://www.arXiv.org/abs/1002.2581}{\texttt{arXiv:1002.2581}}.

\bibitem{Czakon:2011xx}
\hrefCMSnoop {}{M.~Czakon and A.~Mitov, ``\textsc{top++}: a program for the
  calculation of the top-pair cross-section at hadron colliders'',} \textit{
  Comput. Phys. Commun.} \textbf{ 185} (2014) 2930,
  \href{http://dx.doi.org/10.1016/j.cpc.2014.06.021}{\doi{10.1016/j.cpc.2014.06.021}},
  \href{http://www.arXiv.org/abs/1112.5675}{\texttt{arXiv:1112.5675}}.

\bibitem{Beneke:2011mq}
\hrefCMSnoop {}{M.~Beneke, P.~Falgari, S.~Klein, and C.~Schwinn, ``Hadronic
  top-quark pair production with {NNLL} threshold resummation'',} \textit{
  Nucl. Phys. B} \textbf{ 855} (2012) 695,
  \href{http://dx.doi.org/10.1016/j.nuclphysb.2011.10.021}{\doi{10.1016/j.nuclphysb.2011.10.021}},
  \href{http://www.arXiv.org/abs/1109.1536}{\texttt{arXiv:1109.1536}}.

\bibitem{Cacciari:2011hy}
M.~Cacciari\hrefCMSnoop {}{ { et~al.}, ``Top-pair production at hadron
  colliders with next-to-next-to-leading logarithmic soft-gluon resummation'',}
  \textit{ Phys. Lett. B} \textbf{ 710} (2012) 612,
  \href{http://dx.doi.org/10.1016/j.physletb.2012.03.013}{\doi{10.1016/j.physletb.2012.03.013}},
  \href{http://www.arXiv.org/abs/1111.5869}{\texttt{arXiv:1111.5869}}.

\bibitem{Czakon:2012zr}
\hrefCMSnoop {}{M.~Czakon and A.~Mitov, ``{NNLO} corrections to top-pair
  production at hadron colliders: the all-fermionic scattering channels'',}
  \textit{ JHEP} \textbf{ 12} (2012) 054,
  \href{http://dx.doi.org/10.1007/JHEP12(2012)054}{\doi{10.1007/JHEP12(2012)054}},
  \href{http://www.arXiv.org/abs/1207.0236}{\texttt{arXiv:1207.0236}}.

\bibitem{Czakon:2012pz}
\hrefCMSnoop {}{M.~Czakon and A.~Mitov, ``{NNLO} corrections to top pair
  production at hadron colliders: the quark-gluon reaction'',} \textit{ JHEP}
  \textbf{ 01} (2013) 080,
  \href{http://dx.doi.org/10.1007/JHEP01(2013)080}{\doi{10.1007/JHEP01(2013)080}},
  \href{http://www.arXiv.org/abs/1210.6832}{\texttt{arXiv:1210.6832}}.

\bibitem{Czakon:2013goa}
\hrefCMSnoop {}{M.~Czakon, P.~Fiedler, and A.~Mitov, ``Total top-quark
  pair-production cross section at hadron colliders through
  $\mathcal{O}({\alpS}^4)$'',} \textit{ Phys. Rev. Lett.} \textbf{ 110} (2013)
  252004,
  \href{http://dx.doi.org/10.1103/PhysRevLett.110.252004}{\doi{10.1103/PhysRevLett.110.252004}},
  \href{http://www.arXiv.org/abs/1303.6254}{\texttt{arXiv:1303.6254}}.

\bibitem{Czakon:2017wor}
M.~Czakon\hrefCMSnoop {}{ { et~al.}, ``Top-pair production at the {LHC} through
  {NNLO} {QCD} and {NLO} {EW}'',} \textit{ JHEP} \textbf{ 10} (2017) 186,
  \href{http://dx.doi.org/10.1007/JHEP10(2017)186}{\doi{10.1007/JHEP10(2017)186}},
  \href{http://www.arXiv.org/abs/1705.04105}{\texttt{arXiv:1705.04105}}.

\bibitem{Czakon:2017dip}
\hrefCMSnoop {}{M.~Czakon, D.~Heymes, and A.~Mitov, ``{fastNLO} tables for
  {NNLO} top-quark pair differential distributions'',} 2017.
  \href{http://www.arXiv.org/abs/1704.08551}{\texttt{arXiv:1704.08551}}.

\bibitem{Czakon:2015owf}
\hrefCMSnoop {}{M.~Czakon, D.~Heymes, and A.~Mitov, ``High-precision
  differential predictions for top-quark pairs at the {LHC}'',} \textit{ Phys.
  Rev. Lett.} \textbf{ 116} (2016) 082003,
  \href{http://dx.doi.org/10.1103/PhysRevLett.116.082003}{\doi{10.1103/PhysRevLett.116.082003}},
  \href{http://www.arXiv.org/abs/1511.00549}{\texttt{arXiv:1511.00549}}.

\bibitem{Czakon:2016dgf}
\hrefCMSnoop {}{M.~Czakon, D.~Heymes, and A.~Mitov, ``Dynamical scales for
  multi-{\TeVns} top-pair production at the {LHC}'',} \textit{ JHEP} \textbf{
  04} (2017) 071,
  \href{http://dx.doi.org/10.1007/JHEP04(2017)071}{\doi{10.1007/JHEP04(2017)071}},
  \href{http://www.arXiv.org/abs/1606.03350}{\texttt{arXiv:1606.03350}}.

\bibitem{Sjostrand:2014zea}
T.~Sj{\"o}strand\hrefCMSnoop {}{ { et~al.}, ``An introduction to
  {\PYTHIA8.2}'',} \textit{ Comput. Phys. Commun.} \textbf{ 191} (2015) 159,
  \href{http://dx.doi.org/10.1016/j.cpc.2015.01.024}{\doi{10.1016/j.cpc.2015.01.024}},
  \href{http://www.arXiv.org/abs/1410.3012}{\texttt{arXiv:1410.3012}}.

\bibitem{CMS:2015lbj}
\hrefCMSnoop {}{{CMS Collaboration}, ``Measurement of the top quark mass using
  proton-proton data at $\sqrt{s}=7$ and {8\TeV}'',} \textit{ Phys. Rev. D}
  \textbf{ 93} (2016) 072004,
  \href{http://dx.doi.org/10.1103/PhysRevD.93.072004}{\doi{10.1103/PhysRevD.93.072004}},
  \href{http://www.arXiv.org/abs/1509.04044}{\texttt{arXiv:1509.04044}}.

\bibitem{Skands:2014pea}
\hrefCMSnoop {}{P.~Skands, S.~Carrazza, and J.~Rojo, ``Tuning {\PYTHIA8.1}: the
  {Monash} 2013 tune'',} \textit{ Eur. Phys. J. C} \textbf{ 74} (2014) 3024,
  \href{http://dx.doi.org/10.1140/epjc/s10052-014-3024-y}{\doi{10.1140/epjc/s10052-014-3024-y}},
  \href{http://www.arXiv.org/abs/1404.5630}{\texttt{arXiv:1404.5630}}.

\bibitem{Khachatryan:2015pea}
\hrefCMSnoop {}{{CMS Collaboration}, ``Event generator tunes obtained from
  underlying event and multiparton scattering measurements'',} \textit{ Eur.
  Phys. J. C} \textbf{ 76} (2016) 155,
  \href{http://dx.doi.org/10.1140/epjc/s10052-016-3988-x}{\doi{10.1140/epjc/s10052-016-3988-x}},
  \href{http://www.arXiv.org/abs/1512.00815}{\texttt{arXiv:1512.00815}}.

\bibitem{CMS:GEN-17-001}
\hrefCMSnoop {}{{CMS Collaboration}, ``Extraction and validation of a new set
  of {CMS} {\PYTHIA8} tunes from underlying-event measurements'',} \textit{
  Eur. Phys. J. C} \textbf{ 80} (2020) 4,
  \href{http://dx.doi.org/10.1140/epjc/s10052-019-7499-4}{\doi{10.1140/epjc/s10052-019-7499-4}},
  \href{http://www.arXiv.org/abs/1903.12179}{\texttt{arXiv:1903.12179}}.

\bibitem{Ball:2017nwa}
\hrefCMSnoop {}{{NNPDF} Collaboration, ``Parton distributions from
  high-precision collider data'',} \textit{ Eur. Phys. J. C} \textbf{ 77}
  (2017) 663,
  \href{http://dx.doi.org/10.1140/epjc/s10052-017-5199-5}{\doi{10.1140/epjc/s10052-017-5199-5}},
  \href{http://www.arXiv.org/abs/1706.00428}{\texttt{arXiv:1706.00428}}.

\bibitem{Alwall:2007fs}
J.~Alwall\hrefCMSnoop {}{ { et~al.}, ``Comparative study of various algorithms
  for the merging of parton showers and matrix elements in hadronic
  collisions'',} \textit{ Eur. Phys. J. C} \textbf{ 53} (2008) 473,
  \href{http://dx.doi.org/10.1140/epjc/s10052-007-0490-5}{\doi{10.1140/epjc/s10052-007-0490-5}},
  \href{http://www.arXiv.org/abs/0706.2569}{\texttt{arXiv:0706.2569}}.

\bibitem{Frederix:2012ps}
\hrefCMSnoop {}{R.~Frederix and S.~Frixione, ``Merging meets matching in
  {\MCATNLO}'',} \textit{ JHEP} \textbf{ 12} (2012) 061,
  \href{http://dx.doi.org/10.1007/JHEP12(2012)061}{\doi{10.1007/JHEP12(2012)061}},
  \href{http://www.arXiv.org/abs/1209.6215}{\texttt{arXiv:1209.6215}}.

\bibitem{Agostinelli:2002hh}
\hrefCMSnoop {}{{GEANT4} Collaboration, ``{\GEANTfour}---a simulation
  toolkit'',} \textit{ Nucl. Instrum. Meth. A} \textbf{ 506} (2003) 250,
  \href{http://dx.doi.org/10.1016/S0168-9002(03)01368-8}{\doi{10.1016/S0168-9002(03)01368-8}}.

\bibitem{Sirunyan:2018nqx}
\hrefCMSnoop {}{{CMS Collaboration}, ``Measurement of the inelastic
  proton-proton cross section at $\sqrt{s}={13\TeV}$'',} \textit{ JHEP}
  \textbf{ 07} (2018) 161,
  \href{http://dx.doi.org/10.1007/JHEP07(2018)161}{\doi{10.1007/JHEP07(2018)161}},
  \href{http://www.arXiv.org/abs/1802.02613}{\texttt{arXiv:1802.02613}}.

\bibitem{CMS-TDR-15-02}
\hrefCMSnoop {}{{CMS Collaboration}, ``Technical proposal for the {Phase-II}
  upgrade of the {Compact Muon Solenoid}'',} CMS Technical Proposal
  CERN-LHCC-2015-010, CMS-TDR-15-02, 2015.
\newblock
  \href{http://dx.doi.org/10.17181/CERN.VU8I.D59J}{\doi{10.17181/CERN.VU8I.D59J}}.

\bibitem{Sirunyan:2017ulk}
\hrefCMSnoop {}{{CMS Collaboration}, ``Particle-flow reconstruction and global
  event description with the {CMS} detector'',} \textit{ JINST} \textbf{ 12}
  (2017) P10003,
  \href{http://dx.doi.org/10.1088/1748-0221/12/10/P10003}{\doi{10.1088/1748-0221/12/10/P10003}},
  \href{http://www.arXiv.org/abs/1706.04965}{\texttt{arXiv:1706.04965}}.

\bibitem{Electrons:2021}
\hrefCMSnoop {}{{CMS Collaboration}, ``Electron and photon reconstruction and
  identification with the {CMS} experiment at the {CERN} {LHC}'',} \textit{
  JINST} \textbf{ 16} (2021) P05014,
  \href{http://dx.doi.org/10.1088/1748-0221/16/05/P05014}{\doi{10.1088/1748-0221/16/05/P05014}},
  \href{http://www.arXiv.org/abs/2012.06888}{\texttt{arXiv:2012.06888}}.

\bibitem{CMS-DP-2020-021}
\href {https://cds.cern.ch/record/2717925}{{CMS Collaboration}, ``{ECAL} 2016
  refined calibration and \mbox{Run 2} summary plots'',} CMS Detector
  Performance Note CMS-DP-2020-021, 2020.

\bibitem{CMS:2018rym}
\hrefCMSnoop {}{{CMS Collaboration}, ``Performance of the {CMS} muon detector
  and muon reconstruction with proton-proton collisions at
  $\sqrt{s}={13\TeV}$'',} \textit{ JINST} \textbf{ 13} (2018) P06015,
  \href{http://dx.doi.org/10.1088/1748-0221/13/06/P06015}{\doi{10.1088/1748-0221/13/06/P06015}},
  \href{http://www.arXiv.org/abs/1804.04528}{\texttt{arXiv:1804.04528}}.

\bibitem{ECALpaper}
\hrefCMSnoop {}{{CMS Collaboration}, ``Performance of the {CMS} electromagnetic
  calorimeter in ${\Pp\Pp}$ collisions at $\sqrt{s}={13\TeV}$'',} \textit{
  JINST} \textbf{ 19} (2024) P09004,
  \href{http://dx.doi.org/10.1088/1748-0221/19/09/P09004}{\doi{10.1088/1748-0221/19/09/P09004}},
  \href{http://www.arXiv.org/abs/2403.15518}{\texttt{arXiv:2403.15518}}.

\bibitem{CMS-DP-2017-004}
\href {https://cds.cern.ch/record/2255497}{{CMS Collaboration}, ``Electron and
  photon performance in {CMS} with the full 2016 data sample'',} CMS Detector
  Performance Note CMS-DP-2017-004, 2017.

\bibitem{CMS-DP-2020-037}
\href {https://cds.cern.ch/record/2725004}{{CMS Collaboration}, ``Performance
  of electron and photon reconstruction in \mbox{Run 2} with the {CMS}
  experiment'',} CMS Detector Performance Note CMS-DP-2020-037, 2020.

\bibitem{antikt}
\hrefCMSnoop {}{M.~Cacciari, G.~P. Salam, and G.~Soyez, ``The anti-\kt jet
  clustering algorithm'',} \textit{ JHEP} \textbf{ 04} (2008) 063,
  \href{http://dx.doi.org/10.1088/1126-6708/2008/04/063}{\doi{10.1088/1126-6708/2008/04/063}},
  \href{http://www.arXiv.org/abs/0802.1189}{\texttt{arXiv:0802.1189}}.

\bibitem{Cacciari:2011ma}
\hrefCMSnoop {}{M.~Cacciari, G.~P. Salam, and G.~Soyez, ``{\FASTJET} user
  manual'',} \textit{ Eur. Phys. J. C} \textbf{ 72} (2012) 1896,
  \href{http://dx.doi.org/10.1140/epjc/s10052-012-1896-2}{\doi{10.1140/epjc/s10052-012-1896-2}},
  \href{http://www.arXiv.org/abs/1111.6097}{\texttt{arXiv:1111.6097}}.

\bibitem{CMS-PAS-JME-16-003}
\href {https://cds.cern.ch/record/2256875}{{CMS Collaboration}, ``Jet
  algorithms performance in {13\TeV} data'',} CMS Physics Analysis Summary
  CMS-PAS-JME-16-003, 2017.

\bibitem{CMSJetPaper}
\hrefCMSnoop {}{{CMS Collaboration}, ``Determination of jet energy calibration
  and transverse momentum resolution in {CMS}'',} \textit{ JINST} \textbf{ 6}
  (2011) P11002,
  \href{http://dx.doi.org/10.1088/1748-0221/6/11/P11002}{\doi{10.1088/1748-0221/6/11/P11002}},
  \href{http://www.arXiv.org/abs/1107.4277}{\texttt{arXiv:1107.4277}}.

\bibitem{CMS:2017wtu}
\hrefCMSnoop {}{{CMS Collaboration}, ``Identification of heavy-flavour jets
  with the {CMS} detector in ${\Pp\Pp}$ collisions at {13\TeV}'',} \textit{
  JINST} \textbf{ 13} (2018) P05011,
  \href{http://dx.doi.org/10.1088/1748-0221/13/05/P05011}{\doi{10.1088/1748-0221/13/05/P05011}},
  \href{http://www.arXiv.org/abs/1712.07158}{\texttt{arXiv:1712.07158}}.

\bibitem{Bols:2020bkb}
E.~Bols\hrefCMSnoop {}{ { et~al.}, ``Jet flavour classification using
  {DeepJet}'',} \textit{ JINST} \textbf{ 15} (2020) P12012,
  \href{http://dx.doi.org/10.1088/1748-0221/15/12/P12012}{\doi{10.1088/1748-0221/15/12/P12012}},
  \href{http://www.arXiv.org/abs/2008.10519}{\texttt{arXiv:2008.10519}}.

\bibitem{CMS:DP-2023-005}
\href {https://cds.cern.ch/record/2854609}{{CMS Collaboration}, ``Performance
  summary of {AK4} jet {\PQb} tagging with data from proton-proton collisions
  at {13\TeV} with the {CMS} detector'',} CMS Detector Performance Note
  CMS-DP-2023-005, 2023.

\bibitem{CMS:2019ctu}
\hrefCMSnoop {}{{CMS Collaboration}, ``Performance of missing transverse
  momentum reconstruction in proton-proton collisions at $\sqrt{s}={13\TeV}$
  using the {CMS} detector'',} \textit{ JINST} \textbf{ 14} (2019) P07004,
  \href{http://dx.doi.org/10.1088/1748-0221/14/07/P07004}{\doi{10.1088/1748-0221/14/07/P07004}},
  \href{http://www.arXiv.org/abs/1903.06078}{\texttt{arXiv:1903.06078}}.

\bibitem{CMS:2016lmd}
\hrefCMSnoop {}{{CMS Collaboration}, ``Jet energy scale and resolution in the
  {CMS} experiment in ${\Pp\Pp}$ collisions at {8\TeV}'',} \textit{ JINST}
  \textbf{ 12} (2017) P02014,
  \href{http://dx.doi.org/10.1088/1748-0221/12/02/P02014}{\doi{10.1088/1748-0221/12/02/P02014}},
  \href{http://www.arXiv.org/abs/1607.03663}{\texttt{arXiv:1607.03663}}.

\bibitem{CMS:2024irj}
\hrefCMSnoop {}{{CMS Collaboration}, ``Review of top quark mass measurements in
  {CMS}'',} \textit{ Phys. Rept.} \textbf{ 1115} (2025) 116,
  \href{http://dx.doi.org/10.1016/j.physrep.2024.12.002}{\doi{10.1016/j.physrep.2024.12.002}},
  \href{http://www.arXiv.org/abs/2403.01313}{\texttt{arXiv:2403.01313}}.

\bibitem{Tanabashi:2018oca}
\hrefCMSnoop {}{{Particle Data Group}, S.~Navas { et~al.}, ``Review of particle
  physics'',} \textit{ Phys. Rev. D} \textbf{ 110} (2024) 030001,
  \href{http://dx.doi.org/10.1103/PhysRevD.110.030001}{\doi{10.1103/PhysRevD.110.030001}}.

\bibitem{CMS:2019nrx}
\hrefCMSnoop {}{{CMS Collaboration}, ``Measurement of the top quark
  polarization and \ttbar spin correlations using dilepton final states in
  proton-proton collisions at $\sqrt{s}={13\TeV}$'',} \textit{ Phys. Rev. D}
  \textbf{ 100} (2019) 072002,
  \href{http://dx.doi.org/10.1103/PhysRevD.100.072002}{\doi{10.1103/PhysRevD.100.072002}},
  \href{http://www.arXiv.org/abs/1907.03729}{\texttt{arXiv:1907.03729}}.

\bibitem{lumipaper}
\hrefCMSnoop {}{{CMS Collaboration}, ``Precision luminosity measurement in
  proton-proton collisions at $\sqrt{s}={13\TeV}$ in 2015 and 2016 at {CMS}'',}
  \textit{ Eur. Phys. J. C} \textbf{ 81} (2021) 800,
  \href{http://dx.doi.org/10.1140/epjc/s10052-021-09538-2}{\doi{10.1140/epjc/s10052-021-09538-2}},
  \href{http://www.arXiv.org/abs/2104.01927}{\texttt{arXiv:2104.01927}}.

\bibitem{CMS:LUM-17-004}
\href {https://cds.cern.ch/record/2621960}{{CMS Collaboration}, ``{CMS}
  luminosity measurement for the 2017 data-taking period at
  $\sqrt{s}={13\TeV}$'',} CMS Physics Analysis Summary CMS-PAS-LUM-17-004,
  2018.

\bibitem{CMS:LUM-18-002}
\href {https://cds.cern.ch/record/2676164}{{CMS Collaboration}, ``{CMS}
  luminosity measurement for the 2018 data-taking period at
  $\sqrt{s}={13\TeV}$'',} CMS Physics Analysis Summary CMS-PAS-LUM-18-002,
  2019.

\bibitem{Barlow:1993dm}
\hrefCMSnoop {}{R.~Barlow and C.~Beeston, ``Fitting using finite {Monte Carlo}
  samples'',} \textit{ Comput. Phys. Commun.} \textbf{ 77} (1993) 219,
  \href{http://dx.doi.org/10.1016/0010-4655(93)90005-W}{\doi{10.1016/0010-4655(93)90005-W}}.

\bibitem{Ball:2014uwa}
\hrefCMSnoop {}{{NNPDF} Collaboration, ``Parton distributions for the {LHC} run
  {II}'',} \textit{ JHEP} \textbf{ 04} (2015) 040,
  \href{http://dx.doi.org/10.1007/JHEP04(2015)040}{\doi{10.1007/JHEP04(2015)040}},
  \href{http://www.arXiv.org/abs/1410.8849}{\texttt{arXiv:1410.8849}}.

\bibitem{CMS:2021gme}
\hrefCMSnoop {}{{CMS Collaboration}, ``Measurement of the electroweak
  production of ${\PZ\PGg}$ and two jets in proton-proton collisions at
  $\sqrt{s}={13\TeV}$ and constraints on anomalous quartic gauge couplings'',}
  \textit{ Phys. Rev. D} \textbf{ 104} (2021) 072001,
  \href{http://dx.doi.org/10.1103/PhysRevD.104.072001}{\doi{10.1103/PhysRevD.104.072001}},
  \href{http://www.arXiv.org/abs/2106.11082}{\texttt{arXiv:2106.11082}}.

\bibitem{combine_tool}
\hrefCMSnoop {}{{CMS Collaboration}, ``The {CMS} statistical analysis and
  combination tool: \textsc{combine}'',} \textit{ Comput. Softw. Big Sci.}
  \textbf{ 8} (2024) 19,
  \href{http://dx.doi.org/10.1007/s41781-024-00121-4}{\doi{10.1007/s41781-024-00121-4}},
  \href{http://www.arXiv.org/abs/2404.06614}{\texttt{arXiv:2404.06614}}.

\bibitem{Blobel:2203257}
\hrefCMSnoop {}{V.~Blobel, ``Unfolding methods in particle physics'',} in
  \textit{ {Proc. 2011 Workshop on Statistical Issues Related to Discovery
  Claims in Search Experiments and Unfolding (PHYSTAT 2011): Geneva,
  Switzerland, January 17--20, 2011}}, p.~240.
\newblock 2011.
\newblock
  \href{http://dx.doi.org/10.5170/CERN-2011-006}{\doi{10.5170/CERN-2011-006}}.

\bibitem{Collaboration:2267573}
\href {https://cds.cern.ch/record/2267573}{{CMS Collaboration}, ``Object
  definitions for top quark analyses at the particle level'',} CMS Note
  CMS-NOTE-2017-004, 2017.

\bibitem{Cacciari:2008gn}
\hrefCMSnoop {}{M.~Cacciari, G.~P. Salam, and G.~Soyez, ``The catchment area of
  jets'',} \textit{ JHEP} \textbf{ 04} (2008) 005,
  \href{http://dx.doi.org/10.1088/1126-6708/2008/04/005}{\doi{10.1088/1126-6708/2008/04/005}},
  \href{http://www.arXiv.org/abs/0802.1188}{\texttt{arXiv:0802.1188}}.

\bibitem{Stremmer:2024ecl}
\hrefCMSnoop {}{D.~Stremmer and M.~Worek, ``Complete {NLO} corrections to
  top-quark pair production with isolated photons'',} \textit{ JHEP} \textbf{
  07} (2024) 091,
  \href{http://dx.doi.org/10.1007/JHEP07(2024)091}{\doi{10.1007/JHEP07(2024)091}},
  \href{http://www.arXiv.org/abs/2403.03796}{\texttt{arXiv:2403.03796}}.

\bibitem{Bevilacqua:2011xh}
G.~Bevilacqua\hrefCMSnoop {}{ { et~al.}, ``\textsc{helac-nlo}'',} \textit{
  Comput. Phys. Commun.} \textbf{ 184} (2013) 986,
  \href{http://dx.doi.org/10.1016/j.cpc.2012.10.033}{\doi{10.1016/j.cpc.2012.10.033}},
  \href{http://www.arXiv.org/abs/1110.1499}{\texttt{arXiv:1110.1499}}.

\bibitem{Actis:2016mpe}
S.~Actis\hrefCMSnoop {}{ { et~al.}, ``\textsc{recola}---recursive computation
  of one-loop amplitudes'',} \textit{ Comput. Phys. Commun.} \textbf{ 214}
  (2017) 140,
  \href{http://dx.doi.org/10.1016/j.cpc.2017.01.004}{\doi{10.1016/j.cpc.2017.01.004}},
  \href{http://www.arXiv.org/abs/1605.01090}{\texttt{arXiv:1605.01090}}.

\bibitem{Actis:2012qn}
S.~Actis\hrefCMSnoop {}{ { et~al.}, ``Recursive generation of one-loop
  amplitudes in the standard model'',} \textit{ JHEP} \textbf{ 04} (2013) 037,
  \href{http://dx.doi.org/10.1007/JHEP04(2013)037}{\doi{10.1007/JHEP04(2013)037}},
  \href{http://www.arXiv.org/abs/1211.6316}{\texttt{arXiv:1211.6316}}.

\bibitem{Denner:2016kdg}
\hrefCMSnoop {}{A.~Denner, S.~Dittmaier, and L.~Hofer, ``\textsc{collier}: A
  fortran-based complex one-loop library in extended regularizations'',}
  \textit{ Comput. Phys. Commun.} \textbf{ 212} (2017) 220,
  \href{http://dx.doi.org/10.1016/j.cpc.2016.10.013}{\doi{10.1016/j.cpc.2016.10.013}},
  \href{http://www.arXiv.org/abs/1604.06792}{\texttt{arXiv:1604.06792}}.

\bibitem{Bevilacqua:2013iha}
\hrefCMSnoop {}{G.~Bevilacqua, M.~Czakon, M.~Kubocz, and M.~Worek, ``Complete
  {Nagy}--{Soper} subtraction for next-to-leading order calculations in
  {QCD}'',} \textit{ JHEP} \textbf{ 10} (2013) 204,
  \href{http://dx.doi.org/10.1007/JHEP10(2013)204}{\doi{10.1007/JHEP10(2013)204}},
  \href{http://www.arXiv.org/abs/1308.5605}{\texttt{arXiv:1308.5605}}.

\bibitem{Bevilacqua:2022ozv}
\hrefCMSnoop {}{G.~Bevilacqua, M.~Lupattelli, D.~Stremmer, and M.~Worek,
  ``Study of additional jet activity in top quark pair production and decay at
  the {LHC}'',} \textit{ Phys. Rev. D} \textbf{ 107} (2023) 114027,
  \href{http://dx.doi.org/10.1103/PhysRevD.107.114027}{\doi{10.1103/PhysRevD.107.114027}},
  \href{http://www.arXiv.org/abs/2212.04722}{\texttt{arXiv:2212.04722}}.

\bibitem{Czakon:2015cla}
\hrefCMSnoop {}{M.~Czakon, H.~B. Hartanto, M.~Kraus, and M.~Worek, ``Matching
  the {Nagy}--{Soper} parton shower at next-to-leading order'',} \textit{ JHEP}
  \textbf{ 06} (2015) 033,
  \href{http://dx.doi.org/10.1007/JHEP06(2015)033}{\doi{10.1007/JHEP06(2015)033}},
  \href{http://www.arXiv.org/abs/1502.00925}{\texttt{arXiv:1502.00925}}.

\bibitem{Stremmer:2024zhd}
\hrefCMSnoop {}{D.~Stremmer and M.~Worek, ``{NLO} {QCD} predictions for
  ${\ttbar\PGg}$ with realistic photon isolation'',} \textit{ JHEP} \textbf{
  01} (2025) 156,
  \href{http://dx.doi.org/10.1007/JHEP01(2025)156}{\doi{10.1007/JHEP01(2025)156}},
  \href{http://www.arXiv.org/abs/2411.02196}{\texttt{arXiv:2411.02196}}.

\bibitem{CDF:AN5776}
\href
  {http://physics.rockefeller.edu/luc/technical_reports/cdf5776_pulls.pdf}{L.~Demortier
  and L.~Lyons, ``Everything you always wanted to know about pulls'',}
  Technical Report CDF/ANAL/PUBLIC/5776, CDF, 2002.

\bibitem{Gross:2010qma}
\hrefCMSnoop {}{E.~Gross and O.~Vitells, ``Trial factors for the look elsewhere
  effect in high energy physics'',} \textit{ Eur. Phys. J. C} \textbf{ 70}
  (2010) 525,
  \href{http://dx.doi.org/10.1140/epjc/s10052-010-1470-8}{\doi{10.1140/epjc/s10052-010-1470-8}},
  \href{http://www.arXiv.org/abs/1005.1891}{\texttt{arXiv:1005.1891}}.

\end{thebibliography}\endgroup
\cleardoublepage \appendix\section{The CMS Collaboration \label{app:collab}}\begin{sloppypar}\hyphenpenalty=5000\widowpenalty=500\clubpenalty=5000\cmsinstitute{Yerevan Physics Institute, Yerevan, Armenia}
{\tolerance=6000
A.~Hayrapetyan, V.~Makarenko\cmsorcid{0000-0002-8406-8605}, A.~Tumasyan\cmsAuthorMark{1}\cmsorcid{0009-0000-0684-6742}
\par}
\cmsinstitute{Institut f\"{u}r Hochenergiephysik, Vienna, Austria}
{\tolerance=6000
W.~Adam\cmsorcid{0000-0001-9099-4341}, J.W.~Andrejkovic, L.~Benato\cmsorcid{0000-0001-5135-7489}, T.~Bergauer\cmsorcid{0000-0002-5786-0293}, M.~Dragicevic\cmsorcid{0000-0003-1967-6783}, C.~Giordano, P.S.~Hussain\cmsorcid{0000-0002-4825-5278}, M.~Jeitler\cmsAuthorMark{2}\cmsorcid{0000-0002-5141-9560}, N.~Krammer\cmsorcid{0000-0002-0548-0985}, A.~Li\cmsorcid{0000-0002-4547-116X}, D.~Liko\cmsorcid{0000-0002-3380-473X}, M.~Matthewman, I.~Mikulec\cmsorcid{0000-0003-0385-2746}, J.~Schieck\cmsAuthorMark{2}\cmsorcid{0000-0002-1058-8093}, R.~Sch\"{o}fbeck\cmsAuthorMark{2}\cmsorcid{0000-0002-2332-8784}, D.~Schwarz\cmsorcid{0000-0002-3821-7331}, M.~Shooshtari\cmsorcid{0009-0004-8882-4887}, M.~Sonawane\cmsorcid{0000-0003-0510-7010}, W.~Waltenberger\cmsorcid{0000-0002-6215-7228}, C.-E.~Wulz\cmsAuthorMark{2}\cmsorcid{0000-0001-9226-5812}
% ...
\par}
\vskip\cmsinstskip
\dag:~Deceased\\
$^{1}$Also at Yerevan State University, Yerevan, Armenia\\
$^{2}$Also at TU Wien, Vienna, Austria\\
% ...
\end{sloppypar}
%%% END EDITABLE REGION %%%
% skeleton_end
\end{document}